\documentclass[format=acmsmall]{acmart}

\AtBeginDocument{%
  \providecommand\BibTeX{{%
    \normalfont B\kern-0.5em{\scshape i\kern-0.25em b}\kern-0.8em\TeX}}}

\setcopyright{acmcopyright}
\copyrightyear{2023}
\acmYear{2023}
\acmDOI{XXXXXXX.XXXXXXX}

\acmJournal{TOSEM}
\acmVolume{0}
\acmNumber{0}
\acmArticle{X}
\acmMonth{9}

\usepackage{amsmath,amsfonts}
\usepackage{algorithm}
\usepackage{algpseudocode}
\usepackage{multirow}
\usepackage{graphicx}
\usepackage{textcomp}
\usepackage{xcolor}
\usepackage{wrapfig}
\usepackage{tcolorbox}

\usepackage{color}
\newcount\Comments  
\definecolor{darkgreen}{rgb}{0,0.5,0}
\definecolor{purple}{rgb}{1,0,1}
\newcommand{\kibitz}[2]{\ifnum\Comments=1\textcolor{#1}{#2}\fi}

\newcommand{\highlight}[1]  {\kibitz{black}   {#1}}

\usepackage[absolute,overlay]{textpos}
\usepackage{tabularx}
\usepackage{subcaption} 
\usepackage{url}
\usepackage{xpatch}
\usepackage{lscape}
\usepackage[misc,geometry]{ifsym} 
\newtheorem{definition}{Definition}

\newtheorem{remark}{Remark}
\usepackage{multirow}

\newcommand*\diff{\mathop{}\!\mathrm{d}}

\DeclareMathOperator*{\argmax}{arg\,max}

\newcommand{\inputdomain}{\mathcal{X}}

\usepackage{glossaries}
\newacronym{AE}{AE}{Adversarial Example}
\newacronym{OP}{OP}{Operational Profile}
\newacronym{RAM}{RAM}{Reliability Assessment Model}
\newacronym{ACU}{ACU}{Average Cell Unastuteness}
\newacronym{DL}{DL}{Deep Learning}
\newacronym{ML}{ML}{Machine Learning}
\newacronym{VAE}{VAE}{Variational Auto-Encoders}
\newacronym{VnV}{V\&V}{Verification and Validation}
\newacronym{KDE}{KDE}{Kernel Density Estimator}
\newacronym{PDF}{PDF}{Probability Density Function}
\newacronym{GA}{GA}{Genetic Algorithm}
\newacronym{FID}{FID}{Fréchet Inception Distance}
\newacronym{MSE}{MSE}{Mean Squared Error}
\newacronym{OOD}{OOD}{Out-Of-Distribution}
\newacronym{HDA}{HDA}{Hierarchical Distribution-Aware}
\newacronym{GAN}{GAN}{Generative Adversarial Networks}
\newacronym{DNN}{DNN}{Deep Neural Network}
\newacronym{FGSM}{FGSM}{Fast Gradient Sign Method}
\newacronym{PGD}{PGD}{Projected Gradient Descent}

\usepackage[absolute,overlay]{textpos}

\DeclareCaptionLabelFormat{andtable}{#1~#2  \&  \tablename~\thetable}    
\begin{document}

\begin{textblock*}{20cm}(1cm,1cm)
\textcolor{red}{Accepted by ACM Transactions on Software Engineering and Methodology (TOSEM)}
\end{textblock*}

\title{Hierarchical Distribution-Aware Testing of Deep Learning}

\author{Wei Huang}
\affiliation{%
  \institution{Purple Mountain Laboratories}
  \country{China,}
  \institution{University of Liverpool}
  \country{U.K.}}
\email{w.huang23@liverpool.ac.uk}

\author{Xingyu Zhao}
\affiliation{%
  \institution{WMG, University of Warwick}
  \country{U.K.}
}
\email{Xingyu.Zhao@warwick.ac.uk}

\author{Alec Banks}
\affiliation{%
  \institution{Defence Science and Technology Laboratory}
  \country{U.K.}}
\email{abanks@dstl.gov.uk}

\author{Victoria Cox}
\affiliation{%
  \institution{Defence Science and Technology Laboratory}
  \country{U.K.}}
\email{vcox@dstl.gov.uk}

\author{Xiaowei Huang}
\affiliation{%
  \institution{University of Liverpool}
  \country{U.K.}}
\email{Xiaowei.Huang@liverpool.ac.uk}

\begin{abstract}
With its growing use in safety/security-critical applications, Deep Learning (DL) has raised increasing concerns regarding its dependability. In particular, DL has a notorious problem of lacking robustness. Input added with adversarial perturbations, i.e. Adversarial Examples (AEs) are easily mis-predicted by the DL model. Despite recent efforts made in detecting AEs via state-of-the-art attack and testing methods, they are normally input distribution agnostic and/or disregard the perceptual quality of adversarial perturbations. Consequently, the detected AEs are irrelevant inputs in the application context or unrealistic that can be easily noticed by humans. This may lead to a limited effect on improving the DL model's dependability, as the testing budget is likely to be wasted on detecting AEs that are encountered very rarely in its real-life operations.

In this paper, we propose a new robustness testing approach for detecting AEs that considers both the feature level distribution and the pixel level distribution, capturing the perceptual quality of adversarial perturbations. The two considerations are encoded by a novel hierarchical mechanism. First, we select test seeds based on the density of feature level distribution and the vulnerability of adversarial robustness. The vulnerability of test seeds are indicated by the auxiliary information, that are highly correlated with local robustness. Given a test seed, we then develop a novel genetic algorithm based local test case generation method, in which two fitness functions work alternatively to control the perceptual quality of detected AEs. Finally, extensive experiments confirm that our holistic approach considering hierarchical distributions is superior to the state-of-the-arts that either disregard any input distribution or only consider a single (non-hierarchical) distribution, in terms of not only detecting imperceptible AEs but also improving the overall robustness of the DL model under testing.

\end{abstract}


\maketitle

\section{Introduction}
\label{sec_intro}

\gls{DL} is being explored to provide transformational capabilities to many industrial sectors including automotive, healthcare and finance. The reality that DL is not as dependable as required now becomes a major impediment. For instance, key industrial foresight reviews identified that the biggest obstacle to gaining benefits of \gls{DL} is its dependability \cite{lane_new_2016}. There is an urgent need to develop methods to enable the dependable use of \gls{DL}, for which great efforts have been made in recent years in the field of DL \gls{VnV} \cite{huang_survey_2020,zhang_machine_2020}.

\gls{DL} \textbf{robustness} is arguably \textit{the} property in the limelight. Informally, robustness requires that the decision of the DL model is invariant against small perturbations on inputs. That is, all inputs in a small input region (e.g., a norm ball defined in some $L_{p}$-norm distance) should share the same prediction label by the DL model. Inside that region, if an input is predicted differently to the given label, then this input is normally called an \textbf{\gls{AE}}. Most \gls{VnV} methods designed for DL robustness are essentially about detecting AEs, e.g., adversarial attack based methods \cite{DBLP:journals/corr/GoodfellowSS14,DBLP:conf/iclr/MadryMSTV18} and coverage-guided testing \cite{du2019deepstellar,xie_npc_2022,ma2018deepgauge,DBLP:conf/sosp/PeiCYJ17,huang2021coverage,DBLP:conf/icse/SunHKSHA19}.

As recently noticed by the software engineering community, emerging studies on systematically evaluating AEs detected by aforementioned state-of-the-arts have two major drawbacks: (i) they do not take the  \textit{input data distribution} into consideration, therefore it is hard to judge whether the identified AEs are meaningful to the DL application \cite{berend_cats_2020,dola_distribution_aware_2021}; (ii) most detected AEs are of \textit{poor perceptual quality} that are too unnatural/unrealistic \cite{harel_canada_is_2020} to be seen in real-life operations. That said, not all AEs are equal nor significantly contribute to the robustness improvement, given limited resources. A wise strategy is to detect those AEs that are both being ``distribution-aware'' and with natural/realistic pixel-level perturbations, which motivates this work.

Prior to this work, a few notable attempts at distribution-aware testing for DL have been made. Broadly speaking, the field has developed two types of approaches: \textit{\gls{OOD} detector} based \cite{dola_distribution_aware_2021,DBLP:conf/icse/Berend21} and \textit{feature-only} based \cite{toledodistribution,byun_manifold_based_2020}. The former can only detect anomalies/outliers, rather than being ``fully-aware'' of the distribution. While the latter is indeed generating new test cases according to the learnt distribution (in a latent space), it ignores the pixel-level information due to the compression nature of generative models used \cite{zhong2020generative}. 
To this end, our approach is advancing in this direction with the following novelties and contributions:

\textit{a)} We provide a ``divide and conquer'' solution---\gls{HDA} testing---by decomposing the input distribution into two levels (named as \textit{global} and \textit{local}) capturing how the feature-wise and pixel-wise information are distributed, respectively. At the global level, isolated problems of estimating the feature distribution and selecting best test seeds can be solved by dedicated techniques. At the local level where features are fixed, the clear objective is to precisely generate test cases considering perceptual quality\footnote{While determining perceptual quality typically involves subjective assessments from human observers, objective metrics such as peak signal-to-noise ratio (PSNR) and structural similarity index (SSIM) can also be used as measures of perceptual quality \cite{DBLP:journals/tip/WangBSS04}.
	Throughout the paper, perceptual quality is defined as the metrics in Sec.~\ref{sec_per_quatlity_image} that are commonly used in computer vision.}. Our extensive experiments show that such \textit{hierarchical} consideration is more effective to detect high-quality AEs than state-of-the-art that either disregards any data distribution or only considers a single (non-hierarchical) distribution. Consequently, we also show the DL model under testing exhibits higher robustness after ``fixing'' the high-quality AEs detected.

\textit{b)} At the global level, we propose novel methods to select test seeds based on the \textit{approximated feature distribution} of the training data and \textit{predictive robustness indicators}, so that the norm balls of the selected seeds are both from the high-density area of the distribution and relatively unrobust (thus more cost-effective to detect AEs in later stages). Notably, state-of-the-art DL testing methods normally select test seeds \textit{randomly} from the training dataset without any principled rules. Thus, from a software engineering perspective, our test seed selection is more practically useful in the given application context.

\textit{c)} Given a carefully selected test seed, we propose a novel two-step \gls{GA} to generate test cases locally (i.e. within a norm ball) to control the \textit{perceptual quality} of detected AEs.
At this local level, the perceptual quality distribution of data-points inside a norm ball requires pixel-level information that cannot be sufficiently obtained from the training data alone. Thus, we innovatively use common perceptual metrics that quantify image quality as an approximation of such local distribution. Our experiments confirm that the proposed GA is not only effective after being integrated into HDA (as a holistic testing framework), but also outperforms other pixel level AE detectors in terms of perception quality when applied separately.

\textit{d)} We investigate black-box (to the DL model under testing) methods for the main tasks at both levels. Thus, to the best of our knowledge, our HDA approach provides an \textit{end-to-end, black-box} solution, which is the first of its kind and more versatile in software engineering practice.

\textit{e)} A publicly accessible tool of our \gls{HDA} testing framework with all source code, datasets, DL models and experimental results.

\section{Preliminaries and Related Work}
\label{sec_preliminaries_related_work}

\begin{figure}[!ht]
	\centering
	\includegraphics[width=0.8\linewidth]{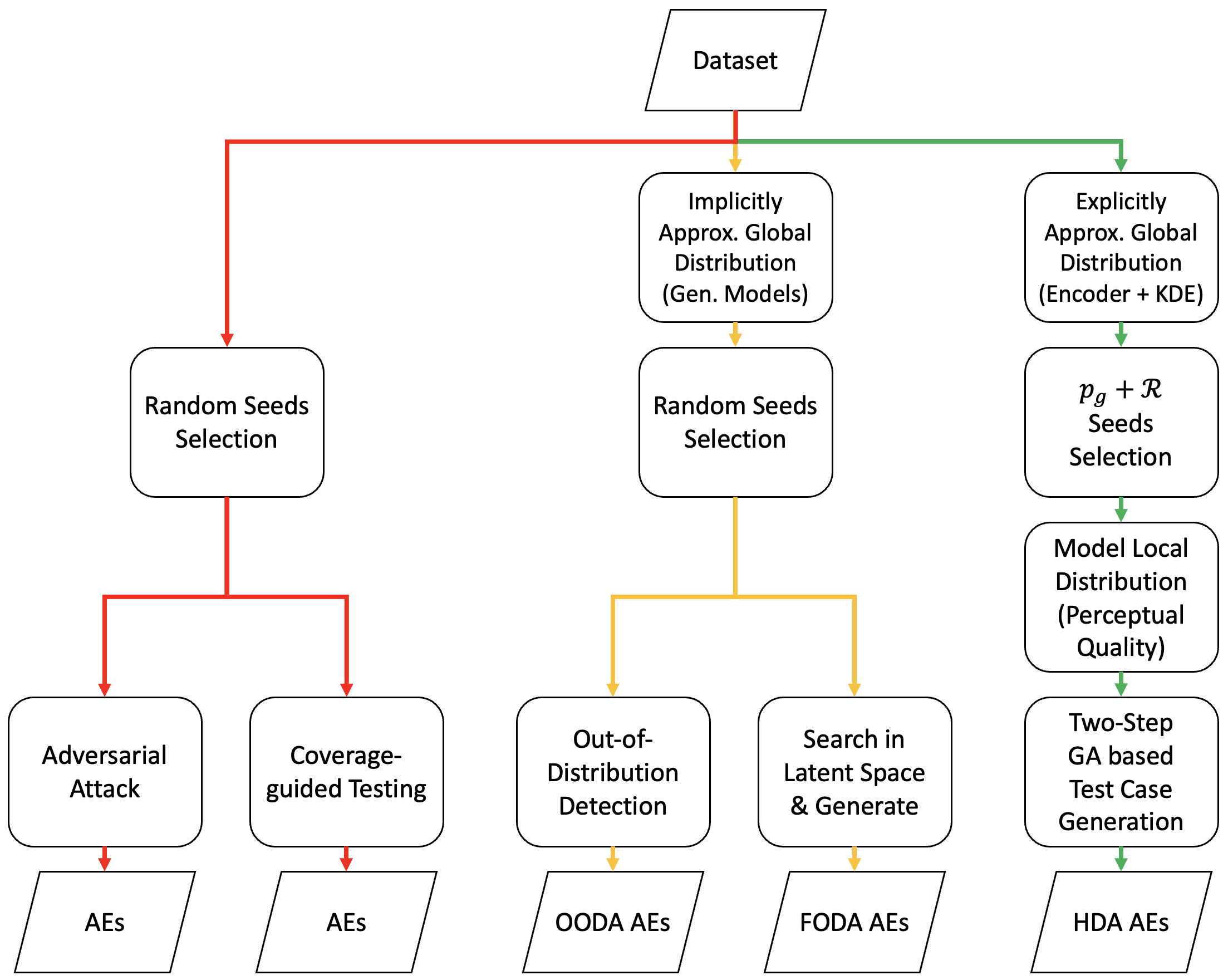}
	\caption{Comparison between our proposed Hierarchical Distribution-Aware (HDA) testing and related works.}
	\label{workflow_op_test}
\end{figure}

In this section, we first introduce preliminaries and related work on DL robustness, together with formal definitions of concepts adopted in our \gls{HDA} approach. Then existing works on distribution-aware testing are discussed. Since our \gls{HDA} testing also considers the naturalness of detected AEs, some common perception quality metrics are introduced. In summary, we present Fig.~\ref{workflow_op_test} to show the stark contrast of our proposed \gls{HDA} testing (the green route) to other related works (the red and amber routes).

\subsection{DL Robustness and Adversarial Examples}
\label{dl_robustness}
We denote the prediction output of DL model as the vector $f(x)$ with size equal to the total number of labels. The predicted label $\hat{f}(x)\!=\!\argmax_{i} f_i(x)$ where $f_i(x)$ is the $i^{th}$ attribute of vector $f(x)$.

DL robustness requires that the decision of the DL model $\hat{f}(x)$ is invariant against small perturbations on input $x$. That is, all inputs in an input region $\eta$ have the same prediction label, where $\eta$ is usually a small norm ball (defined with some $L_{p}$-norm distance\footnote{$p=0, 1 , 2$ and $\infty$. $L_\infty$ norm is more commonly used.}) around an input $x$. If an input $x'$ inside $\eta$ is predicted differently to $x$ by the DL model, then $x'$ is called an \textit{Adversarial Example} (AE).

DL robustness \gls{VnV} can be based on formal methods \cite{huang_safety_2017,ruan_reachability_2018} or statistical approaches \cite{DBLP:conf/iclr/WengZCYSGHD18,webb_statistical_2019}, and normally aims at detecting AEs. In general, we may classify two types of methods (the two branches in the red route of Fig.~\ref{workflow_op_test}) depends on how the test cases are generated: (i) Adversarial attack based methods normally optimise the DL prediction loss to find AEs, which include white-box attack methods like \gls{FGSM} \cite{DBLP:journals/corr/GoodfellowSS14} and \gls{PGD} \cite{DBLP:conf/iclr/MadryMSTV18}, as well as black-box attacks \cite{alzantot2019genattack,DBLP:conf/cec/WuLZXZ21} using \gls{GA} with gradient-free optimisation.
(ii) Coverage-guided testing optimises the certain coverage metrics on the DL model's internal structure, which is inspired by the coverage testing for traditional software. Several popular test metrics, like neuron coverage \cite{DBLP:conf/sosp/PeiCYJ17,ma2018deepgauge}, modified condition/decision coverage \cite{DBLP:conf/icse/SunHKSHA19} for CNNs and temporal coverage \cite{huang2021coverage,du2019deepstellar} for RNNs are proposed. While it is argued that coverage metrics are not strongly correlated with DL robustness \cite{yan_correlations_2020,harel_canada_is_2020}, they are seen as providing insights into the internal behaviours of DL models and hence may guide test selection to find more diverse AEs \cite{huang2021coverage}.

Without loss of generality, we reuse the formal definition of DL robustness in \cite{webb_statistical_2019,weng_proven_2019} in this work:
\begin{definition}[Local Robustness]
	\label{def_lr}
	The local robustness of the DL model $f(x)$, 
	w.r.t. a local region $\eta$ and  a target label $y$, is:
	\begin{equation}
	\label{eq_robust_def}
	\mathcal{R}_l(\eta, y):=\int_{x \in \eta} I(x) p_l(x\mid x \in \eta) \diff{x}
	\end{equation}
	where $p_l(x \mid x\in\eta)$ is the local distribution of region $\eta$ which is precisely the ``input model'' used by both \cite{webb_statistical_2019,weng_proven_2019}. $I(x)$ is an indicator function, and $I(x)=1$ when $\hat{f}(x)=y$, $I(x)=0$ otherwise.
\end{definition}

To detect as many AEs as possible, normally the first question is---which local region shall we search for those AEs? I.e. how to select test seeds?
To be cost-effective, we want to explore unrobust regions, rather than regions where AEs are relatively rare. This requires the local robustness of a region to be known \textit{a priori},
which may imply a paradox (cf. Remark \ref{rmk_paradox} later). In this regard, we can only \textit{predict} the local robustness of some regions before doing the actual testing in those regions. We define:
\begin{definition}[Local Robustness Indicator]
	\label{def_robust_indicator}
	\textbf{Auxiliary information} that strongly \textbf{correlated} with $\mathcal{R}_l(\eta, y)$ (thus can be leveraged in its prediction) is named as a local robustness indicator.
\end{definition}
\noindent We later seek for such indicators (and empirically show their correlation with the local robustness), which forms one of the two key factors considered in selecting test seeds in our method. 

Given a test seed, we search for AEs in a local region $\eta$ (around the test seed) that produce different label from the test seed. This involves the question on what size of $\eta$ should be, for which we later utilise the property of:
\begin{remark}[$r$-separation]
	\label{remark_r_sep}
	For real-world image datasets, any data-points with different ground truth labels are at least distance $2r$ apart in the input (pixel) space, where $r$ is estimated case by case and depending on the dataset.
\end{remark}
\noindent The $r$-separation property was first observed by \cite{yang_closer_2020}: 
intuitively it says, 
there is a minimum distance between two real-world objects of different labels.

Finally, not all AEs are equal in terms of the ``strength of being adversarial'' (stronger AEs may lead to greater robustness improvement in, e.g., adversarial training \cite{DBLP:conf/icml/WangM0YZG19}), for which we define:
\begin{definition}[Prediction Loss]
	\label{def_pred_loss}
	Given a test seed $x$ with label $y$, the prediction loss of an input $x'$, which is obtained by adding perturbations to $x$, is defined as:
	\begin{equation}
	\label{eq_pred_loss}
	\mathcal{J}(f(x'),y) = \max_{i \neq y} (f_i(x') - f_y(x'))
	\end{equation}
	where $f_i(x')$ returns the probability of label $i$ after input $x'$ being processed by the DL model $f$.
\end{definition}
\noindent Note, $\mathcal{J} \geq 0$ implies $\argmax_i f_i(x) \neq y$ and thus $x'$ is an AE of $x$.

Next, to measure the DL models' \textit{overall} robustness across the whole input domain, we introduce a notion of global robustness. Being different to some existing definitions where robustness of local regions are treated equally when calculating global robustness over several regions \cite{wang2021robot,wang2021statistically}, ours is essentially a ``weighted sum'' of the robustness of local regions where each weight is the probability of the associated region on the input data distribution. Defining global robustness in such a ``distribution-aware'' manner aligns with our motivation---as revealed later by empirically estimated global robustness, our \gls{HDA} appears to be more effective in supporting the growth of the overall robustness after ``fixing'' those distribution-aware AEs.
\begin{definition}[Global Robustness]
	\label{def_gr}
	The global robustness of the DL model $f(x)$ is defined as:
	\begin{equation}
	\label{eq_g_robust_def}
	\mathcal{R}_g:=\sum_{\eta \in \inputdomain}  p_g(x\in\eta) \mathcal{R}_l(\eta, y)
	\end{equation}
	where $p_g(x \mid x\in\eta)$ is the global distribution of region $\eta$ (i.e., a pooled probability of all inputs in the region $\eta_z$) and $\mathcal{R}_l(\eta, y)$ is the local robustness of region $\eta$ to the label $y$.
\end{definition}
The estimation of $\mathcal{R}_g$, unfortunately, is very expensive that requires to compute the local robustness $\mathcal{R}_l$ of a large number of regions over the input domain $\inputdomain$. Thus, from a practical standpoint, we adopt an empirical definition of the global robustness in our later experiments, which has been commonly used for DL robustness evaluation in the adversarial training \cite{wang_robot_2021,DBLP:conf/iclr/MadryMSTV18,DBLP:conf/icml/WangM0YZG19,zhang2019theoretically}.
\begin{definition}[Empirical Global Robustness]
	\label{def_emp_gr}
	Given a DL model $f$ and a validation dataset $D_v$, we define the empirical global robustness as $\hat{\mathcal{R}_g}: (f,D_v,T) \rightarrow [0,1]$ where \textit{T} denotes a given type of AE detection method and $\hat{\mathcal{R}_g}$ is the weighted accuracy on AEs obtained by conducting \textit{T} on $\langle f,D_v\rangle$.
\end{definition}
To be ``distribution-aware'', the synthesis of $D_v$ should conform to the global distribution. \highlight{$D_v$ can be sampled from the train/test data according to global distribution. The norm ball around each input data $x$ in $D_v$ represents a region $\eta$. Each region $\eta$ is explicitly assigned a weight to indicate the density on global distribution. For each region $\eta$, we calculate the prediction accuracy on AEs, detect by \textit{T} according to local distribution, to approximate the local robustness $\mathcal{R}_l$. Consequently, the set of AEs for $D_v$ may represent the input distribution and the weighted accuracy on these AEs approximate the global robustness.}


\subsection{Distribution-Aware Testing for DL}
\label{sec_ooda_related_work}
There are increasing amount of DL testing works developed towards being distribution-aware (as summarised in the amber route of Fig.~\ref{workflow_op_test}).
Deep generative models, such as \gls{VAE} and \gls{GAN}, are applied to approximate the training data distribution, since the inputs (like images) to \gls{DNN} are usually in a high dimensional space. Previous works heavily rely on \gls{OOD} detection \cite{dola_distribution_aware_2021,DBLP:conf/icse/Berend21} or synthesising new test cases directly from latent spaces \cite{DBLP:conf/icse/ByunR20a,toledodistribution,dunn2020evaluating,kang2020sinvad,dunn2021exposing}. The former does not comprehensively consider the whole distribution, rather flags outliers, thus a more pertinent name of it should be \textit{out-of-distribution-aware} (OODA) testing. While for both types of methods, another problem arises that the distribution encoded by generative models only contain the \textit{feature-wise} information and easily filter out the \textit{pixel-wise} perturbations \cite{zhong2020generative}. The images added with \textit{pixel-wise} perturbations should still fall into the local region $\eta$, which is decided by $r$-separation property (ref. Remark \ref{remark_r_sep}). \highlight{Although, Dunn et al. \cite{dunn2020evaluating} propose to generate the fine-grained perturbations by perturbing the output of last layers of GAN's generator, the fine-grained perturbations cannot be guaranteed to follow $r$-separation property and thus may change the class of the image.} Consequently, directly searching and generating test cases from the latent space of generative models may \textit{only perturb features}, thus called \textit{Feature-Only Distribution-Aware} (FODA) in this paper (while also named as \textit{semantic} AEs in some literature \cite{DBLP:conf/cvpr/HosseiniP18,DBLP:conf/iclr/ZhaoDS18}). As an example in later experiment results, i.e. Table~\ref{distribution_aware_compare}, FODA will produce AEs, the perturbations of which exceed the $r$-separation limit.
\highlight{Our approach, the green route in Fig.~\ref{workflow_op_test}, differs from aforementioned works by 1) considering both the global (feature level) distribution in latent spaces and the local (pixel level) perceptual quality distribution in the input space; 2) leveraging the robustness indicator $\mathcal{R}$ to select test seeds which is more error-prone and thus easier for detecting AEs; 3) proposing novel GA based test case generation to detect AEs with high perceptual quality. Therefore, when HDA is experimentally compared to the existing works, the selected test seeds of HDA have higher probability density and lower robustness. The detected AEs by HDA also have higher perceptual quality (in terms of those metrics encoded by GA) than others.}

\subsection{Test Input Prioritisation and Generation}
When testing DL based systems, in order to save computation and reduce the cost on labelling data, test input prioritisation strategies are adopted. \cite{weiss2022simple} experimentally confirms that DeepGini outperforms various types of surprise and neuron coverage metrics in terms of the capability to detect misclassifications. \cite{attaoui2022black} propose the unsafe set selection algorithm, leveraging the density-based clustering of error-inducing images. Both test selection methods aim at detecting misclassified images for retraining and improving generalisation performance of DNNs.

There are also some test input generation methods, the main idea of which is to promote diversity of test cases in order to cover more faults of DL systems. \cite{riccio2020model} develops a search-based tool to generate frontier inputs for DL systems. DEEPMETIS \cite{riccio2021deepmetis} augment the test set for DL by increasing the mutation scores.

The above test input generation/prioritisation methods have different goals from our HDA testing. They focus on detecting or generating more misclassified test cases within certain test budget, while our HDA testing targets at generating on-distribution AEs to improve operational robustness of DL systems. Feature distribution learnt from VAE as well as the predicted robustness indicator facilitate the generation of those on-distribution AEs.

\subsection{Perceptual Quality of Images}
\label{sec_per_quatlity_image}
\highlight{Locally, data-points (around the selected seed) sharing the same feature information may exhibit perceptual difference from the selected seed. To capture such distribution, some \textit{perceptual quality} metric can be utilised to compare the perceptual difference between the original image $x$ and perturbed image $x'$. Some \textit{common} metrics for perceptual quality include:
	\begin{itemize}
		\item \textbf{\gls{MSE}}: measures the mean squared difference between the original image $x$ and perturbed image $x'$,
		\begin{equation}
		MSE = \frac{1}{n}\sum_{i=1}^n(x_i-x_i')^2
		\end{equation}
		where $n$ is the image size, and $x_i$ is the value of image pixel $i$.
		\item \textbf{Peak Signal-to-Noise Ratio (PSNR)} \cite{rafael1992gonzalez}: measures the quality of a signal's representation after being subject to noise during transmission or processing. PSNR is expressed in decibels (dB) and is calculated by comparing the maximum possible power of the original signal to the power of the noise that affects its fidelity,
		\begin{equation}
		PSNR = 20*log_{10}\frac{MAX}{\sqrt{MSE}}
		\end{equation}
		where $MAX$ is the maximum possible pixel value of the image. When the pixels are represented using 8 bits per sample, this is 255.
		\item \textbf{Structural Similarity Index Measure (SSIM)} \cite{DBLP:journals/tip/WangBSS04}: measures the similarity between two images based on the idea that the human visual system is highly sensitive to changes in structural information, such as edges, textures, and patterns, rather than just changes in pixel values. It works by comparing the luminance (l), contrast (c), and structural information (s) of two images, and producing a score between 0 (completely dissimilar) and 1 (identical) that indicates their similarity,
		\begin{equation}
		\begin{split}
		SSIM(x, x') =& l(x,x')^{\alpha}\cdot c(x,x')^{\beta}\cdot s(x,x')^{\gamma} \\
		l(x,x') = \frac{2\mu_x \mu_{x'}+c_1}{\mu_x^2+\mu_{x'}^2+c_1}, \; c(x,x') &= \frac{2\sigma_x \sigma_{x'}+c_2}{\sigma_x^2+\sigma_{x'}^2+c_2}, \; s(x,x') = \frac{\sigma_{xx'}+c_3}{\sigma_x \sigma_{x'}+c_3}
		\end{split}
		\end{equation}
		where $\mu_x$ and $\mu_{x'}$ are the mean values of $x$ and $x'$, respectively. $\sigma_x$ and $\sigma_{x'}$ are the standard deviations of $x$ and $x'$, respectively. $\sigma_{xx'}$ is the covariance between $x$ and $x'$. $c_1$, $c_2$, and $c_3$ are constants that prevent division by zero. The constants $\alpha$, $\beta$, and $\gamma$ are typically set to 1, 1, and 1, respectively, although different values may be used depending on the application. Unlike other similarity measures, such as MSE and PSNR, SSIM is able to account for perceptual differences in image quality, and is often considered a more accurate measure of image quality.
		\item \textbf{\gls{FID}} \cite{DBLP:conf/nips/HeuselRUNH17}: compares the distribution between a set of original images and a set of perturbed images. It is calculated by first using a pre-trained Inception-v3 neural network \cite{szegedy2016rethinking} to extract features from both sets of images. Then, the mean and covariance of these features are calculated for each set, and the distance between these mean and covariance matrices is calculated using the Fréchet distance,
		\begin{equation}
		FID = \|\mu - \mu'\|_2^2 + Tr(\Sigma + \Sigma' - 2(\Sigma^{1/2} \cdot \Sigma' \cdot \Sigma^{1/2})^{1/2})
		\end{equation}
		where $\mu$ and $\mu'$ are the mean feature vectors of the original and perturbed image sets, respectively. $\Sigma$ and $\Sigma'$ are the covariance matrices of the original and perturbed image sets, respectively. $Tr()$ denotes the trace of a matrix. Lower FID scores indicate that the perturbed images are closer in terms of perception to the original images, while higher FID scores indicate greater differences between the perturbed and original images.
	\end{itemize}
}

Notably, all these metrics are current standards for assessing the quality of images, as widely used in the experiments of aforementioned related works.



\section{The Proposed Method}
\label{sec_test_method}

\begin{figure*}[!ht]
	\centering
	\includegraphics[width=\linewidth]{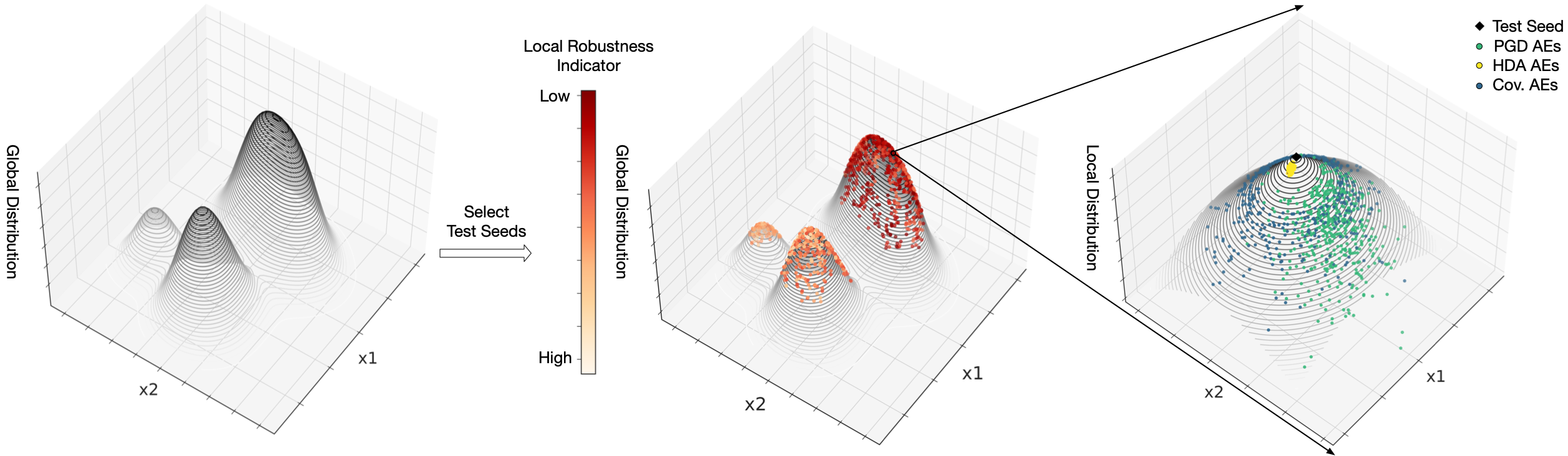}
	\caption{An example of Hierarchical Distribution Aware Testing}
	\label{feature_pixel_op}
\end{figure*}

We first present an overview of our HDA testing, cf. the green route in Fig.~\ref{workflow_op_test}, and then dive into details of how we implement each stage by referring to an illustrative example in Fig.~\ref{feature_pixel_op}.

\subsection{Overview of HDA Testing}
The core of HDA testing is the hierarchical structure of two distributions. We formally define the following two levels of distributions: 

\begin{definition}[Global Distribution]
	\label{def_opg}
	The global distribution captures how \textbf{feature} level information is distributed in some (low-dimensional) latent space after data compression. 
\end{definition}

\begin{definition}[Local Distribution]
	\label{def_opl}
	Given a data-point sampled from the latent space, we consider its norm ball in the input \textbf{pixel} space. The local distribution is a conditional distribution capturing the perceptual quality of all data-points within the norm ball.
\end{definition}
\highlight{Due to the sparsity of data over the high dimensional input space, it is hard to estimate the input distribution. Therefore, we turn to estimate the feature level (global) distribution in latent space, which is the representation of compressed data, in which data points with similar features are closer to each other \cite{liu2019latent}. DNNs, e.g. encoder of VAEs, map any data points in the high dimensional input space to the low dimensional latent space. It infers that the input space can be partitioned into well-defined regions, where each region corresponds to a particular data point in the latent space. latent space is an abstract, multidimensional space where each dimension represents a feature or characteristic of the data \cite{liu2019latent}. The regions in the input space are determined based on the values of these features and how they relate to each other. In other words, we can identify distinct patterns or clusters of data in the input space, which share the same set of features, and maps them to corresponding point in the latent space.} By fitting a global distribution in the latent space, we actually model the distribution of distinct regions over the input space. The local distribution is defined as a conditional distribution within each region. Thus, we propose the following remark.
\begin{remark}[Decompose one distribution into two levels]
	\label{remark_op_decompose}
	Given the definitions of global and local distributions, denoted as $p_g$ and $p_l$ respectively, we may decompose a single distribution over the entire input domain $\inputdomain$ as:
	\begin{equation}
	\label{eq_op_opg_opl}
	p(x)=\int p_l(x|x \in \eta_z)p_g(x \in \eta_z) \diff{z}
	\end{equation}
	where variable $z$ represents a set of features while $\eta_z$ represents a region in the input space that ``maps'' to the $z$ point in the latent space.
\end{remark}
Intuitively, compared to modelling a single distribution, our hierarchical structure of distributions is superior in that the global distribution guides for which regions of the input space to test, while the local distribution can be leveraged to precisely control the quality of test cases. Given the definition of distribution, the goal of HDA testing can be formalised as below:
\begin{remark}[Goal of HDA testing]
	The goal of HDA testing is to detect AE $x'$, which is around the local region of test seed $x$, such that $x'$ lies within the high probability density region of distribution $p(x')$, and $x$ and $x'$ are classified differently.
\end{remark}
To detect such on distribution AEs, HDA testing has the following process, which can divided into three stages, depicted as the green route in Fig.~\ref{workflow_op_test}:

\paragraph{Stage 1: Explicitly Approximate the Global Distribution}
We first extract \textit{feature-level} information from the given dataset by using data compression techniques---the encoder of VAEs in our case, and then explicitly approximate the global distribution in the latent-feature space, using \gls{KDE}. 


\paragraph{Stage 2: Select Test Seeds Based on the Global Distribution and Local Robustness Indicators}
Given the limited testing budget, we want to test in those local input regions that are both more error-prone and representative of the input distribution. Thus, when selecting test seeds, we consider two factors---the local robustness indicators (cf. Definition \ref{def_robust_indicator}) and the global distribution. For the former, we propose several auxiliary information with empirical studies showing their correlation with the local robustness, while the latter has already been quantified in the first stage via \gls{KDE}.

\paragraph{Stage 3: Generate Test Cases Around Test Seeds Considering the Local Distribution and Prediction Loss of AEs}
When searching for AEs locally around a test seed given by the 2nd stage, we develop a two-step \gls{GA} in which the objective function is defined as a \textit{fusion} of the prediction loss (cf. Definition \ref{def_pred_loss}) and the local distribution (modelled by common perceptual quality metrics). Such fusion of two fitness functions allows the trade-off between the ``strength of being adversarial'' and the perceptual quality of the detected AEs. The optimisation is subject to the constraint of only exploring in a norm ball whose central point is the test seed and with a radius smaller than the $r$-separation distance (cf. Remark \ref{remark_r_sep}).

\highlight{While there are some alternatives may also suffice for the purpose of each stage, our chosen technical solutions are the most effective and popular in consideration of the distribution awareness.}

\subsection{Approximation of the Global Distribution}
\label{p_g_cal}
Given the training dataset $\mathcal{D}$, the task of approximating the input distribution is equivalent to estimating a \gls{PDF} over the input domain $\inputdomain$ given $\mathcal{D}$. Despite this is a common problem with many established solutions, it is hard to accurately approximate the distribution due to the relatively sparse data of $\mathcal{D}$, compared to the high dimensionality of the input domain $\inputdomain$. So the practical solution is to do dimensionality reduction and then estimate the global distribution, which indeed is the first step of all existing methods of distribution-aware DL testing.


Specifically, we choose \gls{VAE}-Encoder+\gls{KDE}\footnote{We only use the encoder of VAEs for feature extraction, rather than generate new data from the decoder, which is different to other methods mentioned in Section \ref{sec_ooda_related_work}.} for their simplicity and popularity. To effectively train the \gls{VAE} model, we use a combination of a reconstruction loss and a KL divergence loss to optimise the model. The reconstruction loss measures the difference between the input data and the output of the decoder, while the KL divergence loss measures the difference between the learned latent distribution and the prior distribution. During training, the reconstruction loss and KL divergence loss are calculated and minimised simultaneously. Regularisation techniques such as dropout and weight decay can be used to prevent overfitting.

Assume $\mathcal{D}$ contains $n$ samples and each $x_i \in \mathcal{D}$ is encoded by VAE-Encoder as a Gaussian distribution $z_i$ in the latent space, we can estimate the \gls{PDF} of $z$ (denoted as $Pr(z)$) based on the encoded $\mathcal{D}$. The $Pr(z)$ conforms to the mixture of Gaussian distributions, i.e., $z \sim \mathcal{N}(\mu_{z_i},\sigma_{z_i})$. Notably, this mixture of Gaussian distributions nicely aligns with the gist of adaptive \gls{KDE} \cite{DBLP:conf/snn/LokerseVB95}, which uses the following estimator:
\begin{equation}
\label{eq_kde}
p_g(x\in \eta_z) \propto Pr(z) \simeq \frac{1}{n} \sum_{i = 1}^n K_{h_i} (z-\mu_{z_i}) 
\end{equation}
That is, when choosing a Gaussian kernel for $K$ in Eqn.~\eqref{eq_kde} and adaptively setting the bandwidth parameter $h_i = \sigma_{z_i}$ (i.e., the standard deviation of the Gaussian distribution representing the compressed sample $z_i$), the VAE-Encoder and KDE are combined ``seamlessly''. Finally, our global distribution $p_g(x\in \eta_z)$ (a pooled probability of all inputs in the region $\eta_z$ that corresponds to a point $z$ in the latent space) is proportional to the approximated distribution of $z$ with the PDF $Pr(z)$.

\textit{\textbf{Running Example}}: The left diagram in Fig.~\ref{feature_pixel_op} depicts the global distribution learnt by KDE, after projected to a two-dimensional  space for visualisation. The peaks\footnote{Most training data lie in this region or gather around the region.} are evaluated with highest probability density over the latent space by KDE.



\subsection{Test Seeds Selection}

Selecting test seeds is actually about choosing which norm balls (around the test seeds) to test for AEs. To be cost-effective, we want to test those with higher global probabilities and lower local robustness at the same time.
For the latter requirement, there is potentially a paradox:
\begin{remark}[A Paradox of Selecting Unrobust Norm Balls]
	\label{rmk_paradox}
	To be informative on which norm balls to test for AEs, we need to estimate the local robustness of candidate norm balls (by invoking robustness estimators to quantify $\mathcal{R}_l(\eta, y)$, e.g., \cite{webb_statistical_2019}). However, local robustness evaluation itself is usually about sampling for AEs (then fed into statistical estimators) that consumes the testing resources.
\end{remark}
To this end, instead of \textit{directly evaluating} the local robustness of a norm ball, we can only \textit{indirectly predict} it (i.e., without testing/searching for AEs) via auxiliary information that we call local robustness indicators (cf. Definition \ref{def_robust_indicator}). In doing so, we save all the testing budget for the later stage when generating local test cases.

Given a test seed $x$  with label $y$, we propose two robustness indicators (both relate to the vulnerability of the test seed to adversarial attacks)---the \textit{prediction gradient based score} (denoted as ${S}_{\textit{grad}}$) and the 
\textit{score based on separation distance of the output-layer activation}
(denoted as $S_{\textit{sep}}$):
\begin{equation}\label{equ:robustness_indicator}
\begin{split}
&S_{\textit{grad}} = ||\nabla_x \mathcal{J}(f(x),y)||_{\infty} \\
&S_{\textit{sep}} = \min_{\hat{x}}||f(x)-f(\hat{x})||_{\infty}\;  \quad
\textrm{s.t.}\; y \neq \hat{y}
\end{split}
\end{equation}

These allow prediction of a whole norm ball's local robustness by the limited information of its central point (the test seed). The gradient of a \gls{DNN}'s prediction with respect to the input is a white-box metric, that is widely used in adversarial attacks, such as \gls{FGSM} \cite{DBLP:journals/corr/GoodfellowSS14} and \gls{PGD} \cite{DBLP:conf/iclr/MadryMSTV18} attacks. A greater gradient calculated at a test seed implies that AEs are more likely to be found around it. The activation separation distance is regarded as a black-box metric and refers to the minimum $L_\infty$ norm between the output activations of the test seed and any other data with different labels. Intuitively, a smaller separation distance implies a greater vulnerability of the seed to adversarial attacks. We later show empirically that indeed these two indicators are highly correlated with the local robustness.

After quantifying the two required factors, we combine them in a way that was inspired by \cite{zhao_assessing_2021}. In \cite{zhao_assessing_2021}, the DL reliability metric is formalised as a weighted sum of local robustness where the weights are operational probabilities of local regions. To align with that reliability metric, we do the following \textbf{steps to select test seeds}: 

(i) For each data-point $x_i$ in the test set, we calculate its global probability (i.e., $Pr(z_i)$ where $z_i$ is its compressed point in the VAE latent space) and one of the local robustness indicators (either white-box or black-box, depending on the available information).

(ii) Normalise both quantities to the same scale. 

(iii) Rank all data-points by the product of their global probability and local robustness indicator. 

(iv) Finally we select top-$k$ data-points as our test seeds, and $k$ depends on the testing budget.

\textit{\textbf{Running Example}}: In the middle diagram of Fig.~\ref{feature_pixel_op}, we add in the local robustness indicator results of the training data which are represented by a scale of colours---darker means lower predicted local robustness while lighter means higher predicated local robustness. By our method, test seeds selected are both from the highest peak (high probability density area of the global distribution) and relatively darker ones (lower predicated local robustness).

\subsection{Local Test Cases Generation}
\label{sec:test_case_generation}

Not all AEs are equal in terms of the ``strength of being adversarial'', and stronger AEs are associated with higher prediction loss (cf. Definition \ref{def_pred_loss}). Detecting AEs with higher prediction loss may benefit more when considering the future ``debugging'' step, e.g., by adversarial retraining \cite{DBLP:conf/icml/WangM0YZG19}. Thus, at this stage, we want to search for AEs that exhibit a high degree of adversarial behaviour while also conform to the local distribution. That is, the local test case generation can be formulated as the following optimisation given a seed $(x,y)$:

\begin{equation}
\begin{aligned}
\max_{x'} & \mathcal{J}(f(x'),y) + \alpha \cdot p_l(x'|x' \in \eta_{z_x})
\\
\textrm{s.t.} \, & ||x-x'||_{\infty}  \leq r
\end{aligned}
\label{equ_obj}
\end{equation}
where $\mathcal{J}$ is the prediction loss, $p_l(x'|x' \in \eta_{z_x})$ is the local distribution (note, $z_x$ represents the latent features of test seed $x$), $r$ is the $r$-separation distance, and $\alpha$ is a coefficient to balance the two terms. As what follows, we note two points on Eqn.~\eqref{equ_obj}: why we need the constraint and how we quantify the local distribution.

The constraint in Eqn.~\eqref{equ_obj} determines the right \textit{locality} of local robustness---the ``neighbours'' that should have the same ground truth label $y$ as the test seed. We notice the $r$-separation property of real-world image datasets (cf. Remark \ref{remark_r_sep}) provides a sound basis to the question. Thus, it is formalised as a constraint that the optimiser can only search in a norm ball with a radius smaller than $r$, to guarantee the detected AEs are indeed ``adversarial'' to label $y$.


While the feature level information is captured by the global distribution over a latent space, we only consider how the pixel level information is locally distributed in terms of \textit{perceptual quality}. Three common quantitative metrics---MSE, PSNR and SSIM introduced in Section \ref{sec_per_quatlity_image}---are investigated. We note, those three metrics by no means are the true local distribution representing perceptual quality, rather quantifiable indicators from different aspects. Thus, in the optimisation problem of Eqn.~\eqref{equ_obj}, replacing the local distribution term with them would suffice our purpose. So, we redefine the optimisation problem as:
\begin{equation}
\begin{aligned}
\max_{x'} \mathcal{J}(f(x'),y) \!+\! \alpha\! \cdot\! \mathcal{L}(x,x'), \quad
\textrm{s.t.}\, ||x\!-\!x'||_{\infty} \! \leq\! r
\end{aligned}
\label{equ_obj_redef}
\end{equation}
where $\mathcal{L}(x,x')$ represents those perceptual quality metrics correlated with the local distribution of the seed $x$.
Certainly, implementing $\mathcal{L}(x,x')$ requires some prepossessing, e.g., normalisation and negation, depending on which metric is adopted.

Considering that the second term of the objective function in Eqn.~\eqref{equ_obj_redef} may not be differentiable and/or the DL model's parameters are not always accessible, Non-dominated Sorting Genetic Algorithm II (NSGA-II) \cite{deb2002fast} may be adopted here to solve the multi-objective optimisation. NSGA-II is designed to search for a diverse set of solutions along the Pareto-optimal front, rather than just a single solution. This means that NSGA-II produces a set of solutions that span the entire Pareto-optimal front, providing decision-makers with a range of options to choose from. Therefore, NSGA-II is usually computationally intensive and requires a huge amount of queries of DL model's prediction. Since we are only concern about test cases which have high perceptual quality and conditioned on AEs (prediction loss greater than 0), it is not necessary to produce a range of options for the trades-off between perceptual quality and prediction loss, which may waste a lot of computational resource on generating high perceptual quality but no adversarial examples. For this reason, we propose to scalarize the vector of objectives into one objective by averaging the objectives with weight vector and reduce the weight dependency with alternation mechanism. That is, we develop a \gls{GA} with two fitness functions to effectively and efficiently detect AEs, as shown in Algorithm~\ref{alg:GA}. 

\begin{algorithm}
	\caption{Two-Step GA Based Local Test Cases Generation}\label{alg:GA}
	\begin{algorithmic}[1]
		\Require Test seed $(x, y)$, neural network function $f(x)$, local perceptual quality metric $\mathcal{L}(x,x')$, population size $N$, maximum iterations $T$, norm ball radius $r$, weight parameter $\alpha$, number of generated test cases $m$.
		\Ensure A set of $m$ test cases $\mathcal{T}$ 
		\State $F_1 = \mathcal{J}(f(x'),y)$, $F_2 = \mathcal{J}(f(x'),y) + \alpha \cdot \mathcal{L}(x,x')$
		\For{$i = 1,..., N$} 
		\State $\mathcal{T}[i] = x + \mathit{uniform}(-r, +r)$
		\EndFor
		\While{$t < T$ or $\max(\mathit{fit\_list_2})$ does not converge}
		\State $\mathit{fit\_list_1} = \mathit{cal\_fitness}(F_1,\mathcal{T})$
		\State $\mathit{fit\_list_2} = \mathit{cal\_fitness}(F_2,\mathcal{T})$
		\If{$\mathit{majority}(\mathit{fit\_list_1}<0)$}
		\State $\mathit{parents} = \mathit{selection}(\mathit{fit\_list_1},\mathcal{T})$
		\Else
		\State $\mathit{parents} = \mathit{selection}(\mathit{fit\_list_2},\mathcal{T})$
		\EndIf
		\State $\mathcal{T} = \mathit{crossover}(\mathit{parents}, N)$
		\State $\mathcal{T} = \mathit{mutation}(\mathcal{T}) \cup \mathit{parents}$
		\State $t = t + 1$
		\EndWhile  
		\State $\mathit{fit\_list_2} = \mathit{cal\_fitness}(F_2,\mathcal{T})$
		\State $idx = \argmax(fit\_list_2)[:m]$ 
		\State $\mathcal{T} = \mathcal{T}[idx]$ \\
		\Return test set $\mathcal{T}$
	\end{algorithmic}
\end{algorithm}

\highlight{Algorithm~\ref{alg:GA} presents the \textbf{process of generating a set of $m$ test cases} $\mathcal{T}$ from a given seed $x$ with label $y$ (denoted as $(x,y)$). At line 1, we define two fitness functions (the reason behind it will be discussed later). GA based test case generation consists of 4 steps: initialisation, selection, crossover, and mutation, the last three of which are repeated until the convergence of fitness values or the maximum iterations are reached.
	\paragraph{Initialisation} The initialisation of population is crucial to the quick convergence. Diversity of initial population could promise approximate global optimal \cite{konak2006multi}. We initialise the population by adding uniform noise in range $(-r,+r)$ to the test seed, at line 2-4.
	\paragraph{Selection} The fitness function is defined to select fitted individuals as parents for the latter operations. We use the fitness proportionate selection \cite{lipowski2012roulette} for operator $selection()$.
	\begin{equation}
	p_i = \frac{\mathcal{F}_i}{\sum_{i=1}^{n} \mathcal{F}_i}, \; \mathcal{F}_i \in fit\_list
	\end{equation}
	The fitness value is used to associate a probability of selection $p_i$ for each individuals to maintaining good diversity of population and avoid premature convergence. The fitness function is the objective function to be optimised. At line 6-12, we calculate two defined fitness function values on the population and select individuals based on one of the fitness values according to some judgement (the reason behind it will be discussed later).
	\begin{figure}[!htb]
		\centering
		\includegraphics[width=0.7\linewidth]{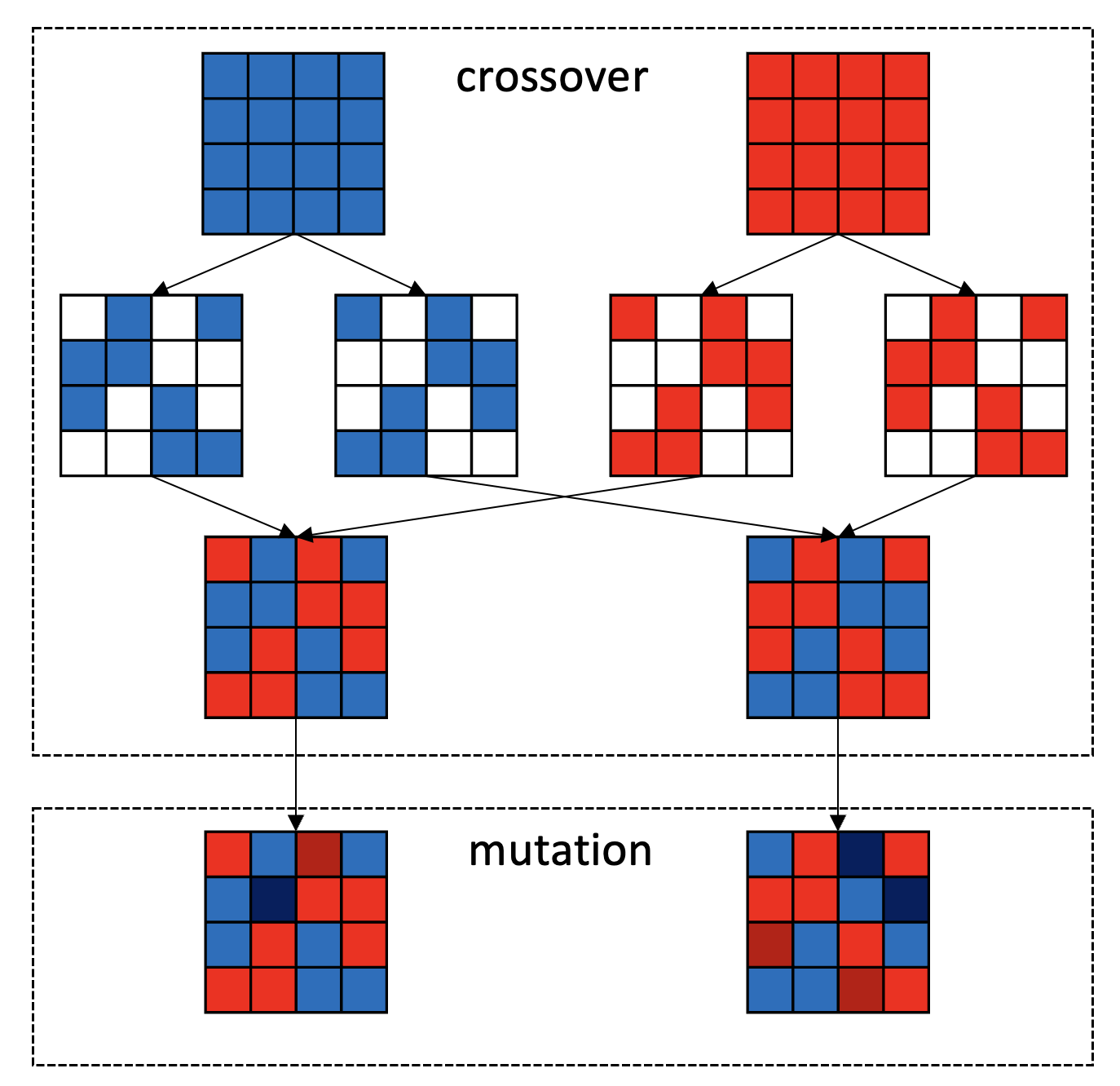}
		\caption{Illustration of crossover and mutation in Two-Step GA Based Local Test Cases Generation}
		\label{fig:ga_crossover_mutation}
	\end{figure}
	\paragraph{Crossover} At line 13, the crossover operator will combine a pair of parents from last step to generate a pair of children, which share many of the characteristics from the parents. The half elements of parents are randomly exchanged as an example shown in Fig.~\ref{fig:ga_crossover_mutation}.
	\paragraph{Mutation} At line 14, some elements of children are randomly altered to add variance in the evolution. It should be noticed that the mutated samples should still fall into the local region $\eta$ around test seed $x$. Finally, the children and parents will be the individuals for the next generation.
	\paragraph{Termination} At line 5, the termination condition of GA is either maximum number of iterations is reached or the highest ranking of fitness reaches a plateau such that successive iterations no longer produce better results.}


As seen above, the main difference between our approach and common GA test case generation is that we use two fitness functions, which work alternatively to guide the selection of parents. The reason why we propose two fitness functions is because, we notice that there is a trade-off between the two objectives $\mathcal{J}$ and $\mathcal{L}$ in the optimisation. Prediction loss $\mathcal{J}$ is related to the adversarial strength, while $\mathcal{L}$ indicates the local distribution. Intuitively, generating the test cases with high local probability tends to add small amount of perturbations to the seed, while a greater perturbation is more likely to induce high prediction loss. To avoid the competition between the two terms that may finally leads to a failure of detecting AEs, we define two fitness functions to precisely control the preference at different stages:
\begin{align}
F_1 = \mathcal{J}(f(x'),y), \quad
F_2 = \mathcal{J}(f(x'),y) + \alpha \cdot \mathcal{L}(x,x')
\end{align}
At early stage, $F_1$ is optimised to quickly direct the generation of AEs, with $F_1>0$ meaning the detection of an AE. When most individuals in the population are AEs, i.e., $\mathit{majority}(\mathit{fit\_list_1} \geq 0)$, the optimisation moves to the second stage, in which $F_1$ is replaced by $F_2$ to optimise the local distribution indicator as well as the prediction loss.  It is possible\footnote{Especially when a large $\alpha$ is used, i.e., with preference on detecting AEs with high local probability than with high adversarial strength, cf. the aforementioned trade-off.} that the prediction loss of most individuals again becomes
negative, then the optimisation will go back to the first stage. With such a mechanism of alternatively using two fitness functions in the optimisation, the proportion of AEs in the population is effectively prevented from decreasing.


Algorithm~\ref{alg:GA} describes the process for generating $m$ local test cases given a single test seed. Suppose $n$ test seeds are selected earlier and in total $M$ local test cases are affordable, we can allocate, for each test seed $x_i$, the number of local test cases $m_i$, according to the $n$ (re-normalised) global probabilities, which emphasises more on the role of distribution in our detected set of AEs.

\textit{\textbf{Running Example}}: The right diagram in Fig.~\ref{feature_pixel_op} plots the local distribution using MSE as its indicator, and visualises the detected AEs by different testing methods. Unsurprisingly, all AEs detected by our proposed HDA testing are located at the high density regions (and very close to the central test seed), given it considers the perceptual quality metric as one of the optimisation objectives in the two-step GA based test case generation. In contrast, other methods (PGD and coverage-guided testing) are less effective.

\section{Evaluation}
\label{sec_evaluation}
We evaluate the proposed HDA testing method by performing extensive experiments to address the following research questions (RQs):




\paragraph {\textbf{RQ1 (Effectiveness): How effective are the methods adopted in the three main stages of HDA?}} Namely, we conduct experiments to \textit{i)} examine the accuracy of combining VAE-Encoder+KDE to approximate the global distribution; \textit{ii)} check the correlation significance of the two proposed local robustness	indicators  with the local robustness; \textit{iii)} investigate the effectiveness of our two-step GA for local test cases generation.

\paragraph {\textbf{RQ2 (AE Quality): How is the quality of AEs detected by HDA?}} Comparing to conventional adversarial attack and coverage-guided testing methods and more recent distribution-aware testing methods, such as OODA and FODA, we introduce a comprehensive set of metrics to evaluate the quality of AEs detected by HDA and others. 

\paragraph {\textbf{RQ3 (Sensitivity): How sensitive is HDA to the DL models under testing?}} We carry out experiments to assess the capability of HDA applied on DL models (adversarially trained) with different levels of robustness.

\paragraph {\textbf{RQ4 (Robustness Growth): How useful is HDA to support robustness growth of the DL model under testing?}} We examine the global robustness of DL models after ``fixing'' the AEs detected by various testing methods.


%


\subsection{Experiment Setup}

We consider five popular benchmark datasets and five diverse model architectures for evaluation. In order to obtain statistical results, we train 10 models for each benchmark dataset, initialising their weights according to the normal distribution. We use Adam optimiser with learning rate $10^{-2}$ and weight decay $10^{-5}$, and train 100 epochs. Details of the datasets and average accuracy (Mean $\pm$ SD) of trained DL models under testing are listed in Table~\ref{table_model_details}. The norm ball radius $r$ is calculated based on the $r$-separation distance (cf. Remark \ref{remark_r_sep}) of each dataset. When comparing different AE detection approaches on 10 models, the evaluation results are dependent on individual model and the difference approximately follow the normal distribution by visualisation and normality test\footnote{Some statistical tests for AE Prop. and \% of Valid AEs  will output ``NaN'' due to the identical evaluation results between different approaches.}. Therefore, we perform \textit{paired two-sample T-test} and discuss the statistical significance in the experiments. The null hypothesis is that different AE detection approaches have identical average values for evaluation metrics. The calculated $p-value>0.05$ indicates that the null hypothesis is true, otherwise it is false. 

In \textbf{RQ1}, we validate the accuracy and effectiveness of HDA on detecting high quality AEs from high probability density region and decide the hyper-parameter settings for the following experiments. That is, for \textbf{RQ2-RQ4}, we use the activation separation distance score ${S}_{\textit{sep}}$ as local robustness indicator and MSE as perceptual quality metric for images. In \textbf{RQ2} we compare the quality of AEs detected by HDA and others. In \textbf{RQ3}, we add the comparison on DL models, enhanced by PGD-based adversarial training, for sensitivity analysis. Table~\ref{table_model_details} also records the accuracy of these adversarially trained models. Adversarial training trades the generalisation accuracy for the robustness as expected (thus a noticeable decrement of the training and testing accuracy) \cite{zhang2019theoretically}. In \textbf{RQ4}, we firstly sample 10000 data points from the global distribution as validation set and detect AEs around them by different methods. Then, we fine-tune the normally trained models with training dataset augmented by these AEs. 10 epochs are taken along with `slow start, fast decay' learning rate schedule \cite{jeddi_simple_2021} to reduce the computational cost while improving the accuracy-drop and robustness. To empirically estimate the global robustness on validation set, we find another set of AEs according to local distribution, different from the fine-tuning data. These AEs, crafted from validation datasets, are miss-classified by normally trained models. Thus, empirical global robustness of normally trained models is set to 0 as the baseline.


When training VAE models, the loss function is a combination of reconstruction loss and KL divergence loss. Two losses are weighted equally and simultaneously optimised. We also use Adam optimiser with learning rate $10^{-2}$ and weight decay $10^{-5}$, and train 100 epochs to obtain VAE models. To avoid the posterior collapse that VAE often converges to a degenerated local optimum, we add Batch Normalisation \cite{zhu-etal-2020-batch} before the output of encoder. The latent dimensions and the reconstruction loss are listed in Table~\ref{pca_vae}.

For readers' convenience, all the metrics used in \textbf{RQ2}, \textbf{RQ3} and \textbf{RQ4} for comparisons are listed in Table~\ref{evaluation_metrics}. The metrics are introduced to comprehensively evaluate the quality of detected AEs and the DL models from different aspects. When comparing HDA testing with PGD attack, coverage guided testing, OODA and FODA, we have the following settings for the tools. Specifically, we use PGD attack with 10 steps for gradient ascent, and step size 2/255; HDA testing with population size $N=1000$, maximum iterations $T=500$, and weight parameter $\alpha=1$; neuron coverage \cite{sun2018concolic} with Gaussian noise $mean=0,std=1$ to generate perturbed data, and 100 iterations to increase the coverage rate by fuzzing; OODA testing \cite{dola_distribution_aware_2021} with 10 steps for gradient ascent, step size 2/255, default hyperparameter for balancing two goals and default reconstruction probability threshold used in the released code \footnote{https://github.com/swa112003/ DistributionAwareDNNTesting.}; FODA testing \cite{byun_manifold_based_2020} with the same latent space encoded by VAE used in HDA, random sampling to search for AEs in the latent space. To achieve the fair comparison, we set the same perturbation radius $r$ for PGD attack, HDA testing and coverage guided testing, since they have the restrictions for the validity of test cases. In addition, we set the same number of steps and step size for PGD attack and OODA testing, and utilise the same latent space across HDA and FODA.


\begin{table}[ht]
	\centering
	\caption{Details of the datasets and DL models under testing.}
	\resizebox{\linewidth}{!}{
		\begin{tabular}{cccccccc}
			\hline
			\multirow{2}{*}{Dataset} & \multirow{2}{*}{Image Size} & \multirow{2}{*}{$r$} & \multirow{2}{*}{DL Model} & \multicolumn{2}{c}{Normal Training} & \multicolumn{2}{c}{Adversarial Training} \\ 
			& & & & Avg. Train Acc. & Avg. Test Acc. & Avg. Train Acc. & Avg. Test Acc.\\ \hline
			MNIST & $1\times32\times32$ & 0.1 & LeNet5 & $99.88\%\pm0.06\%$ & $98.73\%\pm0.12\%$ & $99.77\%\pm0.01\%$ & $98.84\%\pm0.01\%$ \\
			Fashion-MNIST & $1\times32\times32$ & 0.08 & AlexNet & $95.12\%\pm0.85\%$ & $90.70\%\pm0.25\%$ & $86.23\%\pm0.05\%$ & $85.11\%\pm0.05\%$ \\
			SVHN & $3\times32\times32$ & 0.03 & VGG11 & $96.12\%\pm0.45\%$ & $94.86\%\pm0.21\%$ & $89.89\%\pm0.69\%$ & $90.32\%\pm0.81\%$ \\
			CIFAR-10 & $3\times32\times32$ & 0.03 & ResNet20 & $97.81\%\pm0.58\%$ & $88.28\%\pm0.63\%$ & $79.48\%\pm0.72\%$ & $76.42\%\pm1.01\%$ \\
			CelebA & $3\times64\times64$ & 0.05 & MobileNetV1 & $94.19\%\pm1.63\%$ & $90.79\%\pm0.49\%$ & $77.74\%\pm0.22\%$ & $79.72\%\pm0.22\%$ \\ \hline
	\end{tabular}}
	\label{table_model_details}
\end{table}

\begin{table}[ht]
	\centering
	\caption{Evaluation metrics for the quality of detected AEs and DL models}
	\resizebox{1\linewidth}{!}{
		\begin{tabular}{cc}
			\hline
			Metrics & Meanings \\ \hline
			AE Prop. & Proportion of test seeds from which AEs can be detected over the total number of test seeds \\
			Pred. Loss & Adversarial strength of AEs as formally defined by Definition~\ref{def_pred_loss} \\
			$p_g$ & Normalised global probability density of test-seeds/AEs \\
			$\mathcal{R}_l$ & Local robustness to the correct classification label, as formally defined by Definition~\ref{def_lr} \\
			$\hat{\mathcal{R}_g}$ & Empirical global robustness of DL models over input domain as defined in Definition~\ref{def_emp_gr}\\
			FID & Distribution difference between original images (test seeds) and perturbed images (AEs)
			\\
			$\epsilon$ & Average perturbation distance between test seeds and AEs \\
			\% of Valid AEs & Percentage of ``in-distribution'' AEs in all detected AEs \ \\ \hline
	\end{tabular}}
	\label{evaluation_metrics}
\end{table}

All experiments were run on a machine of Ubuntu 18.04.5 LTS x86\_64 with Nvidia A100 GPU and 40G RAM. The source code, DL models, datasets and all experiment results are publicly available at \url{https://github.com/havelhuang/HDA-Testing}.

\subsection{Evaluation Results and Discussions}

\subsubsection{\textbf{RQ1}}
There are 3 sets of experiments in \textbf{RQ1} to examine the accuracy of technical solutions in our tool-chain, corresponding to the 3 main stages respectively.

First, to approximate the global distribution, we essentially proceed in two steps---dimensionality reduction and PDF fitting, for which we adopt the VAE-Encoder+KDE solution. Notably, the VAE trained in this step is for data-compression only (not for generating new data). To reflect the effectiveness of both aforementioned steps, we (i) compare VAE-Encoder with the Principal Component Analysis (PCA), and (ii) measure the FID between the training dataset and a set of random samples drawn from the fitted global distribution by KDE.

PCA is a common approach for dimensionality reduction. We use scikit-learn \cite{scikit-learn} to implement the PCA with 'auto' solver for applying Singular Value Decomposition (SVD). That is, if the input data is larger than 500x500 and the number of components to extract (latent dimensions) is lower than 80\% of the smallest dimension of the data, then the more efficient ‘randomised’ method \cite{scikit-learn} is enabled. Otherwise the exact full SVD is computed and optionally truncated afterwards. To achieve the fair comparison, the latent dimensions of PCA and VAE-Encoder are set to be the same. We compare the performance of VAE-Encoder and PCA from the following two perspectives. The \textit{quality of latent representation} can be measured by the \textit{clustering} and \textit{reconstruction accuracy}. 

To learn the feature level (global) distribution from latent data, we require that latent representations should group together data points that share similar semantic features. To evaluate this clustering ability, we apply K-means clustering to the latent data, which partitions the data points into clusters based on their similarity. Then, we calculate the Completeness Score (CS), Homogeneity Score (HS) and V-measure Score (VS) \cite{DBLP:conf/emnlp/RosenbergH07}. These scores provide a measure of how well the resulting clusters group together data points that share similar semantic features. Specifically, the CS measures how well the clustering captures all data points that belong to the same true class in a single cluster, while the HS measures how well the clustering captures all data points within a cluster that belong to the same true class. The VS is a harmonic mean of the CS and HS, providing an overall measure of the quality of the clustering.

In addition to ensuring that latent representations group together similar data points, we also require that the latent representations can be decoded to reconstruct the original images with minimal information loss. The reconstruction loss is calculated based on the MSE. As is shown in Table~\ref{pca_vae}, VAE-Encoder achieves higher CS, HS, VS scores and less reconstruction loss than PCA. In other words, the latent representations encoded by VAE-Encoder is better in terms of capturing feature information than that of PCA.

\begin{table}[ht!]
	\centering
	\caption{{Quality of Latent Representation in PCA \& VAE-Encoder}\normalsize}
	\resizebox{0.9\linewidth}{!}{
		\begin{tabular}{cccccccccc}
			\hline
			\multirow{3}{*}{Dataset} & \multirow{3}{*}{Latent Dimensions} &\multicolumn{4}{c}{PCA} & \multicolumn{4}{c}{VAE-Encoder} \\
			& & \multicolumn{3}{c}{Clustering} & \multirow{2}{*}{Recon. Loss} & \multicolumn{3}{c}{Clustering} & \multirow{2}{*}{Recon. Loss} \\
			& & CS & HS & VS &  & CS & HS & VS &  \\ \hline
			MNIST & 8 & 0.505 & 0.508 & 0.507 & 44.09 & \textbf{0.564}& \textbf{0.566} & \textbf{0.565} & \textbf{27.13} \\
			F.-MNIST & 4 & 0.497 & 0.520 & 0.508 & 55.56 & \textbf{0.586} & \textbf{0.601} & \textbf{0.594} & \textbf{23.72} \\
			SVHN & 4 & 0.007 & 0.007 & 0.007 & 65.75 & \textbf{0.013} & \textbf{0.011} & \textbf{0.015} & \textbf{62.38} \\
			CIFAR-10 & 8 & 0.084 & 0.085 & 0.085 & 188.22 & \textbf{0.105} & \textbf{0.105} & \textbf{0.105} & \textbf{168.44} \\
			CelebA & 32 & 0.112 & 0.092 & 0.101 & 764.94 & \textbf{0.185} & \textbf{0.150} & \textbf{0.166} & \textbf{590.54} \\ \hline
	\end{tabular}}
	\label{pca_vae}
\end{table}

\begin{figure}[!ht]
	\centering
	\includegraphics[width=0.28\linewidth]{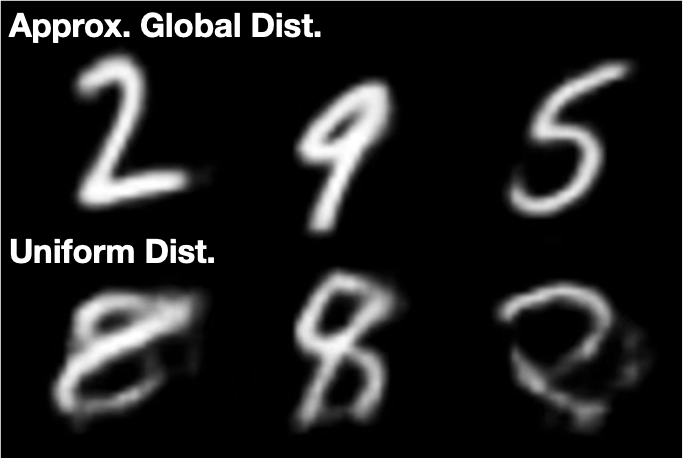}
	\quad
	\resizebox{0.45\linewidth}{!}{
		\begin{tabular}[b]{ccc}
			\hline
			Dataset & Global Dist. & Uni. Dist. \\ \hline
			MNIST & \textbf{0.395} & 13.745 \\
			Fashion-MNIST & \textbf{0.936} & 90.235 \\
			SVHN & \textbf{0.961} & 141.134 \\
			CIFAR-10 & \textbf{0.285} & 12.053 \\
			CelebA & \textbf{0.231} & 8.907 \\ \hline
	\end{tabular}}
	\captionlistentry[table]{..}
	\label{tab:kde_uniform}
	\captionsetup{labelformat=andtable}
	\caption{Samples drawn from the approximated global distribution by KDE and a uniform distribution over the latent feature space (Figure); and FID to the ground truth based on 1000 samples (Table).}
	\label{img:kde_uniform}
\end{figure}

To evaluate the accuracy of using KDE to fit the global distribution, we calculate the FID between a new dataset (with 1000 samples) based on the fitted global distribution by KDE and the training dataset. The new dataset is sampled from the fitted global distribution over latent space and decoded by VAE decoder into images. The FID scores are shown in Table~\ref{tab:kde_uniform}. As a baseline, we also present the results of using a uniform distribution over the latent space. As expected, we observe that all FID scores based on approximated distributions are significantly smaller (better). We further decode the newly generated images for visualisation in Fig.~\ref{img:kde_uniform}, from which we can see that generated images by KDE keep high fidelity while the uniformly sampled images are more difficult to be recognised.

\begin{tcolorbox}
	Answer to \textbf{RQ1} on HDA stage 1: The combination of VAE-Encoder+KDE may accurately approximate the global distribution, since the new sampled data from approximated distribution keep high fidelity.
\end{tcolorbox}

Move on to stage 2, we study the correlations between a norm ball's local robustness and its two indicators proposed earlier---the prediction gradient based score and the score based on separation distance of output-layer activation (cf. Eq.~\ref{equ:robustness_indicator}).

\begin{figure}[ht!]
	\centering
	\includegraphics[width=0.9\linewidth]{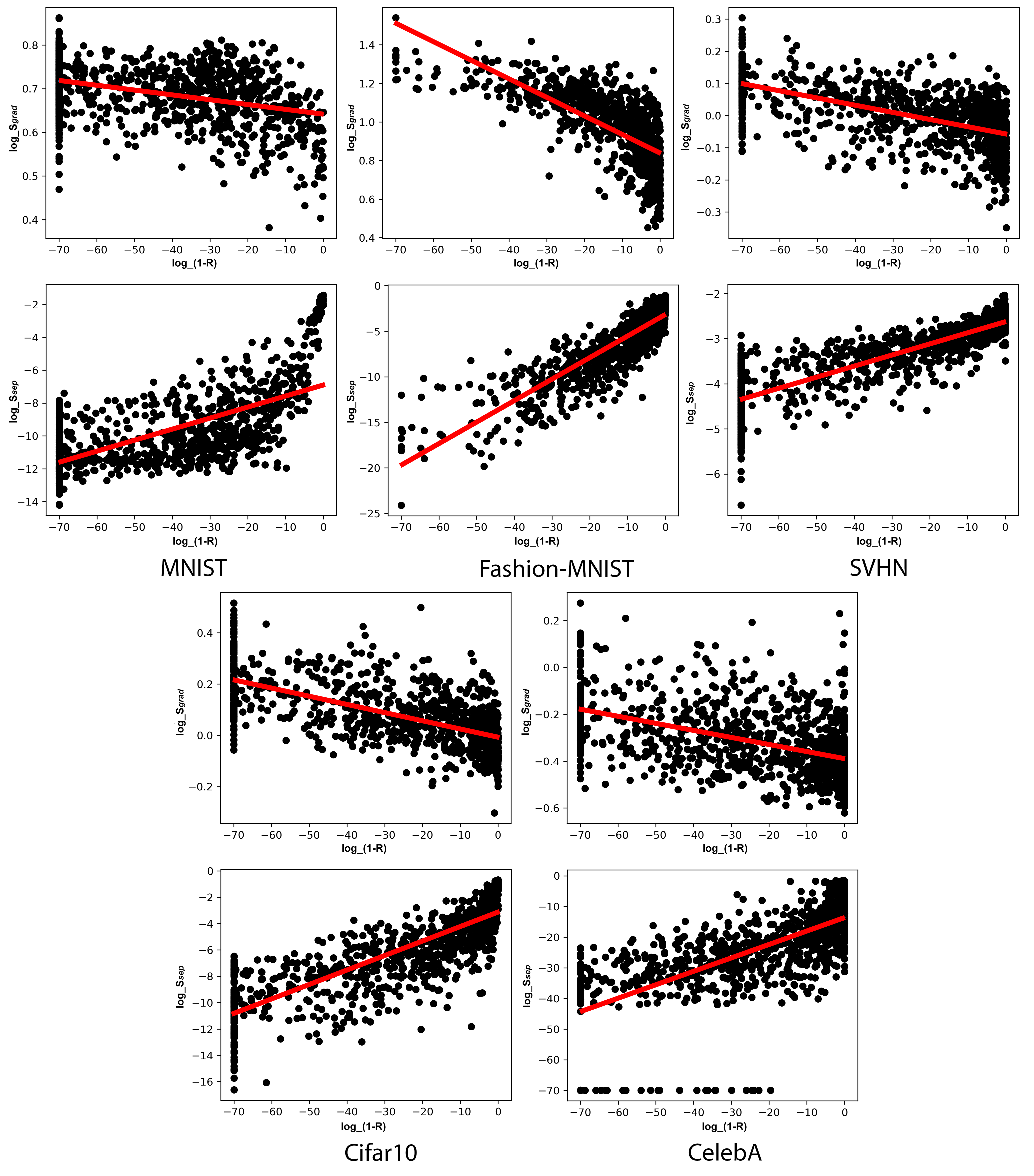}
	\caption{Scatter plots of the local robustness evaluation vs. its two indicators, based on 1000 random norm balls.}
	\label{robustness_aux}
\end{figure}

\begin{table}[ht!]
	\centering
	\caption{Pearson correlation coefficients (in absolute values) between the local robustness \& its two indicators.}
	\resizebox{0.7\linewidth}{!}{
		\begin{tabular}{ccccc}
			\hline
			\multirow{2}{*}{Dataset} & \multirow{2}{*}{$S_{\textit{grad}}$} & \multirow{2}{*}{$S_{\textit{sep}}$} & \multicolumn{2}{c}{$T(S_{\textit{grad}},S_{\textit{sep}})$} \\ 
			& & & $t$ & $p-value$\\\hline
			MNIST & $\mathbf{0.631\pm0.025}$ & $0.564\pm0.019$ & $5.488$ & $5.813\times10^{-4}$\\
			Fashion-MNIST & $0.717\pm0.109$ & $\mathbf{0.789\pm0.056}$ & $-1.793$ & $0.107$ \\
			SVHN & $\mathbf{0.747\pm0.039}$ & $0.745\pm0.030$ & $0.150$ & $0.884$\\
			CIFAR-10 & $0.603\pm0.038$ & $\mathbf{0.668\pm0.065}$ & $-5.466$ & $3.975\times10^{-4}$\\
			CelebA & $0.639\pm0.073$ & $\mathbf{0.728\pm0.077}$ & $-3.872$ & $0.004$\\ \hline
		\end{tabular}
	}
	\label{Pearson_coefficient}
\end{table}

We invoke the tool \cite{webb_statistical_2019} for estimating the local robustness $\mathcal{R}_l$ defined in Definition \ref{def_lr}. Based on 1000 randomly selected data-points from the test set as the central point of 1000 norm balls, we calculate the local robustness of each norm ball\footnote{Radius $r$ is usually small by definition (cf. Remark \ref{remark_r_sep}), yielding very small $log{(1-\mathcal{R}_l)}$.} as well as the two proposed indicators. Then, we do the scatter plots (in log-log scale\footnote{There are dots collapsed on the vertical line of $log{(1-R)}=-70$, due to a limitation of the estimator \cite{webb_statistical_2019}---it terminates with the specified threshold when the estimation is lower than that value. Note, the correlation calculated with such noise is not undermining our conclusion, rather the real correlation would be even higher.}), as shown in Fig.~\ref{robustness_aux}. Apparently, for all 5 datasets, the indicator based on activation separation distance is negatively correlated (1st row), while the gradient based indicator is positively correlated with the estimated local robustness (2nd row). We further quantify the correlation by calculating the Pearson correlation coefficients, as recorded in Table~\ref{Pearson_coefficient}. There is a rule of thumb that Pearson correlation coefficients greater than 0.6 indicate strong correlations \cite{akoglu2018user}. We observe, both indicators are highly correlated with the local robustness, while the separation distance based indicator is slightly better. The statistical test shows that the separation distance based indicator is significantly better than gradient based indicator in CIFAR10 and CelebA datasets. Therefore, for the latter experiment in RQ2-RQ4, we choose separation distance based indicator $S_{\textit{sep}}$ to guide the selection of test seeds when comparing HDA testing with other AE detection methods.

\begin{tcolorbox}
	Answer to \textbf{RQ1} on HDA stage 2: The two proposed local robustness indicators are significantly correlated with the local robustness.
\end{tcolorbox}

\begin{figure*}[htbp!]
	\centering
	\includegraphics[width=0.95\linewidth]{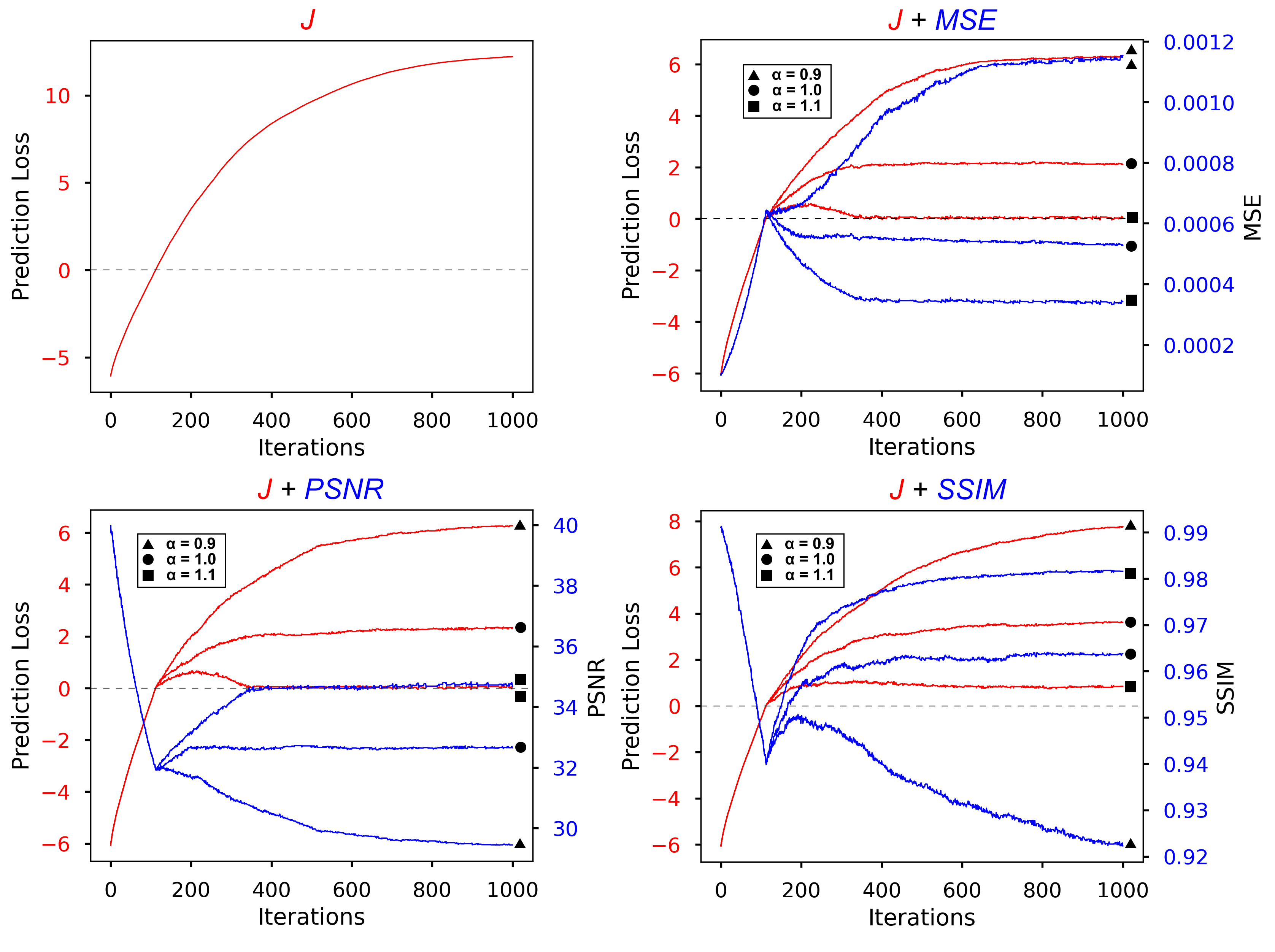}
	\caption{The prediction loss (red) and the three quantified local distribution indicators (blue) of the best fitted test case during the iterations of our two-step GA based local test case generation.}
	\label{ga_exp}
\end{figure*}

For the local test case generation in stage 3, by configuring the parameter $\alpha$ in our two-step GA, we may do trade-off between the ``strength of being adversarial'' (measured by prediction loss $\mathcal{J}$) and the local distribution (measured by a specific perceptual quality metric $\mathcal{L}$, they are MSE, PSNR and SSIM), so that the quality of detected AEs can be optimised.

In Fig.~\ref{ga_exp}, we visualise the changes of the two fitness values as the iterations of the GA. As shown in the first plot, only the prediction loss $\mathcal{J}$ is taken as the fitness function (i.e., $\alpha=0$) during the whole iteration process. The GA can effectively find AEs with maximised adversarial strength, as observed by the convergence of the prediction loss of the best-fitted test case in the population after hundreds of iterations. From the second to the last plot, the fitness function consists not only of prediction loss $\mathcal{J}$, but also of a perceptual quality metric $\mathcal{L}$, representing the local distribution information (i.e., $ \alpha>0$). Intuitively, a smaller MSE or greater PSNR and SSIM implies higher local probability density.

Thanks to the two-step setting of the fitness functions, the prediction loss $\mathcal{J}$ of best-fitted test case goes over 0 quickly in less than 200 iterations, which means it detects a first AE in the population. The $\mathcal{J}$ of the best fitted test case is always quite close to the rest in the population, thus we may confidently claim that many AEs are efficiently detected by the population not long after the first AE was detected. Then, the optimisation goes to the second stage, in which the quantified local distribution indicator $\mathcal{L}$ is pursued. The $\mathcal{J}$ and $\mathcal{L}$ finally converge and achieve a balance between them. If we configure the coefficient $\alpha$, the balance point will change correspondingly. A greater $\alpha$ (e.g., $\alpha=1.1$ in the plots) detects less perceptible AEs (i.e., with higher local probability density), and the price paid is that the detected AEs are with weaker adversarial strength (i.e., with smaller but still \textit{positive} prediction loss). 

\begin{figure}[!htbp]
	\centering
	\includegraphics[width=0.6\linewidth]{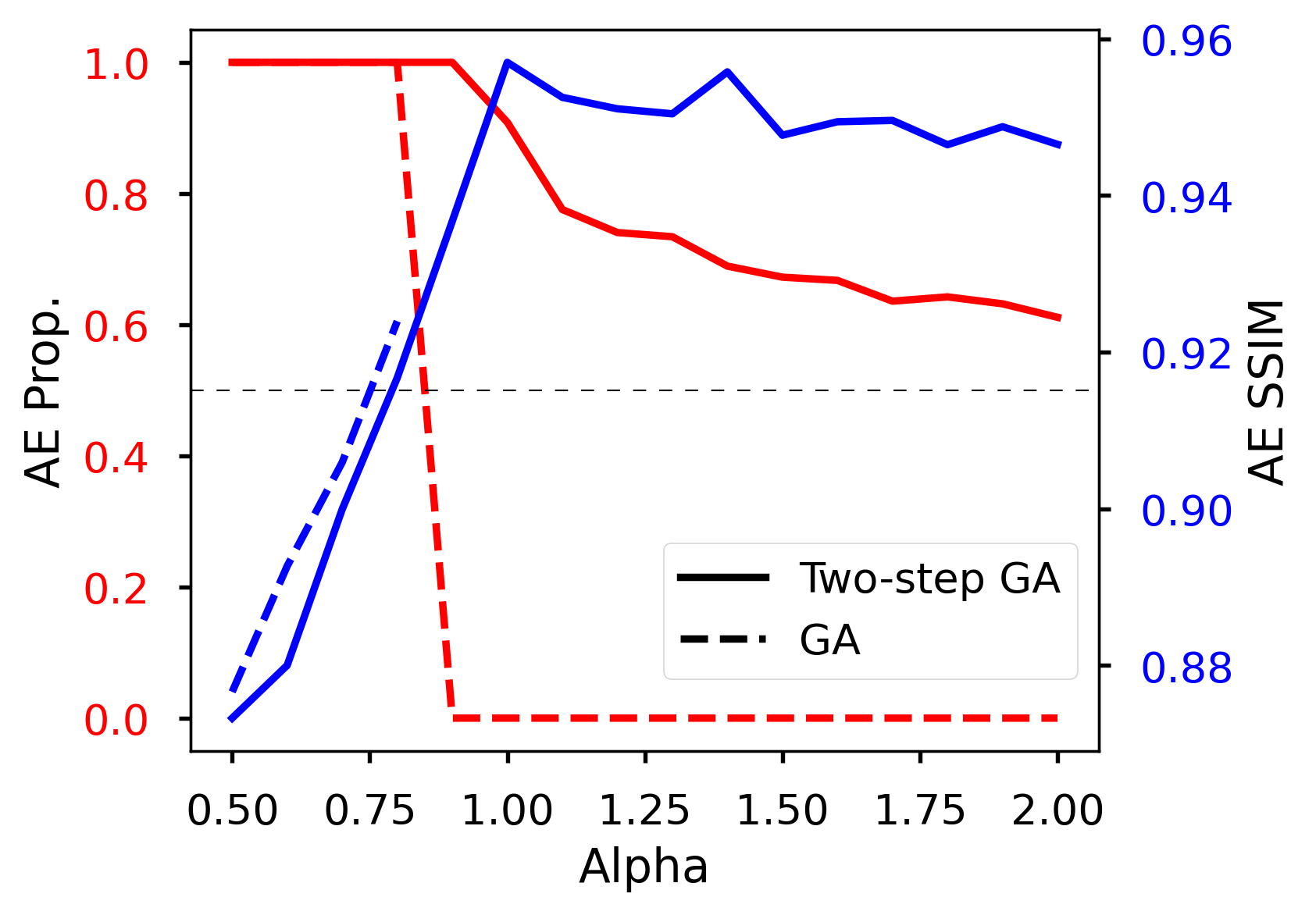}
	\caption{Comparison between regular GA and two-step GA.}
	\label{ga_alpha}
\end{figure}

\begin{figure}[!htbp]
	\centering
	\includegraphics[width=\linewidth]{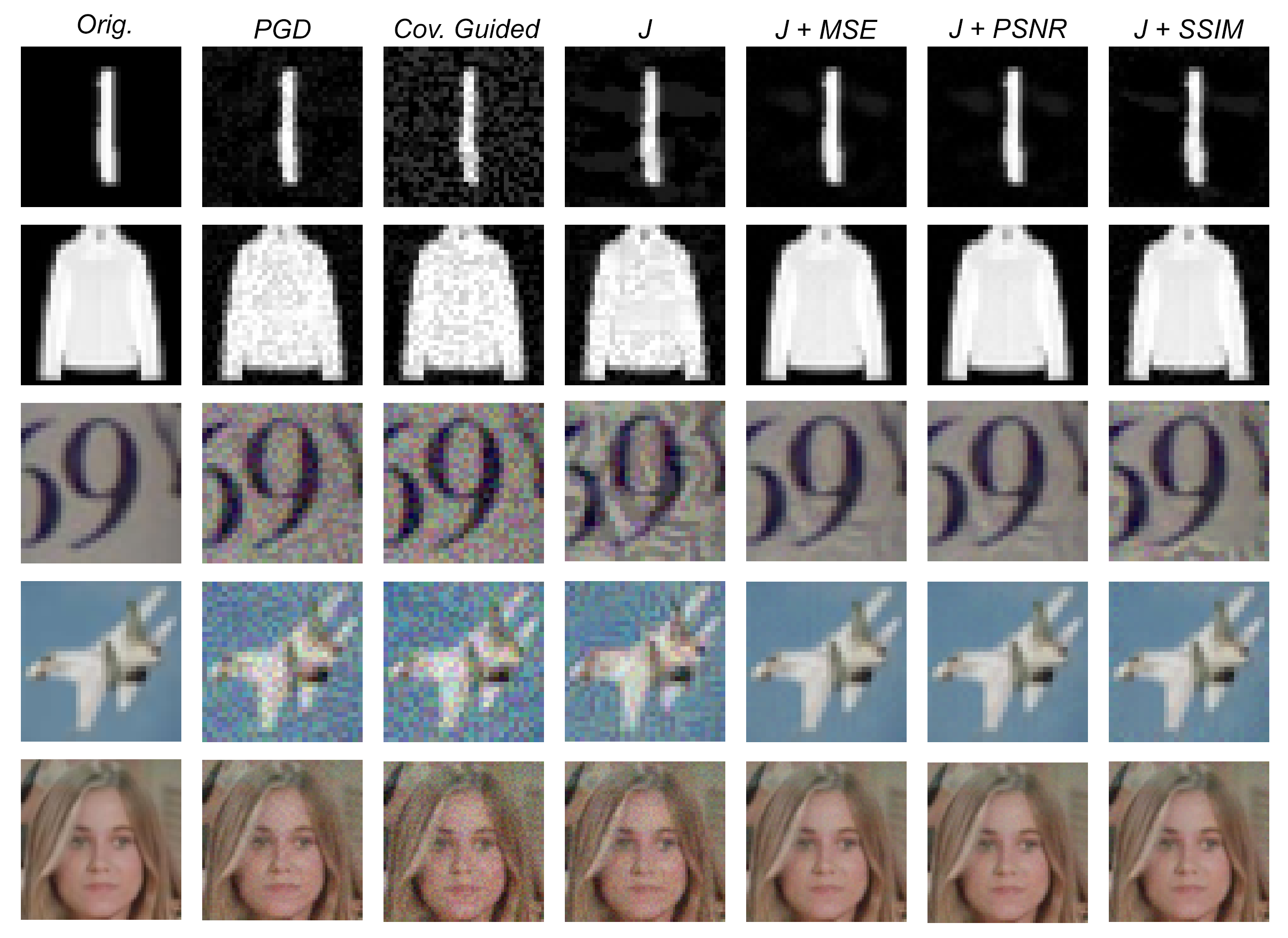}
	\caption{AEs detected by our two-step GA (last 3 columns) \& other methods}
	\label{adv_display}
\end{figure}

We further investigate the advantages of our 2-step GA over the regular GA (using $F_2$ as the objective function). In Fig.~\ref{ga_alpha}, as $\alpha$ increases, the proportion of AEs in the population exhibits a sharp drop to 0 when using the regular GA. In contrast, the two-step GA prevents such decreasing of the AE proportion while preserving it at a high-level of 0.6, even when $\alpha$ is quite large. Moreover, larger $\alpha$ represents the situations when the AEs are less perceptible in terms of perceptual quality metrics---as shown by the blue curves\footnote{The blue dashed line stops earlier as there is no AEs in the population when $\alpha$ is 
	big.}, the imperceptibility ( measured by SSIM in this case) is only sufficiently high when $\alpha$ is big enough. Thus, compared to the regular GA, we may claim our novel 2-step GA is more robust (in detecting AEs) to the choices of $\alpha$ and more suitable in our framework for detecting AEs with high perceptual quality.   

\begin{table}[!htbp]
	\centering
	\caption{Perceptual quality measured by FID between a set of original images and a set of AEs detected by  two-step GA with different fitness functions. }
	\begin{subtable}{\linewidth}
		\centering
		\resizebox{0.8\linewidth}{!}{
			\begin{tabular}{lllll}
				\hline
				Dataset & $J$ & $J+MSE$ & $J+PSNR$ & $J+SSIM$ \\ \hline
				MNIST & $1.185\pm0.157$ & $\mathbf{0.667\pm0.202}$ & $0.668\pm0.203$ & $0.875\pm0.184$\\
				Fashion-MNIST & $2.736\pm0.680$ & $0.138\pm0.106$ & $\mathbf{0.131\pm0.100}$ & $0.679\pm0.182$ \\
				SVHN & $120.172\pm7.529$ & $105.399\pm9.009$ & $\mathbf{104.147\pm9.934}$ & $111.139\pm6.754$ \\
				CIFAR-10 & $96.449\pm5.906$ & $\mathbf{65.727\pm5.672}$ & $67.426\pm7.398$ & $76.293\pm6.493$ \\
				CelebA & $85.658\pm3.334$ & $\mathbf{65.981\pm4.976}$ & $66.599\pm3.312$ & $71.878\pm3.153$ \\ \hline
		\end{tabular}}
		\caption{Results}
	\end{subtable}
	\begin{subtable}{\linewidth}
		\centering
		\resizebox{0.8\linewidth}{!}{
			\begin{tabular}{ccccccc}
				\hline
				\multirow{2}{*}{Dataset} & \multicolumn{2}{c}{$T(J+MSE,J)$} & \multicolumn{2}{c}{$T(J+MSE,J+PSNR)$} & \multicolumn{2}{c}{$T(J+MSE,J+SSIM)$} \\
				& $t$ & $p-value$ & $t$ & $p-value$ & $t$ & $p-value$ \\ \hline
				MNIST & $-13.631$ & $2.581\times10^{-7}$ & $-0.912$ & $0.385$ & $-10.279$ & $2.844\times10^{-6}$ \\
				Fashion-MNIST & $-13.362$ & $3.065\times10^{-7}$ & $2.188$ & $0.056$ & $-11.957$ & $7.934\times10^{-7}$ \\
				SVHN & $-9.214$ & $7.036\times10^{-6}$ & $1.271$ & $0.235$ & $-4.443$ & $0.002$ \\
				CIFAR-10 & $-17.32$ & $3.215\times10^{-8}$ & $-1.363$ & $0.206$ & $-8.125$ & $1.954\times10^{-5}$ \\
				CelebA & $-23.491$ & $2.187\times10^{-9}$ & $-0.511$ & $0.621$ & $-3.931$ & $0.003$ \\ \hline
		\end{tabular}}
		\caption{Statistical Test}
	\end{subtable}
	\label{tab:pq}
\end{table}

Fig.~\ref{adv_display} displays some selected AEs from the five datasets. Same as the PGD attack and the coverage-guided testing, if we only use the prediction loss $\mathcal{J}$ as the objective function in the GA, the perturbations added to the images can be easily recognised. In stark contrast, AEs generated by our two-step GA (with the 3 perceptual quality metrics in the last 3 columns) are of high quality and less distinguishable from the original images (first column). We further calculate the FID to quantify the perceptual quality of detected AEs in Table~\ref{tab:pq}. Since the comparison with PGD attack, coverage guided testing, FODA, and OODA are given in the subsequent experiments, i.e. Table~\ref{test_cases} and Table~\ref{distribution_aware_compare}, we focus on comparing the performance of 3 perceptual quality metrics here. Results show that MSE is significantly better than J and SSIM to guide the generation of high fidelity AEs for the experiment dataset. While MSE has similar performance with PSNR. Therefore, we decide to utilise the J+MSE in the following experiments for the comparison with the state-of-the-art.

\begin{tcolorbox}
	Answer to \textbf{RQ1} on HDA stage 3: Two-step GA based local test case generation can effectively detect AEs with high perception quality (in terms of those metrics encoded by GA).
\end{tcolorbox}

\subsubsection{\textbf{RQ2}}

We compare HDA with the state-of-the-art AE detection methods in two sets of experiments. In the first set, we focus on comparing with the adversarial attack and coverage-guided testing (i.e., the typical PGD attack and neuron coverage metric for brevity, while the conclusion can be generalised to other attacks and coverage metrics, since they all lack the consideration of distribution when detecting AEs). Then in the second set of experiments, we show the advantages of our HDA testing over other distribution-aware testing methods.


In fact, both PGD attack and coverage-guided testing do not contribute to test seeds selection. They simply use randomly sampled data from the test set as test seeds, by default. We also notice that a large amount of test seeds prioritisation metrics are proposed, the typical ones among which are Distance-Based Surprise Adequacy (DSA) \cite{weiss2022simple}, DeepGini \cite{weiss2022simple}. Thus, we compare the randomly selected test seeds, DSA guided test seeds, DeepGini guided test seeds with our ``global distribution probability\footnote{Refer to Section~\ref{p_g_cal} for the calculation. The value of probability density is further normalised by training dataset for a better presentation.} plus local robustness indicated'' test seeds, shown as ```$p_g+\mathcal{R}_l$'' in Table~\ref{seeds_selection}. Specifically, for each test seed, we calculate two metrics---the local robustness $\mathcal{R}_l$ of its norm ball and its corresponding global probability $p_g$. We invoke the estimator of \cite{webb_statistical_2019} to calculate the former ($log{(1-\mathcal{R}_l)}$, to be exact). To reduce the sampling noise, we repeat the test seed selection 100 times and present the averaged results in Table~\ref{seeds_selection}.



\begin{table}[!htbp]
	\centering
	\caption{Comparison between randomly selected test seeds, DSA guided test seeds, DeepGini guided test seeds and our ``$p_g+\mathcal{R}_l$ indicated'' test seeds (based on 100 test seeds).}
	\begin{subtable}{\linewidth}
		\centering
		\resizebox{\linewidth}{!}{
			\begin{tabular}{ccccccccc}
				\hline
				\multirow{2}{*}{Dataset} & \multicolumn{2}{c}{Random Test Seeds} &  \multicolumn{2}{c}{DSA Test Seeds} &  \multicolumn{2}{c}{DeepGini Test Seeds} & \multicolumn{2}{c}{$p_g + \mathcal{R}_l$ Test Seeds} \\
				& $log{(1-\mathcal{R}_l)}$ & $p_g$ & $log{(1-\mathcal{R}_l)}$ & $p_g$ & $log{(1-\mathcal{R}_l)}$ & $p_g$ & $log{(1-\mathcal{R}_l)}$ & $p_g$ \\ \hline
				MNIST & $-64.67\pm1.27$ & $0.0034\pm0.0005$ & $-64.59\pm1.15$ & $0.0016\pm0.0003$ & $-65.16\pm1.31$ & $0.0018\pm0.0002$ & $\mathbf{-22.12\pm3.01}$ & $\mathbf{0.0293\pm0.0072}$\\
				Fashion-MNIST & $-14.98\pm4.50$ & $0.0033\pm0.0009$ & $-15.74\pm4.82$ & $0.0016\pm0.0003$ & $-15.29\pm4.89$ & $0.0017\pm0.0003$ & $\mathbf{-1.24\pm0.48}$ & $\mathbf{0.0226\pm0.0088}$\\
				SVHN & $-53.34\pm3.98$ & $0.0032\pm0.0002$ & $-53.44\pm2.10$ & $0.0013\pm0.0002$ & $-53.62\pm2.32$ & $0.0008\pm0.0001$ & $\mathbf{-8.27\pm2.43}$ & $\mathbf{0.0132\pm0.0021}$\\
				CIFAR-10 & $-13.77\pm1.24$ & $0.0033\pm0.0009$ & $-14.25\pm2.26$ & $0.0018\pm0.0003$ & $-14.82\pm0.86$ & $0.0018\pm0.0003$ & $\mathbf{-2.49\pm0.88}$ & $\mathbf{0.0397\pm0.0121}$\\
				CelebA & $-17.68\pm5.00$ & $0.0034\pm0.0002$ & $-17.27\pm5.44$ & $0.0010\pm0.0001$ & $-17.28\pm5.37$ & $0.0005\pm0.0001$ &$\mathbf{-1.59\pm0.54}$ & $\mathbf{0.0118\pm0.0005}$\\ \hline
			\end{tabular}
		}
		\caption{Results}
	\end{subtable}
	\begin{subtable}{\linewidth}
		\centering
		\resizebox{0.8\linewidth}{!}{
			\begin{tabular}{cccccccc}
				\hline
				\multirow{2}{*}{Dataset} & \multirow{2}{*}{Metric} & \multicolumn{2}{c}{$T(p_g + \mathcal{R}_l, \text{Random})$} & \multicolumn{2}{c}{$T(p_g + \mathcal{R}_l, \text{DSA})$} & \multicolumn{2}{c}{$T(p_g + \mathcal{R}_l, \text{DeepGini})$} \\
				&  & $t$ & $p-value$ & $t$ & $p-value$ & $t$ & $p-value$ \\ \hline
				\multirow{2}{*}{MNIST} & $log{(1-\mathcal{R}_l)}$ & $41.186$ & $2.890\times10^{-19}$ & $41.680$ & $2.337\times10^{-19}$ & $41.461$ & $2.568\times10^{-19}$ \\
				& $p_g$ & $11.348$ & $1.234\times10^{-9}$ & $12.155$ & $4.101\times{-10}$ & $12.073$ & $4.574\times10^{-10}$ \\ \hline
				\multirow{2}{*}{Fashion-MNIST} & $log{(1-\mathcal{R}_l)}$ & $9.601$ & $1.665\times10^{-8}$ & $9.466$ & $2.063\times10^{-8}$ & $9.042$ & $4.106\times10^{-8}$ \\
				& $p_g$ & $6.899$ & $1.885\times10^{-6}$ & $7.542$ & $5.616\times10^{-7}$ & $7.506$ & $6.001\times10^{-7}$ \\ \hline
				\multirow{2}{*}{SVHN} & $log{(1-\mathcal{R}_l)}$ & $30.564$ & $5.769\times10^{-17}$ & $44.475$ & $7.345\times10^{-20}$ & $42.686$ & $1.528\times10^{-19}$ \\
				& $p_g$ & $14.991$ & $1.303\times10^{-11}$ & $17.839$ & $6.867\times10^{-13}$ & $18.651$ & $3.203\times10^{-13}$ \\ \hline
				\multirow{2}{*}{CIFAR-10} & $log{(1-\mathcal{R}_l)}$ & $23.459$ & $6.033\times10^{-15}$ & $15.334$ & $8.916\times10^{-12}$ & $31.688$ & $3.045\times10^{-17}$ \\
				& $p_g$ & $9.487$ & $1.997\times10^{-8}$ & $9.902$ & $1.039\times10^{-8}$ & $9.902$ & $1.039\times10^{-8}$ \\ \hline
				\multirow{2}{*}{CelebA} & $log{(1-\mathcal{R}_l)}$ & $10.117$ & $7.461\times10^{-9}$ & $9.071$ & $3.922\times10^{-8}$ & $9.193$ & $3.207\times10^{-8}$ \\
				& $p_g$ & $49.326$ & $1.156\times10^{-20}$ & $66.979$ & $4.830\times10^{-23}$ & $70.079$ & $2.145\times10^{-23}$ \\ \hline
			\end{tabular}
		}
		\caption{Statistical Test}
	\end{subtable}
	\label{seeds_selection}
\end{table}

From Table~\ref{seeds_selection}, we observe: (i) test seeds selected by our method have much higher global probability density, meaning their norm balls are much more representative of the data distribution; (ii) the norm balls of our test seeds have worse local robustness, meaning it is more \textit{cost-effective} to detect AEs in them. Both metrics of HDA are significantly different from those of other test seeds selection methods, as indicated by large t-scores and small p-values. This is unsurprising because we have explicitly considered the distribution and local robustness information in the test seed selection. (iii) DSA and DeepGini guided test seeds are even worse than random test seeds, since they target at prioritising the test seeds which are prone to be misclassified. These misclassified test seeds are usually out of distribution.

Finally, the overall evaluation on the generated test cases and the detected AEs by them are shown in Table~\ref{test_cases}. The results are presented in two dimensions---3 types of testing methods versus 2 ways of test seeds selection, yielding 6 combinations (although by default, PGD attack and coverage-guided methods are using random seeds, while our method is using the ``$p_g+\mathcal{R}_l$'' seeds). For each combination, we study 4 metrics (cf. Table \ref{evaluation_metrics} for meanings behind them): (i) the AE proportion; (ii) the average prediction loss; (iii) the FID\footnote{To show how close the perturbed test cases are to the test seeds in the latent space, we use the last convolutional layer of InceptionV3 to extract the latent representations of colour images for FID. InceptionV3 is a well-trained CNN and commonly used to show FID that captures the perturbation levels, e.g., in \cite{DBLP:conf/nips/HeuselRUNH17}. While InceptionV3 is used for colour images, VAE is used for grey-scale datasets MNIST and Fashion-MNIST.} of the test set quantifying the image quality; and (iv) the computational time (and an additional coverage rate for coverage-guided testing). We note the observations on these 4 metrics in the following paragraphs.

\begin{table}[!htbp]
	\centering
	\caption{Evaluation of the generated test cases and detected AEs by \gls{PGD} Attack, coverage-guided testing and the proposed HDA testing (all results are averaged over 100 seeds). HDA can detect more high perception quality AEs by $p_g$+$\mathcal{R}_l$ seeds selection, which is evidenced by high AE Prop. and low FID. }
	\begin{subtable}{\linewidth}
		\centering
		\resizebox{\linewidth}{!}{
			\begin{tabular}{c|c|c|ccccc}
				\hline
				\begin{tabular}[c]{@{}c@{}}AE Detection\\ Method\end{tabular} & Test Seeds & Metric & MNIST & F.-MNIST & SVHN & CIFAR-10 & CelebA \\ \hline
				\multirow{8}{*}{PGD Attack} & \multirow{4}{*}{Random Seeds} & AE Prop. & $\mathbf{0.422\pm0.046}$ & $\mathbf{1.000\pm0.000}$ & $\mathbf{0.951\pm0.031}$ & $\mathbf{1.000\pm0.000}$ & $0.989\pm0.035$ \\
				&  & Pred. Loss & $\mathbf{6.362\pm1.089}$ & $\mathbf{29.200\pm15.754}$ & $\mathbf{6.911\pm0.571}$ & $\mathbf{46.099\pm4.037}$ & $\mathbf{46.502\pm48.154}$ \\
				&  & FID & $0.705\pm0.047$ & $2.938\pm0.249$ & $106.581\pm1.275$ & $96.149\pm2.265$ & $88.401\pm1.459$ \\
				&  & Time(s) & $\mathbf{0.038\pm0.070}$ & $\mathbf{0.035\pm0.013}$ & $\mathbf{9.632\pm0.274}$ & $\mathbf{9.294\pm0.532}$ & $\mathbf{8.998\pm0.646}$ \\ \cline{2-8} 
				& \multirow{4}{*}{$p_g$+$\mathcal{R}_l$ Seeds} & AE Prop. & $\mathbf{0.785\pm0.103}$ & $\mathbf{1.000\pm0.000}$ & $\mathbf{0.989\pm0.006}$ & $\mathbf{1.000\pm0.000}$ & $\mathbf{1.000\pm0.000}$ \\
				&  & Pred. Loss & $\mathbf{10.184\pm1.135}$ & $\mathbf{36.139\pm18.205}$ & $\mathbf{9.385\pm0.763}$ & $\mathbf{44.816\pm3.221}$ & $\mathbf{52.136\pm58.991}$ \\
				&  & FID & $0.696\pm0.051$ & $1.351\pm0.208$ & $103.269\pm1.153$ & $98.592\pm1.187$ & $83.709\pm1.095$ \\
				&  & Time(s) & $\mathbf{0.031\pm0.024}$ & $\mathbf{0.035\pm0.014}$ & $\mathbf{9.541\pm0.499}$ & $\mathbf{9.739\pm0.763}$ & $\mathbf{9.473\pm0.529}$ \\ \hline
				\multirow{10}{*}{\begin{tabular}[c]{@{}c@{}}Cov. Guided\\ Testing\end{tabular}} & \multirow{5}{*}{Random Seeds} & Cov. Rate & $0.980\pm0.002$ & $0.946\pm0.023$ & $0.966\pm0.002$ & $0.979\pm0.004$ & $0.985\pm0.002$ \\
				&  & AE Prop. & $0.0002\pm0.0006$ & $0.153\pm0.021$ & $0.014\pm0.005$ & $0.150\pm0.016$ & $0.080\pm0.033$ \\
				&  & Pred. Loss & $0.019\pm0.001$ & $2.404\pm1.512$ & $0.230\pm0.045$ & $3.732\pm1.098$ & $2.841\pm4.311$ \\
				&  & FID & $0.802\pm0.026$ & $3.771\pm0.357$ & $101.034\pm1.014$ & $87.899\pm1.807$ & $85.918\pm1.109$ \\
				&  & Time(s) & $248.902\pm11.088$ & $1189.87\pm223.93$ & $4416.95\pm144.93$ & $6678.35\pm690.49$ & $972.17\pm83.57$ \\ \cline{2-8} 
				& \multirow{5}{*}{$p_g$+$\mathcal{R}_l$ Seeds} & Cov. Rate & $0.977\pm0.003$ & $0.915\pm0.025$ & $0.921\pm0.008$ & $0.978\pm0.004$ & $0.978\pm0.002$ \\
				&  & AE Prop. & $0.030\pm0.017$ & $0.485\pm0.068$ & $0.138\pm0.030$ & $0.449\pm0.049$ & $0.286\pm0.056$ \\
				&  & Pred. Loss & $1.184\pm0.632$ & $2.867\pm1.244$ & $0.198\pm0.039$ & $3.607\pm0.459$ & $2.137\pm2.283$ \\
				&  & FID & $0.804\pm0.043$ & $1.816\pm0.285$ & $94.043\pm1.786$ & $90.491\pm1.071$ & $81.243\pm1.162$ \\
				&  & Time(s) & $81.33\pm7.09$ & $216.59\pm152.62$ & $3073.97\pm598.92$ & $1822.64\pm102.24$ & $2065.70\pm32.68$ \\ \hline
				\multirow{8}{*}{HDA} & \multirow{4}{*}{Random Seeds} & AE Prop. & $0.208\pm0.042$ & $0.999\pm0.003$ & $0.843\pm0.046$ & $1.000\pm0.000$ & $\mathbf{1.000\pm0.000}$ \\
				&  & Pred. Loss & $1.076\pm0.341$ & $5.214\pm3.325$ & $2.784\pm0.198$ & $32.251\pm2.287$ & $16.020\pm15.306$ \\
				&  & FID & $\mathbf{0.094\pm0.009}$ & $\mathbf{0.119\pm0.043}$ & $\mathbf{91.431\pm1.118}$ & $\mathbf{62.587\pm2.144}$ & $\mathbf{59.162\pm1.469}$ \\
				&  & Time(s) & $74.211\pm1.71$ & $136.652\pm22.71$ & $153.956\pm62.92$ & $351.992\pm74.62$ & $191.313\pm68.35$ \\ \cline{2-8} 
				& \multirow{4}{*}{$p_g$+$\mathcal{R}_l$ Seeds} & AE Prop. & $0.676\pm0.127$ & $\mathbf{1.000\pm0.000}$ & $\mathbf{0.989\pm0.006}$ & $\mathbf{1.000\pm0.000}$ & $\mathbf{1.000\pm0.000}$ \\
				&  & Pred. Loss & $1.591\pm0.306$ & $10.020\pm5.961$ & $3.925\pm0.433$ & $33.361\pm1.873$ & $20.702\pm21.941$ \\
				&  & FID & $\mathbf{0.042\pm0.011}$ & $\mathbf{0.016\pm0.004}$ & $\mathbf{91.102\pm1.342}$ & $\mathbf{65.829\pm2.154}$ & $\mathbf{55.478\pm1.399}$ \\
				&  & Time(s) & $204.002\pm85.96$ & $61.793\pm1.69$ & $237.882\pm98.26$ & $708.832\pm269.81$ & $305.499\pm82.47$ \\ \hline
			\end{tabular}
		}
		\caption{Results}
	\end{subtable}
	\begin{subtable}{\linewidth}
		\centering
		\resizebox{\linewidth}{!}{
			\begin{tabular}{cc|cccc|ccll}
				\hline
				\multirow{3}{*}{Dataset} & \multirow{3}{*}{Metric} & \multicolumn{4}{c|}{Random Seeds} & \multicolumn{4}{c}{$p_g$+$\mathcal{R}_l$ Seeds} \\ \cline{3-10} 
				&  & \multicolumn{2}{c}{$T(\text{HDA}, \text{PGD})$} & \multicolumn{2}{c|}{$T(\text{HDA}, \text{Cov.})$} & \multicolumn{2}{c}{$T(\text{HDA}, \text{PGD})$} & \multicolumn{2}{c}{$T(\text{HDA}, \text{Cov.})$} \\
				&  & $t$ & $p-value$ & $t$ & $p-value$ & $t$ & $p-value$ & \multicolumn{1}{c}{$t$} & \multicolumn{1}{c}{$p-value$} \\ \hline
				\multirow{2}{*}{MNIST} & AE Prop. & $-10.864$ & $2.458\times10^{-9}$ & $15.644$ & $6.365\times10^{-12}$ & $-2.108$ & $0.049$ & $15.943$ & $4.627\times10^{-12}$ \\
				& FID & $-40.376$ & $4.118\times10^{-19}$ & $-81.374$ & $1.469\times10^{-24}$ & $-39.640$ & $5.715\times10^{-19}$ & $-54.290$ & $2.080\times10^{-21}$ \\ \hline
				\multirow{2}{*}{Fashion-MNIST} & AE Prop. & $-1.054$ & $0.306$ & $126.114$ & $5.595\times10^{-28}$ & $nan$ & $nan$ & $23.949$ & $4.204\times10^{-15}$ \\
				& FID & $-35.279$ & $4.541\times10^{-18}$ & $-32.117$ & $2.400\times10^{-17}$ & $-20.293$ & $7.495\times10^{-14}$ & $-19.970$ & $9.882\times10^{-14}$ \\ \hline
				\multirow{2}{*}{SVHN} & AE Prop. & $-6.157$ & $8.186\times10^{-6}$ & $56.656$ & $9.695\times10^{-22}$ & $nan$ & $nan$ & $87.961$ & $3.629\times10^{-25}$ \\
				& FID & $-28.252$ & $2.312\times10^{-16}$ & $-20.119$ & $8.690\times10^{-14}$ & $-21.746$ & $2.259\times10^{-14}$ & $-4.163$ & $5.843\times10^{-4}$ \\ \hline
				\multirow{2}{*}{CIFAR-10} & AE Prop. & $nan$ & $nan$ & $167.996$ & $3.221\times10^{-30}$ & $nan$ & $nan$ & $35.559$ & $3.945\times10^{-18}$ \\
				& FID & $-34.030$ & $8.609\times10^{-18}$ & $-28.547$ & $1.925\times10^{-16}$ & $-42.126$ & $1.933\times10^{-19}$ & $-32.419$ & $2.033\times10^{-17}$ \\ \hline
				\multirow{2}{*}{CelebA} & AE Prop. & $0.994$ & $0.333$ & $88.160$ & $3.485\times10^{-25}$ & $nan$ & $nan$ & $40.319$ & $4.223\times10^{-19}$ \\
				& FID & $-44.658$ & $6.825\times10^{-20}$ & $-45.968$ & $4.073\times10^{-20}$ & $-50.251$ & $8.295\times10^{-21}$ & $-44.801$ & $6.449\times10^{-20}$ \\ \hline
			\end{tabular}
		}
		\caption{Statistical Test}
	\end{subtable}
	\label{test_cases}
\end{table}

Regarding the AE proportion in the set of generated test cases, the default setting of our proposed approach ($p_g$+$\mathcal{R}_l$ Seeds with HDA) has comparable performance with PGD attack, as indicated by identical results and $nan$ output by statistical test. Both our novel $p_g$+$\mathcal{R}_l$ test seed selection and two-step GA local test case generation methods contribute to the comparable performance. This is evident from the decreased AE proportion when using random seeds in our method, but still the result is relatively higher than most combinations. PGD attack, as a white-box approach using the gradient information, is definitely quite efficient in detecting AEs, especially when paired with our new test seed selection method. On the other hand, coverage-guided testing is comparatively less effective in detecting AEs (even with high coverage rate), but our test seed selection method can improve it.

As per the results of prediction loss, PGD attack, as a gradient based attack method, unsurprisingly finds the AEs with the largest prediction loss. With better test seeds considering local robustness information by our method, the prediction loss of PGD attack can be even higher. Both coverage-guided testing and our HDA testing can detect AEs with relatively lower prediction loss, meaning the AEs are with ``weaker adversarial strength''. The reason for the low prediction loss of AEs detected by our approach is that two-step GA makes the trade-off and sacrifices it for higher local probabilities (i.e., perceptual quality). This can be seen through the significantly smaller FID of test set generated by HDA, compared with PGD attack and coverage-guided testing. PGD attack has relatively high FID scores, as well as coverage-guided testing.

On the computational overheads, we observe PGD is the most efficient, given it is by nature a white-box approach using the gradient information. While, our approach is an end-to-end black-box approach (if without using the gradient based indicator when selecting test seeds) requiring less information and being more generic, at the price of being relatively less efficient. That said, the computational time of our approach is still acceptable and better than coverage-guided testing.

\begin{tcolorbox}
	Answer to \textbf{RQ2} on comparing with adversarial attack and coverage-guided testing: HDA can select more significant test seeds, which are from high probability density region and lack of robustness. HDA can generate higher perception quality AEs, which are measured with smaller FID values.
\end{tcolorbox}

\begin{table}[!htbp]
	\centering
	\caption{Evaluation of AEs detected by OODA, FODA and our HDA testing methods (based on 100 test seeds).}
	\begin{subtable}{\linewidth}
		\centering
		\resizebox{0.8\linewidth}{!}{
			\begin{tabular}{cccccc}
				\hline
				Dataset & \begin{tabular}[c]{@{}c@{}} AE Detection\\Method \end{tabular} & $p_g$ & \begin{tabular}[c]{@{}c@{}}\% of\\ Valid AEs\end{tabular} & $\epsilon$ & FID \\ \hline
				\multirow{3}{*}{MNIST} & OODA & $0.0018\pm0.0002$ & $22.5\pm8.3$ & $0.917\pm0.031$ & $3.463\pm0.729$ \\
				& FODA & $0.0032\pm0.0006$ & $98.5\pm0.4$ & $0.558\pm0.009$ & $0.158\pm0.017$ \\
				& HDA & $\mathbf{0.0292\pm0.0071}$ & $\mathbf{99.3\pm0.5}$ & $\mathbf{0.076\pm0.003}$ & $\mathbf{0.042\pm0.011}$ \\
				\hline
				\multirow{3}{*}{SVHN} & OODA & $0.0026\pm0.0004$ & $11.3\pm1.2$ & $0.811\pm0.009$ & $125.126\pm2.838$ \\
				& FODA & $0.0034\pm0.0002$ & $100\pm0$ & $0.252\pm0.012$ & $111.694\pm1.461$ \\
				& HDA & \textbf{$\mathbf{0.0132\pm0.0021}$} & $\mathbf{100\pm0}$ & $\mathbf{0.029\pm0.001}$ & $\mathbf{91.102\pm1.342}$ \\ \hline
			\end{tabular}
		}
		\caption{Results}
	\end{subtable}
	\begin{subtable}{\linewidth}
		\centering
		\resizebox{0.7\linewidth}{!}{
			\begin{tabular}{cccccc}
				\hline
				\multirow{2}{*}{Dataset} & \multirow{2}{*}{Metric} & \multicolumn{2}{c}{$T(\text{HDA}, \text{OODA})$} & \multicolumn{2}{c}{$T(\text{HDA}, \text{FODA})$} \\
				&  & $t$ & $p-value$ & $t$ & $p-value$ \\ \hline
				\multirow{4}{*}{MNIST} & $p_g$ & $12.199$ & $3.871\times10^{-10}$ & $11.539$ & $9.456\times10^{-10}$ \\
				& \% of Valid AEs & $29.208$ & $1.286\times10^{-16}$ & $3.951$ & $9.368\times10^{-4}$ \\
				& $\epsilon$ & $-85.391$ & $6.183\times10^{-25}$ & $-160.667$ & $7.186\times10^{-30}$ \\
				& FID & $-14.838$ & $1.546\times10^{-11}$ & $-18.116$ & $5.275\times10^{-13}$ \\ \hline
				\multirow{4}{*}{SVHN} & $p_g$ & $15.680$ & $6.124\times10^{-12}$ & $14.691$ & $1.826\times10^{-11}$ \\
				& \% of Valid AEs & $233.745$ & $8.456\times10^{-33}$ & $nan$ & $nan$ \\
				& $\epsilon$ & $-273.086$ & $5.146\times10^{-34}$ & $-58.563$ & $5.359\times10^{-22}$ \\
				& FID & $-34.273$ & $7.588\times10^{-18}$ & $-32.825$ & $1.632\times10^{-17}$ \\ \hline
			\end{tabular}
		}
		\caption{Statistical Test}
	\end{subtable}
	\label{distribution_aware_compare}
\end{table}

Next, we try to answer the difference between HDA testing and other distribution-aware testing as summarised earlier (the amber route of Fig.~\ref{workflow_op_test}). We not only study the common evaluation metrics in earlier RQs, but also the input validation method in \cite{dola_distribution_aware_2021}, which flags the validity of AEs according to a user-defined reconstruction probability threshold. Overall, HDA is significantly better than OODA and FODA in four evaluation metrics, as observed from the results of statistical test.


As shown in Table~\ref{distribution_aware_compare}, HDA can select test seeds from much higher density region on the global distribution and find more valid AEs than OODA. The reason behind this is that OODA aims at detecting outliers---only AEs with lower reconstruction probabilities (from the test seed) than the given threshold will be marked as invalid test cases. While, HDA explicitly explores the high density meanwhile error-prone regions by combining the global distribution and local robustness indicators. In other words, HDA performs priority ordering (according to the global distribution and local robustness) and then selects the best, while OODA rules out the worst. As expected, FODA performs similarly poorly as OODA in terms of $p_g$, since both use randomly selected seeds. However, FODA has high proportion of valid AEs since the test cases are directly sampled from the distribution in latent space.

Regarding the perceptual quality of detected AEs, HDA can always find AEs with small pixel-level perturbations ($\epsilon$) in consideration of the $r$-separation constraint, and with small FID thanks to the use of perceptual quality metrics (MSE in this case) as objective functions. While OODA only utilises the reconstruction probability (from VAE) to choose AEs, and FODA directly samples test cases from VAE without any restrictions (thus may suffer from the oracle problem, cf. Remark \ref{rm_oracle_prob_FODA_OODA} later). Due to the compression nature of generative models---they are good at extracting feature level information but ignore pixel level information \cite{zhong2020generative}, AEs detected by OODA and FODA are all distant to the original test seeds, yielding large $\epsilon$ and FID scores. Notably, the average distance $\epsilon$ between test seeds and AEs detected by OODA and FODA are much (7$\sim$28 times) greater than the $r$-separation constraints (cf. Table~\ref{table_model_details}), leading to the potential oracle issues of those AEs, for which we have the following remark:
\begin{remark}[Oracle Issues of AEs Detected by OODA and FODA]
	AEs detected by OODA and FODA are normally distant to the test seeds with a perturbation distance even greater than the $r$-separation constraint. Consequently, there is the risk that the perturbed image may not share the same ground truth label of the test seed, and thus hard to determine the ground truth label of the ``AE''\footnote{In quotes, because the perturbed image could be a ``benign example'' with a \textit{correct} predicted label (but different to the test seed).}.
	\label{rm_oracle_prob_FODA_OODA}
\end{remark}

\begin{figure}[!htbp]
	\centering
	\includegraphics[width=\linewidth]{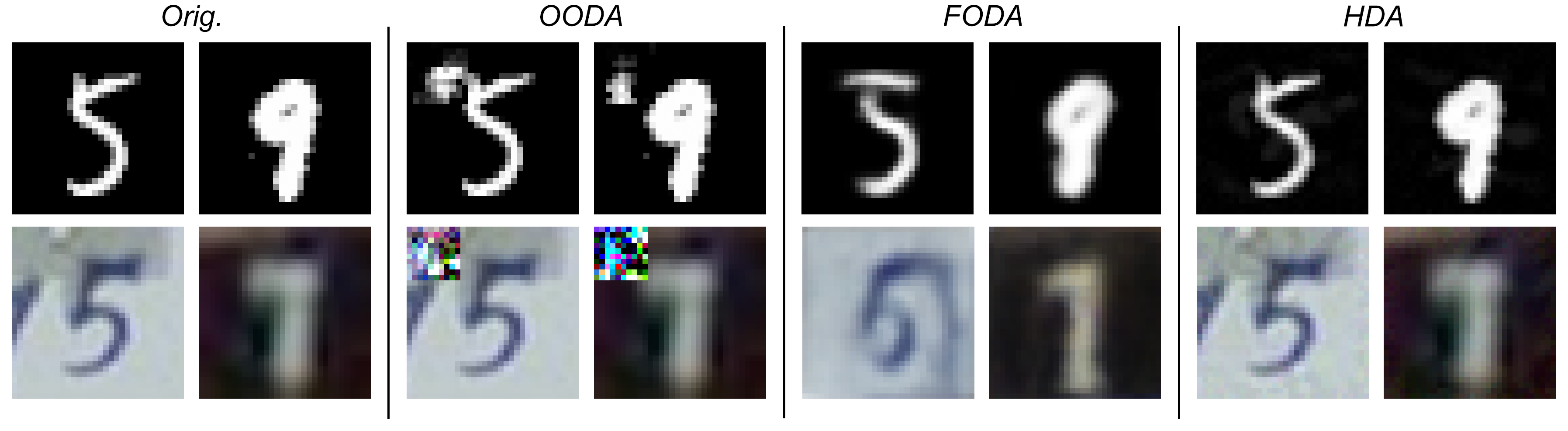}
	\caption{{\small Example AEs detected by different distribution-aware testing methods. AEs detected by our HDA are indistinguishable from the original images, while AEs detected by FODA and OODA are of low perceptual quality and subject to the oracle issues noted by Remark \ref{rm_oracle_prob_FODA_OODA}. }\normalsize}
	\label{dist_aware_adv_display}
\end{figure}

To visualise the difference between AEs detected by HDA, FODA and OODA, we present 4 examples in Fig.~\ref{dist_aware_adv_display}. We may observe the AEs detected by HDA are almost indistinguishable from the original images. Moreover, the AEs by FODA is a set of concrete evidence for Remark \ref{rm_oracle_prob_FODA_OODA}---it is actually quite hard to tell what is the ground truth label of some perturbed image (e.g., the bottom left one), while others appear to have a different label of the seed (e.g., the bottom right one should be with a label ``1'' instead of ``7'').


\begin{tcolorbox}
	Answer to \textbf{RQ2} on comparing with other distribution-aware testing: Compared to OODA and FODA, the proposed HDA testing can detect more valid AEs, free of oracle issues, with higher global probabilities and perception quality.
\end{tcolorbox}

\subsubsection{\textbf{RQ3}}

In earlier RQs, we have varied the datasets and model architectures to check the effectiveness of HDA. In this \textbf{RQ3}, we are concern about HDA's sensitivity to DL models with different levels of robustness. Adversarial training may greatly improve the robustness of DL models and is widely used as the defence to adversarial attack. To this end, we apply HDA on both normally and \textit{adversarially trained} models (by \cite{DBLP:conf/iclr/MadryMSTV18} to be exact), and then compare with three most representative adversarial attack methods---the most classic FGSM, the most popular PGD, and the most advanced AutoAttack \cite{DBLP:conf/icml/Croce020a}. Experimental results are presented in Table~\ref{new_exp}.

\begin{table}[!htbp]
	\centering
	\caption{Evaluation of AEs generated by \gls{FGSM}, PGD, AutoAttack and HDA on normally and adversarially trained DL models (all results are averaged over 100 test seeds).}
	\begin{subtable}{\linewidth}
		\centering
		\resizebox{1\linewidth}{!}{
			\begin{tabular}{c|c|c|ccccc}
				\hline
				Model & \begin{tabular}[c]{@{}c@{}}AE Detection\\ Method\end{tabular} & Eval. Metric & MNIST & F.-MNIST & SVHN & CIFAR-10 & CelebA \\ \hline
				\multirow{12}{*}{\begin{tabular}[c]{@{}c@{}}Normally\\ Trained\end{tabular}} & \multirow{3}{*}{FGSM} & $p_g$ & $0.0045\pm0.0005$ & $0.0034\pm0.0008$ & $0.0031\pm0.0002$ & $0.0028\pm0.0004$ & $0.0034\pm0.0002$ \\
				&  & AE Prop. & $0.413\pm0.093$ & $0.797\pm0.039$ & $0.727\pm0.069$ & $0.893\pm0.016$ & $0.989\pm0.035$ \\
				&  & FID & $1.022\pm0.053$ & $4.134\pm0.322$ & $114.256\pm2.931$ & $104.877\pm1.832$ & $88.401\pm1.459$ \\ \cline{2-8} 
				& \multirow{3}{*}{PGD} & $p_g$ & $0.0040\pm0.0005$ & $0.0034\pm0.0008$ & $0.0032\pm0.0002$ & $0.0028\pm0.0004$ & $0.0034\pm0.0002$ \\
				&  & AE Prop. & $0.422\pm0.046$ & $1.000\pm0.000$ & $0.984\pm0.011$ & $1.000\pm0.000$ & $0.989\pm0.035$ \\
				&  & FID & $0.705\pm0.047$ & $2.938\pm0.248$ & $107.370\pm1.206$ & $101.157\pm1.774$ & $88.401\pm1.459$ \\ \cline{2-8} 
				& \multirow{3}{*}{AutoAttack} & $p_g$ & $0.0042\pm0.0004$ & $0.0034\pm0.0008$ & $0.0032\pm0.0002$ & $0.0028\pm0.0004$ & $0.0035\pm0.0002$ \\
				&  & AE Prop. & $\mathbf{0.787\pm0.075}$ & $1.000\pm0.000$ & $0.984\pm0.011$ & $1.000\pm0.000$ & $1.000\pm0.000$ \\
				&  & FID & $0.795\pm0.099$ & $4.962\pm0.359$ & $106.780\pm1.297$ & $101.173\pm1.543$ & $90.385\pm1.305$ \\ \cline{2-8} 
				& \multirow{3}{*}{HDA} & $p_g$ & $\mathbf{0.0292\pm0.0071}$ & $\mathbf{0.0224\pm0.0087}$ & $\mathbf{0.0132\pm0.0021}$ & $\mathbf{0.0406\pm0.012}$ & $\mathbf{0.0124\pm0.001}$ \\
				&  & AE Prop. & $0.676\pm0.127$ & $\mathbf{1.000\pm0.000}$ & $\mathbf{0.989\pm0.006}$ & $\mathbf{1.000\pm0.000}$ & $\mathbf{1.000\pm0.000}$ \\
				&  & FID & $\mathbf{0.042\pm0.011}$ & $\mathbf{0.016\pm0.004}$ & $\mathbf{91.102\pm1.342}$ & $\mathbf{65.829\pm2.154}$ & $\mathbf{55.478\pm1.399}$ \\ \hline
				\multirow{12}{*}{\begin{tabular}[c]{@{}c@{}}Adversarially\\ Trained\end{tabular}} & \multirow{3}{*}{FGSM} & $p_g$ & $0.0042\pm0.0005$ & $0.0033\pm0.0004$ & $0.0036\pm0.0004$ & $0.0028\pm0.0005$ & $0.0031\pm0.0003$ \\
				&  & AE Prop. & $0.033\pm0.021$ & $0.208\pm0.037$ & $0.323\pm0.061$ & $0.455\pm0.039$ & $0.404\pm0.045$ \\
				&  & FID & $1.064\pm0.055$ & $4.898\pm0.503$ & $135.730\pm4.321$ & $113.106\pm3.989$ & $108.070\pm4.368$ \\ \cline{2-8} 
				& \multirow{3}{*}{PGD} & $p_g$ & $0.0038\pm0.0005$ & $0.0032\pm0.0004$ & $0.0034\pm0.0004$ & $0.0028\pm0.0005$ & $0.0032\pm0.0003$ \\
				&  & AE Prop. & $0.014\pm0.017$ & $0.201\pm0.039$ & $0.507\pm0.049$ & $0.455\pm0.039$ & $0.372\pm0.036$ \\
				&  & FID & $0.762\pm0.040$ & $3.871\pm0.407$ & $124.388\pm2.613$ & $113.106\pm3.989$ & $111.931\pm2.539$ \\ \cline{2-8} 
				& \multirow{3}{*}{AutoAttack} & $p_g$ & $0.0031\pm0.0004$ & $0.0031\pm0.0004$ & $0.0034\pm0.0004$ & $0.0028\pm0.0005$ & $0.0031\pm0.0003$ \\
				&  & AE Prop. & $0.049\pm0.021$ & $0.268\pm0.038$ & $0.505\pm0.049$ & $0.419\pm0.039$ & $0.405\pm0.045$ \\
				&  & FID & $0.035\pm0.009$ & $0.128\pm0.046$ & $125.382\pm2.315$ & $117.257\pm3.750$ & $107.742\pm4.169$ \\ \cline{2-8} 
				& \multirow{3}{*}{HDA} & $p_g$ & $\mathbf{0.0214\pm0.0075}$ & $\mathbf{0.0118\pm0.0032}$ & $\mathbf{0.0154\pm0.0021}$ & $\mathbf{0.0261\pm0.0112}$ & $\mathbf{0.0103\pm0.0007}$ \\
				&  & AE Prop. & $\mathbf{0.368\pm0.071}$ & $\mathbf{0.903\pm0.021}$ & $\mathbf{0.960\pm0.029}$ & $\mathbf{0.865\pm0.051}$ & $\mathbf{0.968\pm0.012}$ \\
				&  & FID & $\mathbf{0.029\pm0.003}$ & $\mathbf{0.058\pm0.010}$ & $\mathbf{92.389\pm1.152}$ & $\mathbf{62.804\pm2.398}$ & $\mathbf{58.558\pm0.895}$ \\ \hline
			\end{tabular}
		}
		\caption{Results}
	\end{subtable}
	\begin{subtable}{\linewidth}
		\centering
		\resizebox{0.8\linewidth}{!}{
			\begin{tabular}{c|c|c|cccccc}
				\hline
				\multirow{2}{*}{Model} & \multirow{2}{*}{Dataset} & \multirow{2}{*}{Metric} & \multicolumn{2}{c}{$T(\text{HDA}, \text{FGSM})$} & \multicolumn{2}{c}{$T(\text{HDA}, \text{PGD})$} & \multicolumn{2}{c}{$T(\text{HDA}, \text{AutoAttack})$}  \\
				& &  & $t$ & $p-value$ & $t$ & $p-value$ & $t$ & $p-value$  \\ \hline
				\multirow{15}{*}{\begin{tabular}[c]{@{}c@{}}Normally\\ Trained\end{tabular}} & \multirow{3}{*}{MNIST} & $p_g$ & $10.97$ & $2.10\times10^{-9}$ & $11.20$ & $1.53\times10^{-9}$ & $11.12$ & $1.71\times10^{-9}$  \\
				& & AE Prop. & $5.28$ & $5.05\times10^{-5}$ & $5.95$ & $1.26\times10^{-5}$ & $-2.38$ & $0.03$  \\
				& & FID & $-57.25$ & $8.04\times10^{-22}$ & $-43.43$ & $1.12\times10^{-19}$ & $-23.91$ & $4.34\times10^{-15}$ \\ \cline{2-9}
				& \multirow{3}{*}{F.-MNIST} & $p_g$ & $6.88$ & $1.97\times10^{-6}$ & $6.88$ & $1.97\times10^{-6}$ & $6.88$ & $1.97\times10^{-6}$ \\
				& & AE Prop. & $16.46$ & $2.70\times10^{-12}$ & $nan$ & $nan$ & $nan$ & $nan$ \\
				& & FID & $-40.44$ & $4.01\times10^{-19}$ & $-37.25$ & $1.72\times10^{-18}$ & $-43.56$ & $1.06\times10^{-19}$ \\ \cline{2-9}
				& \multirow{3}{*}{SVHN} & $p_g$ & $15.14$ & $1.10\times10^{-11}$ & $14.99$ & $1.30\times10^{-11}$ & $14.99$ & $1.30\times10^{-11}$  \\
				& & AE Prop. & $11.96$ & $5.31\times10^{-10}$ & $1.26$ & $0.22$ & $1.26$ & $0.22$  \\
				& & FID & $-22.71$ & $1.05\times10^{-14}$ & $-28.51$ & $1.97\times10^{-16}$ & $-26.56$ & $6.84\times10^{-16}$ \\ \cline{2-9}
				& \multirow{3}{*}{CIFAR-10} & $p_g$ & $9.96$ & $9.56\times10^{-9}$ & $9.96$ & $9.56\times10^{-9}$ & $9.96$ & $9.56\times10^{-9}$ \\
				& & AE Prop. & $21.15$ & $3.67\times10^{-14}$ & $nan$ & $nan$ & $nan$ & $nan$  \\
				& & FID & $-43.67$ & $1.02\times10^{-19}$ & $-40.03$ & $4.79\times10^{-19}$ & $-42.18$ & $1.89\times10^{-19}$ \\ \cline{2-9}
				& \multirow{3}{*}{CelebA} & $p_g$ & $27.91$ & $2.87\times10^{-16}$ & $27.91$ & $2.87\times10^{-16}$ & $27.59$ & $3.49\times10^{-16}$ \\
				& & AE Prop. & $0.99$ & $0.33$ & $0.99$ & $0.33$ & $nan$ & $nan$ \\
				& & FID & $-51.51$ & $5.34\times10^{-21}$ & $-51.51$ & $5.34\times10^{-21}$ & $-57.69$ & $6.99\times10^{-22}$ \\ \hline
				\multirow{15}{*}{\begin{tabular}[c]{@{}c@{}}Adversarially\\ Trained\end{tabular}} & \multirow{3}{*}{MNIST} & $p_g$ & $7.24$ & $9.92\times10^{-7}$ & $7.40$ & $7.24\times10^{-7}$ & $7.71$ & $4.17\times10^{-7}$ \\
				& & AE Prop. & $14.31$ & $2.83\times10^{-11}$ & $15.33$ & $8.9\times10^{-12}$ & $13.62$ & $6.37\times10^{-11}$  \\
				& & FID & $-59.42$ & $4.12\times10^{-22}$ & $-57.79$ & $6.81\times10^{-22}$ & $-2.00$ & $0.06$  \\ \cline{2-9}
				& \multirow{3}{*}{F.-MNIST} & $p_g$ & $8.33$ & $1.36\times10^{-7}$ & $8.43$ & $1.15\times10^{-7}$ & $8.53$ & $9.70\times10^{-8}$ \\
				& & AE Prop. & $51.66$ & $5.06\times10^{-21}$ & $50.12$ & $8.70\times10^{-21}$ & $46.25$ & $3.65\times10^{-20}$ \\
				& & FID &$-30.42$ & $6.26\times10^{-17}$ & $-29.62$ & $1.06\times10^{-16}$ & $-4.70$ & $1.78\times10^{-4}$  \\ \cline{2-9}
				& \multirow{3}{*}{SVHN} & $p_g$ & $17.46$ & $9.95\times10^{-13}$ & $17.75$ & $7.47\times10^{-13}$ & $17.75$ & $7.47\times10^{-13}$ \\
				& & AE Prop. & $29.82$ & $8.90\times10^{-17}$ & $25.16$ & $1.77\times10^{-15}$ & $25.27$ & $1.64\times10^{-15}$  \\
				& & FID & $-30.65$ & $5.50\times10^{-17}$ & $-35.43$ & $4.19\times10^{-18}$ & $-40.35$ & $4.17\times10^{-19}$ \\ \cline{2-9}
				& \multirow{3}{*}{CIFAR-10} & $p_g$ & $6.57$ & $3.57\times10^{-6}$ & $6.57$ & $3.57\times10^{-6}$ & $6.57$ & $3.57\times10^{-6}$ \\
				& & AE Prop. & $20.19$ & $8.15\times10^{-14}$ & $20.19$ & $8.15\times10^{-14}$ & $21.97$ & $1.89\times10^{-14}$ \\
				& & FID &$-34.18$ & $7.98\times10^{-18}$ & $-34.18$ & $7.98\times10^{-18}$ & $-38.69$ & $8.82\times10^{-19}$  \\ \cline{2-9}
				& \multirow{3}{*}{CelebA} & $p_g$ & $29.90$ & $8.52\times10^{-17}$ & $29.48$ & $1.09\times10^{-16}$ & $29.90$ & $8.52\times10^{-17}$  \\
				& & AE Prop. & $38.30$ & $1.05\times10^{-18}$ & $49.67$ & $1.02\times10^{-20}$ & $38.23$ & $1.09\times10^{-18}$ \\
				& & FID & $-35.12$ & $4.93\times10^{-18}$ & $-62.69$ & $1.58\times10^{-22}$ & $-36.48$ & $2.51\times10^{-18}$ \\ \hline
			\end{tabular}
		}
		\caption{Statistical Test}
	\end{subtable}
	\label{new_exp}
\end{table}

As expected, after the adversarial training by \cite{DBLP:conf/iclr/MadryMSTV18}, the robustness of all five DL models are greatly improved. This can be observed from the metric of AE Prop.: For all four methods, the proportion of AEs detected in the set of test case is sharply decreased for adversarially trained models. Nevertheless, the AE detection performance of HDA is less affected by the adversarial training. HDA testing can maintain significantly higher proportion of AEs in the test set for adversarially trained models, compared with the state of art adversarial attack. This is evident from the fact that there is an insignificant difference in AE Prop. between HDA and other adversarial attacks for normally trained models, while the difference is significant for adversarially trained models.

In terms of the probability density $p_g$ and perception quality measured by FID on generated test cases, HDA significantly outperforms others for both normally and adversarially trained models, as observed from the large t-scores and p-values $<<0.05$ . This is unsurprising, since the rationales behind the three adversarial attacks disregard the consideration of input data distribution and perception quality. 

Finally, we find that the measured $p_g$ on test cases detected by HDA changes due to the variations in local robustness before and after the adversarial training, yet it remained much higher than all other attack methods.





\begin{tcolorbox}
	Answer to \textbf{RQ3}: HDA is shown to be capable and superior to common adversarial attacks when applied on DL models with different levels of robustness.
\end{tcolorbox}

\subsubsection{\textbf{RQ4}}
The ultimate goal of developing HDA testing is to improve the global robustness of DL models. To this end, we refer to a validation set of 10000 test seeds. We fine-tune \cite{jeddi_simple_2021} the DL models with AEs detected for validation set from different methods. Then, we calculate the train accuracy, test accuracy and empirical global robustness before and after the adversarial fine-tuning. Empirical global robustness is measured on a new set of on-distribution AEs for validation set, different from the fine-tuning data. Results are presented in Table~\ref{rob_compare}.

\begin{table}[!htbp]
	\centering
	\caption{Evaluation of DL models' train accuracy, test accuracy, and empirical global robustness (based on 10000 on-distribution AEs) after adversarial fine-tuning, using different number (N) of test cases.}
	\begin{subtable}{\linewidth}
		\centering
		\resizebox{1\linewidth}{!}{
			\begin{tabular}{cc|ccc|ccc}
				\hline
				\multirow{2}{*}{\begin{tabular}[c]{@{}c@{}}AE Detection\\ Method\end{tabular}} & \multirow{2}{*}{Metric} & \multicolumn{3}{c|}{MNIST} & \multicolumn{3}{c}{SVHN} \\
				&  & N = 500 & N = 5000 & N = 50000 & N = 500 & N = 5000 & N = 50000 \\ \hline
				\multirow{3}{*}{PGD Attack} & Train Acc. & $98.26\%\pm0.05\%$ & $98.10\%\pm0.07\%$ & $97.70\%\pm0.05\%$ & $94.85\%\pm0.43\%$ & $92.76\%\pm0.39\%$ & $94.02\%\pm0.40\%$ \\
				& Test Acc. & $97.64\%\pm0.11\%$ & $97.05\%\pm0.09\%$ & $96.90\%\pm0.08\%$ & $93.32\%\pm0.21\%$ & $81.43\%\pm0.31\%$ & $63.81\%\pm0.28\%$ \\
				& $\mathcal{R}_g$ & $46.27\%\pm0.02\%$ & $84.09\%\pm0.05\%$ & $90.52\%\pm0.05\%$ & $46.43\%\pm0.11\%$ & $72.88\%\pm0.10\%$ & $70.09\%\pm0.12\%$ \\ \hline
				\multirow{3}{*}{Cov. Guided Testing} & Train Acc. & $99.89\%\pm0.05\%$ & $99.61\%\pm0.07\%$ & $99.12\%\pm0.05\%$ & $95.73\%\pm0.44\%$ & $93.65\%\pm0.42\%$ & $95.76\%\pm0.43\%$ \\
				& Test Acc. & $98.93\%\pm0.08\%$ & $98.77\%\pm0.08\%$ & $98.41\%\pm0.07\%$ & $94.11\%\pm0.19\%$ & $85.66\%\pm0.19\%$ & $76.43\%\pm0.19\%$ \\
				& $\mathcal{R}_g$ & $36.71\%\pm0.04\%$ & $48.91\%\pm0.04\%$ & $71.12\%\pm0.05\%$ & $16.21\%\pm0.11\%$ & $38.59\%\pm0.16\%$ & $56.58\%\pm0.12\%$ \\ \hline
				\multirow{3}{*}{OODA} & Train Acc. & $98.45\%\pm0.06\%$ & $98.01\%\pm0.06\%$ & $97.66\%\pm0.07\%$ & $94.12\%\pm0.38\%$ & $92.11\%\pm0.39\%$ & $93.21\%\pm0.38\%$ \\
				& Test Acc. & $98.12\%\pm0.10\%$ & $97.87\%\pm0.09\%$ & $97.12\%\pm0.08\%$ & $94.75\%\pm0.19\%$ & $80.23\%\pm0.18\%$ & $72.19\%\pm0.19\%$ \\
				& $\mathcal{R}_g$ & $40.21\%\pm0.05\%$ & $45.69\%\pm0.04\%$ & $51.12\%\pm0.05\%$ & $11.21\%\pm0.11\%$ & $16.23\%\pm0.12\%$ & $18.21\%\pm0.11\%$ \\ \hline
				\multirow{3}{*}{FODA} & Train Acc. & $98.44\%\pm0.05\%$ & $98.18\%\pm0.05\%$ & $97.32\%\pm0.06\%$ & $94.66\%\pm0.42\%$ & $92.75\%\pm0.41\%$ & $94.11\%\pm0.41\%$ \\
				& Test Acc. & $97.87\%\pm0.10\%$ & $97.43\%\pm0.10\%$ & $97.11\%\pm0.11\%$ & $92.13\%\pm0.22\%$ & $82.32\%\pm0.21\%$ & $78.19\%\pm0.22\%$ \\
				& $\mathcal{R}_g$ & $37.71\%\pm0.04\%$ & $47.26\%\pm0.05\%$ & $55.37\%\pm0.05\%$ & $12.21\%\pm0.12\%$ & $20.56\%\pm0.12\%$ & $23.98\%\pm0.11\%$ \\ \hline
				\multirow{3}{*}{HDA} & Train Acc. & $99.56\%\pm0.07\%$ & $99.10\%\pm0.07\%$ & $98.89\%\pm0.06\%$ & $95.00\%\pm0.33\%$ & $92.83\%\pm0.39\%$ & $94.40\%\pm0.40\%$ \\
				& Test Acc. & $98.52\%\pm0.09\%$ & $98.42\%\pm0.08\%$ & $98.21\%\pm0.08\%$ & $93.86\%\pm0.24\%$ & $88.67\%\pm0.25\%$ & $80.60\%\pm0.22\%$ \\
				& $\mathcal{R}_g$ & $\mathbf{89.67\%\pm0.03\%}$ & $\mathbf{96.71\%\pm0.06\%}$ & $\mathbf{99.12\%\pm0.05\%}$ & $\mathbf{51.15\%\pm0.09\%}$ & $\mathbf{86.88\%\pm0.09\%}$ & $\mathbf{91.26\%\pm0.05\%}$ \\ \hline
			\end{tabular}
		}
		\caption{Results}
	\end{subtable}
	\begin{subtable}{\linewidth}
		\centering
		\resizebox{0.95\linewidth}{!}{
			\begin{tabular}{c|c|c|cccccccc}
				\hline
				\multirow{2}{*}{Dataset} & \multirow{2}{*}{\begin{tabular}[c]{@{}c@{}}No. of\\ Test Cases\end{tabular}} & \multirow{2}{*}{Metric} & \multicolumn{2}{c}{T(HDA, PGD)} & \multicolumn{2}{c}{T(HDA, Cov.)} & \multicolumn{2}{c}{T(HDA, OODA)} & \multicolumn{2}{c}{T(HDA, FODA)} \\
				&  &  & $t$ & $p-value$ & $t$ & $p-value$ & $t$ & $p-value$ & $t$ & $p-value$ \\ \hline
				\multirow{9}{*}{MNIST} & \multirow{3}{*}{N=500} & Train Acc. & $47.79$ & $2.04\times10^{-20}$ & $-12.13$ & $4.24\times10^{-10}$ & $38.07$ & $1.17\times10^{-18}$ & $41.17$ & $2.91\times10^{-19}$ \\
				&  & Test Acc. & $19.58$ & $1.39\times10^{-13}$ & $-10.77$ & $2.83\times10^{-9}$ & $9.40$ & $2.29\times10^{-8}$ & $15.28$ & $9.47\times10^{-12}$ \\
				&  & $\mathcal{R}_g$ & $3806.43$ & $1.31\times10^{-54}$ & $3349.48$ & $1.31\times10^{-53}$ & $2682.35$ & $7.12\times10^{-52}$ & $3286.24$ & $1.84\times10^{-53}$ \\ \cline{2-11} 
				& \multirow{3}{*}{N=5000} & Train Acc. & $31.94$ & $2.64\times10^{-17}$ & $-16.29$ & $3.21\times10^{-12}$ & $37.39$ & $1.62\times10^{-18}$ & $33.82$ & $9.61\times10^{-18}$ \\
				&  & Test Acc. & $35.98$ & $3.20\times10^{-18}$ & $-9.78$ & $1.25\times10^{-8}$ & $14.44$ & $2.42\times10^{-11}$ & $24.45$ & $2.94\times10^{-15}$ \\
				&  & $\mathcal{R}_g$ & $510.97$ & $6.52\times10^{-39}$ & $2096.17$ & $6.03\times10^{-50}$ & $2237.38$ & $1.86\times10^{-50}$ & $2002.17$ & $1.38\times10^{-49}$ \\ \cline{2-11} 
				& \multirow{3}{*}{N=50000} & Train Acc. & $48.18$ & $1.76\times10^{-20}$ & $-9.31$ & $2.64\times10^{-8}$ & $42.19$ & $1.88\times10^{-19}$ & $58.51$ & $5.45\times10^{-22}$ \\
				&  & Test Acc. & $36.62$ & $2.35\times10^{-18}$ & $-5.95$ & $1.25\times10^{-5}$ & $30.47$ & $6.10\times10^{-17}$ & $25.57$ & $1.33\times10^{-15}$ \\
				&  & $\mathcal{R}_g$ & $384.60$ & $1.08\times10^{-36}$ & $1252.19$ & $6.42\times10^{-46}$ & $2146.63$ & $3.93\times10^{-50}$ & $1956.56$ & $2.08\times10^{-49}$ \\ \hline
				\multirow{9}{*}{SVHN} & \multirow{3}{*}{N=500} & Train Acc. & $0.88$ & $0.39$ & $-4.19$ & $5.42\times10^{-4}$ & $5.53$ & $2.99\times10^{-5}$ & $2.01$ & $0.06$ \\
				&  & Test Acc. & $5.35$ & $4.34\times10^{-5}$ & $-2.58$ & $0.02$ & $-9.19$ & $3.20\times10^{-8}$ & $16.80$ & $1.90\times10^{-12}$ \\
				&  & $\mathcal{R}_g$ & $105.02$ & $1.50\times10^{-26}$ & $777.40$ & $3.42\times10^{-42}$ & $888.65$ & $3.08\times10^{-43}$ & $820.93$ & $1.28\times10^{-42}$ \\ \cline{2-11} 
				& \multirow{3}{*}{N=5000} & Train Acc. & $0.40$ & $0.69$ & $-4.52$ & $2.63\times10^{-4}$ & $4.13$ & $6.31\times10^{-4}$ & $0.45$ & $0.66$ \\
				&  & Test Acc. & $57.49$ & $7.46\times10^{-22}$ & $30.31$ & $6.67\times10^{-17}$ & $86.64$ & $4.77\times10^{-25}$ & $61.50$ & $2.23\times10^{-22}$ \\
				&  & $\mathcal{R}_g$ & $329.07$ & $1.79\times10^{-35}$ & $831.84$ & $1.01\times10^{-42}$ & $1489.43$ & $2.83\times10^{-47}$ & $1398.15$ & $8.83\times10^{-47}$ \\ \cline{2-11} 
				& \multirow{3}{*}{N=50000} & Train Acc. & $2.12$ & $0.05$ & $-7.32$ & $8.43\times10^{-7}$ & $6.82$ & $2.19\times10^{-6}$ & $1.60$ & $0.13$ \\
				&  & Test Acc. & $149.10$ & $2.75\times10^{-29}$ & $45.36$ & $5.16\times10^{-20}$ & $91.49$ & $1.79\times10^{-25}$ & $24.50$ & $1.83\times10^{-15}$ \\
				&  & $\mathcal{R}_g$ & $514.96$ & $5.67\times10^{-39}$ & $843.60$ & $7.86\times10^{-43}$ & $1911.81$ & $3.16\times10^{-49}$ & $1760.80$ & $1.39\times10^{-48}$ \\ \hline
			\end{tabular}
		}
		\caption{Statistical Test}
	\end{subtable}
	\label{rob_compare}
\end{table}

We first observe that adversarial fine-tuning is effective to improve the DL models' empirical global robustness, measured by the prediction accuracy on AEs, detected from normally trained models ($\mathcal{R}_g$), while compromising the train/test accuracy as expected (in contrast to normal training in Table~\ref{table_model_details}). In most cases, DL models enhanced by HDA testing suffer from the least drop of generalisation, compared with PGD attack, OODA and FODA. This can been from significant difference of Test Acc. between HDA and others. The reason behind this is that HDA testing targets at AEs from high density regions on distributions, usually with small prediction loss, shown in Fig.~\ref{ga_exp}. Thus, eliminating AEs detected by HDA testing requires relatively minor adjustment to DL's models, the generalisation of which can be easily tampered during the fine-tuning with new samples.  

In terms of empirical global robustness, HDA testing detects AEs around test seeds from the high global distribution region, which are more significant to the global robustness improvement. When comparing HDA with other methods, the p-values of $\mathcal{R}_g$ is way smaller than 0.05, indicating HDA contributes more to the global robustness improvement than others. It also can be seen that with 5000 test cases generated by utilising 1000 test seeds, the HDA testing can improve empirical global robustness to nearly or over $90\%$, very closed to the fine-tuning with 50000 test cases from 10000 test seeds. This means the distribution-based test seeds selection is more efficient than random test seeds selection. Moreover, even fine-tuning with 50000 test cases, leveraging all the test seeds in the validation set, HDA is still better than others, due to the consideration of local distributions (approximated by perceptual quality metrics). We notice that PGD-based adversarial fine-tuning minimises the maximum prediction loss within the local region, which is also effective to eliminate the high perceptual quality AEs, but sacrificing more train/test accuracy. DL models fined-tuned with HDA testing achieve the best balance between the generalisation and global robustness.

\begin{tcolorbox}
	Answer to \textbf{RQ4}: Compared with adversarial attack and coverage-guide testing, HDA contributes more to the growth of global robustness, while mitigating the drop of train/test accuracy during adversarial fine-tuning.
\end{tcolorbox}


\section{Threats to Validity}
\label{sec_treats_validity}


\subsection{Internal Validity}
Threats may arise due to bias in establishing cause-effect relationships, simplifications and assumptions made in our experiments. In what follows, we list the main threats of each research question and discuss how we mitigate them. 

\subsubsection{Threats from HDA Techniques}

In \textbf{RQ1}, both the performance of the VAE-Encoder and KDE are threats. For the former, it is mitigated by using four established quality metrics (in Table \ref{pca_vae}) on evaluating dimensionality reduction techniques and compared to the common PCA method. It is known that KDE performs poorly with high-dimensional data and works well when the data dimension is modest \cite{scott1991feasibility,liu2007sparse}.
The data dimensions in our experiments are relatively low given the datasets have been compressed by VAE-Encoder, which mitigates the second threat. When studying the local robustness indicators, quantifying both the indicators and the local robustness may subject to errors, for which we reduce them by carefully inspecting the correctness of the script on calculating the indicators and invoking a reliable local robustness estimator \cite{webb_statistical_2019} with fine-tuned hyper-parameters. For using two-step GA to generate local test cases, a threat arises by the calculation of norm ball radius, which has been mitigated by $r$-separation distance presented in the paper \cite{yang_closer_2020}. Also, the threat related to estimating the local distribution is mitigated by quantifying its three indicators (MSE, PSNR and SSIM) that are typically used in representing image-quality by human-perception. 

\subsubsection{Threats from AEs' Quality Measurement}
A threat for \textbf{RQ1}, \textbf{RQ2} and \textbf{RQ3} (when examining how effective our method models the global distribution and local distribution respectively) is the use of FID as a metric, quantifying how ``similar'' two image datasets are. Given FID is currently the standard metric for this purpose, this threat is sufficiently mitigated now and can be further mitigated with new metrics in future. \textbf{RQ2} includes the method of validating AEs developed in \cite{dola_distribution_aware_2021}, which utilises generative models and OOD techniques to flag valid AEs with reconstruction probabilities greater than a threshold. The determination of this threshold is critical, thus poses a threat to \textbf{RQ2}. To mitigate it, we use same settings across all the experiments for fair comparisons.

\subsubsection{Threats from Adversarial Training and Fine-Tuning} In \textbf{RQ3} and \textbf{RQ4}, the first threat rises from the fact that adversarial training and adversarial fine-tuning will sacrifice the DL model's generalisation for robustness. Since the training process is data-driven and of black-box nature, it is hard to know how the predication of a single data-point will be affected, while it is meaningless to study the robustness of an incorrectly predicted seed.
To mitigate this threat when we compare the robustness before and after adversarial training/fine-tuning, we select enough number of seeds and check the prediction of each selected seed (filtering out incorrect ones if necessary) to make sure test seeds are always predicted correctly.
For the global robustness computation in \textbf{RQ4}, we refer to a validation dataset, where a threat may arise if the empirical result based on the validation dataset cannot represent the global robustness. To mitigate it, we synthesise the validation set with enough data---10000 inputs sampled from global distribution. We further attack the validation dataset to find an AE per seed according to the local distribution. Thus, DL models' prediction accuracy on this dataset empirically represents the global robustness as defined. For the training/fine-tuning to be effective, we need a sufficient number of AEs to augment the training dataset. A threat may arise due to a small proportion of AEs in the augmented training dataset (the DL model will be dominated by the original training data during the training/fine-tuning). To mitigate such a threat, we generate a large proportion of AEs in our experiments.


\subsection{External Validity}
Threats might challenge the generalisability of our findings, e.g. the number of models and datasets considered for experimentation; thus we mitigate these threats as follows. All our experiments are conducted on 5 popular benchmark datasets, covering 5 typical types of DL models, cf. Table \ref{table_model_details}. Experimental results on the effectiveness of each stage in our framework are all based on averaging a large number of samples, reducing the random noise in the experiments. In two-step GA based test case generation, a wide range of the $\alpha$ parameter has been studied showing converging trends. Finally, we enable replication by making all experimental results publicly available/reproducible on our project website to further mitigate the threat.

\section{Conclusion \& Future Work}
\label{sec_conclusion}

In this paper, we propose a HDA testing approach for detecting AEs that considers both the data distribution (thus with higher operational impact assuming the training data statistically representing the future inputs) and perceptual quality (thus looks natural and realistic to humans). The key novelty lies in the hierarchical consideration of two levels of distributions. To the best of our knowledge, it is the first DL testing approach that explicitly and collectively models both (i) the feature-level information when selecting test seeds and (ii) pixel-level information when generating local test cases. To this end, we have developed a tool chain that provides technical solutions for each stage of our HDA testing. Our experiments not only show the effectiveness of each testing stage, but also the overall advantages of HDA testing over state-of-the-arts. From a software engineering's perspective, HDA is cost-effective (by focusing on practically meaningful AEs), flexible (with end-to-end, black-box technical solutions) and may effectively contribute to the robustness growth of the DL software under testing.

The purpose of detecting AEs is to fix them. 
Although existing DL retraining/repairing techniques (e.g. \cite{jeddi_simple_2021} used in \textbf{RQ4} and \cite{DBLP:conf/icml/WangM0YZG19,9508369}) may satisfy the purpose to some extent, \textit{bespoke} ``debugging'' methods with more emphasise on the feature-distribution and perceptual quality can be integrated into our framework in a more efficient way. To this end, our important future work is to close the loop of ``detect-fix-assess'' as depicted in \cite{zhao_detecting_2021} and then organise all generated evidence as safety cases \cite{zhao_safety_2020,dong_reliability_2023}. Finally, same as other distribution-aware testing methods, we assume the input data distribution is same as the training data distribution. To relax this assumption, we plan to take distribution-shift into consideration in future versions of HDA. Distribution-aware testing for systematically detecting explanation AEs \cite{huang2023safari} will also be explored.

\begin{acks}
	This work is supported by the U.K. DSTL (through the project of Safety Argument for Learning-enabled Autonomous Underwater Vehicles) and the U.K. EPSRC (through End-to-End Conceptual Guarding of Neural Architectures [EP/T026995/1]). Xingyu Zhao and Alec Banks’ contribution to the work is partially supported through Fellowships at the Assuring Autonomy International Programme. 
	This project has received funding from the European Union’s Horizon 2020 research and innovation programme under grant agreement No 956123. We thank all three anonymous reviewers whose comments helped improve the paper.
	
	This document is an overview of U.K. MOD (part) sponsored research and is released for informational purposes only. The contents of this document should not be interpreted as representing the views of the U.K. MOD, nor should it be assumed that they reflect any current or future U.K. MOD policy. The information contained in this document cannot supersede any statutory or contractual requirements or liabilities and is offered without prejudice or commitment. 
\end{acks}


\bibliographystyle{ACM-Reference-Format}
\bibliography{ref}


\begin{thebibliography}{66}


\ifx \showCODEN    \undefined \def \showCODEN     #1{\unskip}     \fi
\ifx \showDOI      \undefined \def \showDOI       #1{#1}\fi
\ifx \showISBNx    \undefined \def \showISBNx     #1{\unskip}     \fi
\ifx \showISBNxiii \undefined \def \showISBNxiii  #1{\unskip}     \fi
\ifx \showISSN     \undefined \def \showISSN      #1{\unskip}     \fi
\ifx \showLCCN     \undefined \def \showLCCN      #1{\unskip}     \fi
\ifx \shownote     \undefined \def \shownote      #1{#1}          \fi
\ifx \showarticletitle \undefined \def \showarticletitle #1{#1}   \fi
\ifx \showURL      \undefined \def \showURL       {\relax}        \fi
\providecommand\bibfield[2]{#2}
\providecommand\bibinfo[2]{#2}
\providecommand\natexlab[1]{#1}
\providecommand\showeprint[2][]{arXiv:#2}

\bibitem[\protect\citeauthoryear{Akoglu}{Akoglu}{2018}]%
        {akoglu2018user}
\bibfield{author}{\bibinfo{person}{Haldun Akoglu}.}
  \bibinfo{year}{2018}\natexlab{}.
\newblock \showarticletitle{User's guide to correlation coefficients}.
\newblock \bibinfo{journal}{\emph{Turkish journal of emergency medicine}}
  \bibinfo{volume}{18}, \bibinfo{number}{3} (\bibinfo{year}{2018}),
  \bibinfo{pages}{91--93}.
\newblock


\bibitem[\protect\citeauthoryear{Alzantot, Sharma, Chakraborty, Zhang, Hsieh,
  and Srivastava}{Alzantot et~al\mbox{.}}{2019}]%
        {alzantot2019genattack}
\bibfield{author}{\bibinfo{person}{Moustafa Alzantot}, \bibinfo{person}{Yash
  Sharma}, \bibinfo{person}{Supriyo Chakraborty}, \bibinfo{person}{Huan Zhang},
  \bibinfo{person}{Cho-Jui Hsieh}, {and} \bibinfo{person}{Mani~B Srivastava}.}
  \bibinfo{year}{2019}\natexlab{}.
\newblock \showarticletitle{Genattack: Practical black-box attacks with
  gradient-free optimization}. In \bibinfo{booktitle}{\emph{Proceedings of the
  Genetic and Evolutionary Computation Conference}}.
  \bibinfo{pages}{1111--1119}.
\newblock


\bibitem[\protect\citeauthoryear{Attaoui, Fahmy, Pastore, and Briand}{Attaoui
  et~al\mbox{.}}{2023}]%
        {attaoui2022black}
\bibfield{author}{\bibinfo{person}{Mohammed Attaoui}, \bibinfo{person}{Hazem
  Fahmy}, \bibinfo{person}{Fabrizio Pastore}, {and} \bibinfo{person}{Lionel
  Briand}.} \bibinfo{year}{2023}\natexlab{}.
\newblock \showarticletitle{Black-Box Safety Analysis and Retraining of DNNs
  Based on Feature Extraction and Clustering}.
\newblock \bibinfo{journal}{\emph{ACM Trans. Softw. Eng. Methodol.}}
  \bibinfo{volume}{32}, \bibinfo{number}{3}, Article \bibinfo{articleno}{79}
  (\bibinfo{year}{2023}), \bibinfo{numpages}{40}~pages.
\newblock


\bibitem[\protect\citeauthoryear{Berend}{Berend}{2021}]%
        {DBLP:conf/icse/Berend21}
\bibfield{author}{\bibinfo{person}{David Berend}.}
  \bibinfo{year}{2021}\natexlab{}.
\newblock \showarticletitle{Distribution Awareness for {AI} System Testing}. In
  \bibinfo{booktitle}{\emph{43rd {IEEE/ACM} Int. Conf. on Software Engineering:
  Companion Proceedings, {ICSE} Companion 2021, Madrid, Spain, May 25-28,
  2021}}. \bibinfo{publisher}{{IEEE}}, \bibinfo{pages}{96--98}.
\newblock


\bibitem[\protect\citeauthoryear{Berend, Xie, Ma, Zhou, Liu, Xu, and
  Zhao}{Berend et~al\mbox{.}}{2020}]%
        {berend_cats_2020}
\bibfield{author}{\bibinfo{person}{David Berend}, \bibinfo{person}{Xiaofei
  Xie}, \bibinfo{person}{Lei Ma}, \bibinfo{person}{Lingjun Zhou},
  \bibinfo{person}{Yang Liu}, \bibinfo{person}{Chi Xu}, {and}
  \bibinfo{person}{Jianjun Zhao}.} \bibinfo{year}{2020}\natexlab{}.
\newblock \showarticletitle{Cats {Are} {Not} {Fish}: {Deep} {Learning}
  {Testing} {Calls} for out-of-{Distribution} {Awareness}}. In
  \bibinfo{booktitle}{\emph{Proc. of the 35th {IEEE}/{ACM} {Int.} {Conference}
  on {Automated} {Software} {Engineering}}}
  \emph{(\bibinfo{series}{{ASE}'20})}. \bibinfo{publisher}{ACM},
  \bibinfo{address}{New York, NY, USA}, \bibinfo{pages}{1041--1052}.
\newblock
\showISBNx{978-1-4503-6768-4}
\urldef\tempurl%
\url{https://doi.org/10.1145/3324884.3416609}
\showDOI{\tempurl}


\bibitem[\protect\citeauthoryear{Byun and Rayadurgam}{Byun and
  Rayadurgam}{2020}]%
        {DBLP:conf/icse/ByunR20a}
\bibfield{author}{\bibinfo{person}{Taejoon Byun} {and} \bibinfo{person}{Sanjai
  Rayadurgam}.} \bibinfo{year}{2020}\natexlab{}.
\newblock \showarticletitle{Manifold-based Test Generation for Image
  Classifiers}. In \bibinfo{booktitle}{\emph{{ICSE} '20: 42nd Int. Conference
  on Software Engineering, Workshops}}. \bibinfo{publisher}{{ACM}},
  \bibinfo{pages}{221}.
\newblock
\urldef\tempurl%
\url{https://doi.org/10.1145/3387940.3391460}
\showDOI{\tempurl}


\bibitem[\protect\citeauthoryear{Byun, Vijayakumar, Rayadurgam, and Cofer}{Byun
  et~al\mbox{.}}{2020}]%
        {byun_manifold_based_2020}
\bibfield{author}{\bibinfo{person}{Taejoon Byun}, \bibinfo{person}{Abhishek
  Vijayakumar}, \bibinfo{person}{Sanjai Rayadurgam}, {and}
  \bibinfo{person}{Darren Cofer}.} \bibinfo{year}{2020}\natexlab{}.
\newblock \showarticletitle{Manifold-based {Test} {Generation} for {Image}
  {Classifiers}}. In \bibinfo{booktitle}{\emph{Int. {Conf}. {On} {Artificial}
  {Intelligence} {Testing} ({AITest})}}. \bibinfo{publisher}{IEEE},
  \bibinfo{address}{Oxford, UK}, \bibinfo{pages}{15--22}.
\newblock


\bibitem[\protect\citeauthoryear{Croce and Hein}{Croce and Hein}{2020}]%
        {DBLP:conf/icml/Croce020a}
\bibfield{author}{\bibinfo{person}{Francesco Croce} {and}
  \bibinfo{person}{Matthias Hein}.} \bibinfo{year}{2020}\natexlab{}.
\newblock \showarticletitle{Reliable evaluation of adversarial robustness with
  an ensemble of diverse parameter-free attacks}. In
  \bibinfo{booktitle}{\emph{Proc. of the 37th Int. Conf. on Machine Learning
  (ICML'20)}}, Vol.~\bibinfo{volume}{119}. \bibinfo{publisher}{PMLR},
  \bibinfo{pages}{2206--2216}.
\newblock


\bibitem[\protect\citeauthoryear{Deb, Pratap, Agarwal, and Meyarivan}{Deb
  et~al\mbox{.}}{2002}]%
        {deb2002fast}
\bibfield{author}{\bibinfo{person}{Kalyanmoy Deb}, \bibinfo{person}{Amrit
  Pratap}, \bibinfo{person}{Sameer Agarwal}, {and} \bibinfo{person}{TAMT
  Meyarivan}.} \bibinfo{year}{2002}\natexlab{}.
\newblock \showarticletitle{A fast and elitist multiobjective genetic
  algorithm: NSGA-II}.
\newblock \bibinfo{journal}{\emph{IEEE transactions on evolutionary
  computation}} \bibinfo{volume}{6}, \bibinfo{number}{2}
  (\bibinfo{year}{2002}), \bibinfo{pages}{182--197}.
\newblock


\bibitem[\protect\citeauthoryear{Dola, Dwyer, and Soffa}{Dola
  et~al\mbox{.}}{2021}]%
        {dola_distribution_aware_2021}
\bibfield{author}{\bibinfo{person}{Swaroopa Dola}, \bibinfo{person}{Matthew~B.
  Dwyer}, {and} \bibinfo{person}{Mary~Lou Soffa}.}
  \bibinfo{year}{2021}\natexlab{}.
\newblock \showarticletitle{Distribution-{Aware} {Testing} of {Neural}
  {Networks} {Using} {Generative} {Models}}. In
  \bibinfo{booktitle}{\emph{{IEEE}/{ACM} 43rd {Int.} {Conference} on {Software}
  {Engineering}}} \emph{(\bibinfo{series}{{ICSE}'21})}.
  \bibinfo{publisher}{IEEE}, \bibinfo{address}{Madrid, Spain},
  \bibinfo{pages}{226--237}.
\newblock


\bibitem[\protect\citeauthoryear{Dong, Huang, Bharti, Cox, Banks, Wang, Zhao,
  Schewe, and Huang}{Dong et~al\mbox{.}}{2023}]%
        {dong_reliability_2023}
\bibfield{author}{\bibinfo{person}{Yi Dong}, \bibinfo{person}{Wei Huang},
  \bibinfo{person}{Vibhav Bharti}, \bibinfo{person}{Victoria Cox},
  \bibinfo{person}{Alec Banks}, \bibinfo{person}{Sen Wang},
  \bibinfo{person}{Xingyu Zhao}, \bibinfo{person}{Sven Schewe}, {and}
  \bibinfo{person}{Xiaowei Huang}.} \bibinfo{year}{2023}\natexlab{}.
\newblock \showarticletitle{Reliability {Assessment} and {Safety} {Arguments}
  for {Machine} {Learning} {Components} in {System} {Assurance}}.
\newblock \bibinfo{journal}{\emph{ACM Trans. Embed. Comput. Syst.}}
  \bibinfo{volume}{22}, \bibinfo{number}{3} (\bibinfo{year}{2023}).
\newblock


\bibitem[\protect\citeauthoryear{Du, Xie, Li, Ma, Liu, and Zhao}{Du
  et~al\mbox{.}}{2019}]%
        {du2019deepstellar}
\bibfield{author}{\bibinfo{person}{Xiaoning Du}, \bibinfo{person}{Xiaofei Xie},
  \bibinfo{person}{Yi Li}, \bibinfo{person}{Lei Ma}, \bibinfo{person}{Yang
  Liu}, {and} \bibinfo{person}{Jianjun Zhao}.} \bibinfo{year}{2019}\natexlab{}.
\newblock \showarticletitle{Deepstellar: Model-based quantitative analysis of
  stateful deep learning systems}. In \bibinfo{booktitle}{\emph{Proc. of the
  27th ACM Joint Meeting on European Software Engineering Conference and
  Symposium on the Foundations of Software Engineering}}.
  \bibinfo{pages}{477--487}.
\newblock


\bibitem[\protect\citeauthoryear{Dunn, Hanu, Pouget, Kroening, and Melham}{Dunn
  et~al\mbox{.}}{2020}]%
        {dunn2020evaluating}
\bibfield{author}{\bibinfo{person}{Isaac Dunn}, \bibinfo{person}{Laura Hanu},
  \bibinfo{person}{Hadrien Pouget}, \bibinfo{person}{Daniel Kroening}, {and}
  \bibinfo{person}{Tom Melham}.} \bibinfo{year}{2020}\natexlab{}.
\newblock \showarticletitle{Evaluating robustness to context-sensitive feature
  perturbations of different granularities}.
\newblock \bibinfo{journal}{\emph{arXiv preprint arXiv:2001.11055}}
  (\bibinfo{year}{2020}).
\newblock


\bibitem[\protect\citeauthoryear{Dunn, Pouget, Kroening, and Melham}{Dunn
  et~al\mbox{.}}{2021}]%
        {dunn2021exposing}
\bibfield{author}{\bibinfo{person}{Isaac Dunn}, \bibinfo{person}{Hadrien
  Pouget}, \bibinfo{person}{Daniel Kroening}, {and} \bibinfo{person}{Tom
  Melham}.} \bibinfo{year}{2021}\natexlab{}.
\newblock \showarticletitle{Exposing Previously Undetectable Faults in Deep
  Neural Networks}. In \bibinfo{booktitle}{\emph{ACM SIGSOFT Int. Symposium on
  Software Testing and Analysis (ISSTA'21)}}.
\newblock
\newblock
\shownote{in press.}


\bibitem[\protect\citeauthoryear{Gonzalez and Woods}{Gonzalez and
  Woods}{1992}]%
        {rafael1992gonzalez}
\bibfield{author}{\bibinfo{person}{Rafael~C. Gonzalez} {and}
  \bibinfo{person}{Richard~E. Woods}.} \bibinfo{year}{1992}\natexlab{}.
\newblock \bibinfo{booktitle}{\emph{Digital image processing}}.
\newblock 793 pages.
\newblock


\bibitem[\protect\citeauthoryear{Goodfellow, Shlens, and Szegedy}{Goodfellow
  et~al\mbox{.}}{2015}]%
        {DBLP:journals/corr/GoodfellowSS14}
\bibfield{author}{\bibinfo{person}{Ian~J. Goodfellow},
  \bibinfo{person}{Jonathon Shlens}, {and} \bibinfo{person}{Christian
  Szegedy}.} \bibinfo{year}{2015}\natexlab{}.
\newblock \showarticletitle{Explaining and Harnessing Adversarial Examples}. In
  \bibinfo{booktitle}{\emph{3rd Int. Conference on Learning Representations,
  {ICLR} 2015, San Diego, CA, USA, Conference Track Proceedings}}.
\newblock


\bibitem[\protect\citeauthoryear{Harel-Canada, Wang, Gulzar, Gu, and
  Kim}{Harel-Canada et~al\mbox{.}}{2020}]%
        {harel_canada_is_2020}
\bibfield{author}{\bibinfo{person}{Fabrice Harel-Canada},
  \bibinfo{person}{Lingxiao Wang}, \bibinfo{person}{Muhammad~Ali Gulzar},
  \bibinfo{person}{Quanquan Gu}, {and} \bibinfo{person}{Miryung Kim}.}
  \bibinfo{year}{2020}\natexlab{}.
\newblock \showarticletitle{Is {Neuron} {Coverage} a {Meaningful} {Measure} for
  {Testing} {Deep} {Neural} {Networks}?}. In \bibinfo{booktitle}{\emph{Proc. of
  the 28th {ACM} {Joint} {Meeting} on {European} {Software} {Engineering}
  {Conference} and {Symposium} on the {Foundations} of {Software}
  {Engineering}}}. \bibinfo{publisher}{ACM}, \bibinfo{pages}{851--862}.
\newblock


\bibitem[\protect\citeauthoryear{Heusel, Ramsauer, Unterthiner, Nessler, and
  Hochreiter}{Heusel et~al\mbox{.}}{2017}]%
        {DBLP:conf/nips/HeuselRUNH17}
\bibfield{author}{\bibinfo{person}{Martin Heusel}, \bibinfo{person}{Hubert
  Ramsauer}, \bibinfo{person}{Thomas Unterthiner}, \bibinfo{person}{Bernhard
  Nessler}, {and} \bibinfo{person}{Sepp Hochreiter}.}
  \bibinfo{year}{2017}\natexlab{}.
\newblock \showarticletitle{GANs Trained by a Two Time-Scale Update Rule
  Converge to a Local Nash Equilibrium}. In \bibinfo{booktitle}{\emph{Advances
  in Neural Information Processing Systems 30: Annual Conference on Neural
  Information Processing Systems 2017}}. \bibinfo{pages}{6626--6637}.
\newblock


\bibitem[\protect\citeauthoryear{Hosseini and Poovendran}{Hosseini and
  Poovendran}{2018}]%
        {DBLP:conf/cvpr/HosseiniP18}
\bibfield{author}{\bibinfo{person}{Hossein Hosseini} {and}
  \bibinfo{person}{Radha Poovendran}.} \bibinfo{year}{2018}\natexlab{}.
\newblock \showarticletitle{Semantic Adversarial Examples}. In
  \bibinfo{booktitle}{\emph{2018 {IEEE} Conference on Computer Vision and
  Pattern Recognition Workshops, {CVPR} Workshops 2018}}.
  \bibinfo{pages}{1614--1619}.
\newblock


\bibitem[\protect\citeauthoryear{Huang, Sun, Zhao, Sharp, Ruan, Meng, and
  Huang}{Huang et~al\mbox{.}}{2022}]%
        {huang2021coverage}
\bibfield{author}{\bibinfo{person}{Wei Huang}, \bibinfo{person}{Youcheng Sun},
  \bibinfo{person}{Xingyu Zhao}, \bibinfo{person}{James Sharp},
  \bibinfo{person}{Wenjie Ruan}, \bibinfo{person}{Jie Meng}, {and}
  \bibinfo{person}{Xiaowei Huang}.} \bibinfo{year}{2022}\natexlab{}.
\newblock \showarticletitle{Coverage-{Guided} {Testing} for {Recurrent}
  {Neural} {Networks}}.
\newblock \bibinfo{journal}{\emph{IEEE Transactions on Reliability}}
  \bibinfo{volume}{71}, \bibinfo{number}{3} (\bibinfo{year}{2022}),
  \bibinfo{pages}{1191--1206}.
\newblock


\bibitem[\protect\citeauthoryear{Huang, Zhao, Jin, and Huang}{Huang
  et~al\mbox{.}}{2023}]%
        {huang2023safari}
\bibfield{author}{\bibinfo{person}{Wei Huang}, \bibinfo{person}{Xingyu Zhao},
  \bibinfo{person}{Gaojie Jin}, {and} \bibinfo{person}{Xiaowei Huang}.}
  \bibinfo{year}{2023}\natexlab{}.
\newblock \showarticletitle{{SAFARI}: Versatile and Efficient Evaluations for
  Robustness of Interpretability}. In \bibinfo{booktitle}{\emph{IEEE/CVF Int.
  Conf. on Computer Vision (ICCV'23)}}.
\newblock


\bibitem[\protect\citeauthoryear{Huang, Kroening, Ruan, and et~al}{Huang
  et~al\mbox{.}}{2020}]%
        {huang_survey_2020}
\bibfield{author}{\bibinfo{person}{Xiaowei Huang}, \bibinfo{person}{Daniel
  Kroening}, \bibinfo{person}{Wenjie Ruan}, {and} \bibinfo{person}{et al}.}
  \bibinfo{year}{2020}\natexlab{}.
\newblock \showarticletitle{A survey of safety and trustworthiness of deep
  neural networks: {Verification}, testing, adversarial attack and defence, and
  interpretability}.
\newblock \bibinfo{journal}{\emph{Computer Science Review}}
  \bibinfo{volume}{37} (\bibinfo{year}{2020}), \bibinfo{pages}{100270}.
\newblock
\showISSN{1574-0137}


\bibitem[\protect\citeauthoryear{Huang, Kwiatkowska, Wang, and Wu}{Huang
  et~al\mbox{.}}{2017}]%
        {huang_safety_2017}
\bibfield{author}{\bibinfo{person}{Xiaowei Huang}, \bibinfo{person}{Marta
  Kwiatkowska}, \bibinfo{person}{Sen Wang}, {and} \bibinfo{person}{Min Wu}.}
  \bibinfo{year}{2017}\natexlab{}.
\newblock \showarticletitle{Safety verification of deep neural networks}. In
  \bibinfo{booktitle}{\emph{Computer {Aided} {Verification}}}
  \emph{(\bibinfo{series}{{LNCS}}, Vol.~\bibinfo{volume}{10426})}.
  \bibinfo{publisher}{Springer International Publishing},
  \bibinfo{address}{Cham}, \bibinfo{pages}{3--29}.
\newblock


\bibitem[\protect\citeauthoryear{Jeddi, Shafiee, and Wong}{Jeddi
  et~al\mbox{.}}{2021}]%
        {jeddi_simple_2021}
\bibfield{author}{\bibinfo{person}{Ahmadreza Jeddi},
  \bibinfo{person}{Mohammad~Javad Shafiee}, {and} \bibinfo{person}{Alexander
  Wong}.} \bibinfo{year}{2021}\natexlab{}.
\newblock \showarticletitle{A simple fine-tuning is all you need: {Towards}
  robust deep learning via adversarial fine-tuning}. In
  \bibinfo{booktitle}{\emph{Workshop on {Adversarial} {Machine} {Learning} in
  {Real}-{World} {Computer} {Vision} {Systems} and {Online} {Challenges}
  ({AML}-{CV}) @ {CVPR}'21}}. \bibinfo{pages}{1--5}.
\newblock


\bibitem[\protect\citeauthoryear{Kang, Feldt, and Yoo}{Kang
  et~al\mbox{.}}{2020}]%
        {kang2020sinvad}
\bibfield{author}{\bibinfo{person}{Sungmin Kang}, \bibinfo{person}{Robert
  Feldt}, {and} \bibinfo{person}{Shin Yoo}.} \bibinfo{year}{2020}\natexlab{}.
\newblock \showarticletitle{Sinvad: Search-based image space navigation for dnn
  image classifier test input generation}. In \bibinfo{booktitle}{\emph{Proc.
  of the IEEE/ACM 42nd Int. Conf. on Software Engineering Workshops}}.
  \bibinfo{pages}{521--528}.
\newblock


\bibitem[\protect\citeauthoryear{Konak, Coit, and Smith}{Konak
  et~al\mbox{.}}{2006}]%
        {konak2006multi}
\bibfield{author}{\bibinfo{person}{Abdullah Konak}, \bibinfo{person}{David~W
  Coit}, {and} \bibinfo{person}{Alice~E Smith}.}
  \bibinfo{year}{2006}\natexlab{}.
\newblock \showarticletitle{Multi-objective optimization using genetic
  algorithms: A tutorial}.
\newblock \bibinfo{journal}{\emph{Reliability engineering \& system safety}}
  \bibinfo{volume}{91}, \bibinfo{number}{9} (\bibinfo{year}{2006}),
  \bibinfo{pages}{992--1007}.
\newblock


\bibitem[\protect\citeauthoryear{Lane, Bisset, Buckingham, Pegman, and
  Prescott}{Lane et~al\mbox{.}}{2016}]%
        {lane_new_2016}
\bibfield{author}{\bibinfo{person}{David Lane}, \bibinfo{person}{David Bisset},
  \bibinfo{person}{Rob Buckingham}, \bibinfo{person}{Geoff Pegman}, {and}
  \bibinfo{person}{Tony Prescott}.} \bibinfo{year}{2016}\natexlab{}.
\newblock \bibinfo{booktitle}{\emph{New foresight review on robotics and
  autonomous systems}}.
\newblock \bibinfo{type}{{T}echnical {R}eport} No. 2016.1.
  \bibinfo{institution}{LRF}. \bibinfo{pages}{65} pages.
\newblock


\bibitem[\protect\citeauthoryear{Lipowski and Lipowska}{Lipowski and
  Lipowska}{2012}]%
        {lipowski2012roulette}
\bibfield{author}{\bibinfo{person}{Adam Lipowski} {and} \bibinfo{person}{Dorota
  Lipowska}.} \bibinfo{year}{2012}\natexlab{}.
\newblock \showarticletitle{Roulette-wheel selection via stochastic
  acceptance}.
\newblock \bibinfo{journal}{\emph{Physica A: Statistical Mechanics and its
  Applications}} \bibinfo{volume}{391}, \bibinfo{number}{6}
  (\bibinfo{year}{2012}), \bibinfo{pages}{2193--2196}.
\newblock


\bibitem[\protect\citeauthoryear{Liu, Lafferty, and Wasserman}{Liu
  et~al\mbox{.}}{2007}]%
        {liu2007sparse}
\bibfield{author}{\bibinfo{person}{Han Liu}, \bibinfo{person}{John Lafferty},
  {and} \bibinfo{person}{Larry Wasserman}.} \bibinfo{year}{2007}\natexlab{}.
\newblock \showarticletitle{Sparse nonparametric density estimation in high
  dimensions using the rodeo}. In \bibinfo{booktitle}{\emph{Artificial
  Intelligence and Statistics}}. PMLR, \bibinfo{pages}{283--290}.
\newblock


\bibitem[\protect\citeauthoryear{Liu, Jun, Li, and Heer}{Liu
  et~al\mbox{.}}{2019}]%
        {liu2019latent}
\bibfield{author}{\bibinfo{person}{Yang Liu}, \bibinfo{person}{Eunice Jun},
  \bibinfo{person}{Qisheng Li}, {and} \bibinfo{person}{Jeffrey Heer}.}
  \bibinfo{year}{2019}\natexlab{}.
\newblock \showarticletitle{Latent space cartography: Visual analysis of vector
  space embeddings}. In \bibinfo{booktitle}{\emph{Computer graphics forum}},
  Vol.~\bibinfo{volume}{38}. Wiley Online Library, \bibinfo{pages}{67--78}.
\newblock


\bibitem[\protect\citeauthoryear{Lokerse, Veelenturf, and Beltman}{Lokerse
  et~al\mbox{.}}{1995}]%
        {DBLP:conf/snn/LokerseVB95}
\bibfield{author}{\bibinfo{person}{S.~H. Lokerse}, \bibinfo{person}{L.~P.~J.
  Veelenturf}, {and} \bibinfo{person}{J.~G. Beltman}.}
  \bibinfo{year}{1995}\natexlab{}.
\newblock \showarticletitle{Density Estimation Using {SOFM} and Adaptive
  Kernels}. In \bibinfo{booktitle}{\emph{Neural Networks: Artificial
  Intelligence and Industrial Applications - Proceedings of the Third Annual
  {SNN} Symposium on Neural Networks, Nijmegen, The Netherlands, September
  14-15, 1995}}. \bibinfo{publisher}{Springer}, \bibinfo{pages}{203--206}.
\newblock


\bibitem[\protect\citeauthoryear{Ma, Juefei-Xu, Zhang, Sun, Xue, Li, Chen, Su,
  Li, Liu, et~al\mbox{.}}{Ma et~al\mbox{.}}{2018}]%
        {ma2018deepgauge}
\bibfield{author}{\bibinfo{person}{Lei Ma}, \bibinfo{person}{Felix Juefei-Xu},
  \bibinfo{person}{Fuyuan Zhang}, \bibinfo{person}{Jiyuan Sun},
  \bibinfo{person}{Minhui Xue}, \bibinfo{person}{Bo Li},
  \bibinfo{person}{Chunyang Chen}, \bibinfo{person}{Ting Su},
  \bibinfo{person}{Li Li}, \bibinfo{person}{Yang Liu}, {et~al\mbox{.}}}
  \bibinfo{year}{2018}\natexlab{}.
\newblock \showarticletitle{Deepgauge: Multi-granularity testing criteria for
  deep learning systems}. In \bibinfo{booktitle}{\emph{Proce. of the 33rd
  ACM/IEEE Int. Conference on Automated Software Engineering (ASE'18)}}.
  \bibinfo{pages}{120--131}.
\newblock


\bibitem[\protect\citeauthoryear{Madry, Makelov, Schmidt, Tsipras, and
  Vladu}{Madry et~al\mbox{.}}{2018}]%
        {DBLP:conf/iclr/MadryMSTV18}
\bibfield{author}{\bibinfo{person}{Aleksander Madry},
  \bibinfo{person}{Aleksandar Makelov}, \bibinfo{person}{Ludwig Schmidt},
  \bibinfo{person}{Dimitris Tsipras}, {and} \bibinfo{person}{Adrian Vladu}.}
  \bibinfo{year}{2018}\natexlab{}.
\newblock \showarticletitle{Towards Deep Learning Models Resistant to
  Adversarial Attacks}. In \bibinfo{booktitle}{\emph{6th Int. Conference on
  Learning Representations, {ICLR} 2018, Vancouver, BC, Canada, April 30 - May
  3, 2018, Conference Track Proceedings}}. \bibinfo{publisher}{OpenReview.net}.
\newblock


\bibitem[\protect\citeauthoryear{Pedregosa, Varoquaux, Gramfort, Michel,
  Thirion, Grisel, Blondel, Prettenhofer, Weiss, Dubourg, Vanderplas, Passos,
  Cournapeau, Brucher, Perrot, and Duchesnay}{Pedregosa et~al\mbox{.}}{2011}]%
        {scikit-learn}
\bibfield{author}{\bibinfo{person}{F. Pedregosa}, \bibinfo{person}{G.
  Varoquaux}, \bibinfo{person}{A. Gramfort}, \bibinfo{person}{V. Michel},
  \bibinfo{person}{B. Thirion}, \bibinfo{person}{O. Grisel},
  \bibinfo{person}{M. Blondel}, \bibinfo{person}{P. Prettenhofer},
  \bibinfo{person}{R. Weiss}, \bibinfo{person}{V. Dubourg}, \bibinfo{person}{J.
  Vanderplas}, \bibinfo{person}{A. Passos}, \bibinfo{person}{D. Cournapeau},
  \bibinfo{person}{M. Brucher}, \bibinfo{person}{M. Perrot}, {and}
  \bibinfo{person}{E. Duchesnay}.} \bibinfo{year}{2011}\natexlab{}.
\newblock \showarticletitle{Scikit-learn: Machine Learning in {P}ython}.
\newblock \bibinfo{journal}{\emph{Journal of Machine Learning Research}}
  \bibinfo{volume}{12} (\bibinfo{year}{2011}), \bibinfo{pages}{2825--2830}.
\newblock


\bibitem[\protect\citeauthoryear{Pei, Cao, Yang, and Jana}{Pei
  et~al\mbox{.}}{2017}]%
        {DBLP:conf/sosp/PeiCYJ17}
\bibfield{author}{\bibinfo{person}{Kexin Pei}, \bibinfo{person}{Yinzhi Cao},
  \bibinfo{person}{Junfeng Yang}, {and} \bibinfo{person}{Suman Jana}.}
  \bibinfo{year}{2017}\natexlab{}.
\newblock \showarticletitle{DeepXplore: Automated Whitebox Testing of Deep
  Learning Systems}. In \bibinfo{booktitle}{\emph{Proceedings of the 26th
  Symposium on Operating Systems Principles}}. \bibinfo{publisher}{{ACM}},
  \bibinfo{pages}{1--18}.
\newblock


\bibitem[\protect\citeauthoryear{Riccio, Humbatova, Jahangirova, and
  Tonella}{Riccio et~al\mbox{.}}{2021}]%
        {riccio2021deepmetis}
\bibfield{author}{\bibinfo{person}{Vincenzo Riccio}, \bibinfo{person}{Nargiz
  Humbatova}, \bibinfo{person}{Gunel Jahangirova}, {and} \bibinfo{person}{Paolo
  Tonella}.} \bibinfo{year}{2021}\natexlab{}.
\newblock \showarticletitle{Deepmetis: Augmenting a deep learning test set to
  increase its mutation score}. In \bibinfo{booktitle}{\emph{36th IEEE/ACM Int.
  Conf. on Automated Software Engineering}}. \bibinfo{pages}{355--367}.
\newblock


\bibitem[\protect\citeauthoryear{Riccio and Tonella}{Riccio and
  Tonella}{2020}]%
        {riccio2020model}
\bibfield{author}{\bibinfo{person}{Vincenzo Riccio} {and}
  \bibinfo{person}{Paolo Tonella}.} \bibinfo{year}{2020}\natexlab{}.
\newblock \showarticletitle{Model-based exploration of the frontier of
  behaviours for deep learning system testing}. In
  \bibinfo{booktitle}{\emph{Proceedings of the 28th ACM Joint Meeting on
  European Software Engineering Conference and Symposium on the Foundations of
  Software Engineering}}. \bibinfo{pages}{876--888}.
\newblock


\bibitem[\protect\citeauthoryear{Rosenberg and Hirschberg}{Rosenberg and
  Hirschberg}{2007}]%
        {DBLP:conf/emnlp/RosenbergH07}
\bibfield{author}{\bibinfo{person}{Andrew Rosenberg} {and}
  \bibinfo{person}{Julia Hirschberg}.} \bibinfo{year}{2007}\natexlab{}.
\newblock \showarticletitle{V-Measure: {A} Conditional Entropy-Based External
  Cluster Evaluation Measure}. In \bibinfo{booktitle}{\emph{EMNLP-CoNLL 2007,
  Proceedings of the 2007 Joint Conference on Empirical Methods in Natural
  Language Processing and Computational Natural Language Learning, June 28-30,
  2007, Prague, Czech Republic}}. \bibinfo{publisher}{{ACL}},
  \bibinfo{pages}{410--420}.
\newblock


\bibitem[\protect\citeauthoryear{Ruan, Huang, and Kwiatkowska}{Ruan
  et~al\mbox{.}}{2018}]%
        {ruan_reachability_2018}
\bibfield{author}{\bibinfo{person}{Wenjie Ruan}, \bibinfo{person}{Xiaowei
  Huang}, {and} \bibinfo{person}{Marta Kwiatkowska}.}
  \bibinfo{year}{2018}\natexlab{}.
\newblock \showarticletitle{Reachability {Analysis} of {Deep} {Neural}
  {Networks} with {Provable} {Guarantees}}. In
  \bibinfo{booktitle}{\emph{Proceedings of the {Twenty}-{Seventh} {Int.}
  {Joint} {Conference} on {Artificial} {Intelligence}, {IJCAI}-18}}.
  \bibinfo{publisher}{Int. Joint Conferences on Artificial Intelligence
  Organization}, \bibinfo{pages}{2651--2659}.
\newblock


\bibitem[\protect\citeauthoryear{Scott}{Scott}{1991}]%
        {scott1991feasibility}
\bibfield{author}{\bibinfo{person}{David~W Scott}.}
  \bibinfo{year}{1991}\natexlab{}.
\newblock \showarticletitle{Feasibility of multivariate density estimates}.
\newblock \bibinfo{journal}{\emph{Biometrika}} \bibinfo{volume}{78},
  \bibinfo{number}{1} (\bibinfo{year}{1991}), \bibinfo{pages}{197--205}.
\newblock


\bibitem[\protect\citeauthoryear{Sun, Huang, Kroening, Sharp, Hill, and
  Ashmore}{Sun et~al\mbox{.}}{2019}]%
        {DBLP:conf/icse/SunHKSHA19}
\bibfield{author}{\bibinfo{person}{Youcheng Sun}, \bibinfo{person}{Xiaowei
  Huang}, \bibinfo{person}{Daniel Kroening}, \bibinfo{person}{James Sharp},
  \bibinfo{person}{Matthew Hill}, {and} \bibinfo{person}{Rob Ashmore}.}
  \bibinfo{year}{2019}\natexlab{}.
\newblock \showarticletitle{DeepConcolic: testing and debugging deep neural
  networks}. In \bibinfo{booktitle}{\emph{Proceedings of the 41st Int.
  Conference on Software Engineering: Companion Proceedings, {ICSE} 2019,
  Montreal, QC, Canada, May 25-31, 2019}}. \bibinfo{publisher}{{IEEE} / {ACM}},
  \bibinfo{pages}{111--114}.
\newblock


\bibitem[\protect\citeauthoryear{Sun, Wu, Ruan, Huang, Kwiatkowska, and
  Kroening}{Sun et~al\mbox{.}}{2018}]%
        {sun2018concolic}
\bibfield{author}{\bibinfo{person}{Youcheng Sun}, \bibinfo{person}{Min Wu},
  \bibinfo{person}{Wenjie Ruan}, \bibinfo{person}{Xiaowei Huang},
  \bibinfo{person}{Marta Kwiatkowska}, {and} \bibinfo{person}{Daniel
  Kroening}.} \bibinfo{year}{2018}\natexlab{}.
\newblock \showarticletitle{Concolic testing for deep neural networks}. In
  \bibinfo{booktitle}{\emph{Proc. of the 33rd ACM/IEEE Int. Conf. on Automated
  Software Engineering}}. \bibinfo{pages}{109--119}.
\newblock


\bibitem[\protect\citeauthoryear{Szegedy, Vanhoucke, Ioffe, Shlens, and
  Wojna}{Szegedy et~al\mbox{.}}{2016}]%
        {szegedy2016rethinking}
\bibfield{author}{\bibinfo{person}{Christian Szegedy}, \bibinfo{person}{Vincent
  Vanhoucke}, \bibinfo{person}{Sergey Ioffe}, \bibinfo{person}{Jon Shlens},
  {and} \bibinfo{person}{Zbigniew Wojna}.} \bibinfo{year}{2016}\natexlab{}.
\newblock \showarticletitle{Rethinking the inception architecture for computer
  vision}. In \bibinfo{booktitle}{\emph{Proc. of the IEEE conference on
  computer vision and pattern recognition}}. \bibinfo{pages}{2818--2826}.
\newblock


\bibitem[\protect\citeauthoryear{Toledo, Shriver, Elbaum, and Dwyer}{Toledo
  et~al\mbox{.}}{2021}]%
        {toledodistribution}
\bibfield{author}{\bibinfo{person}{Felipe Toledo}, \bibinfo{person}{David
  Shriver}, \bibinfo{person}{Sebastian Elbaum}, {and}
  \bibinfo{person}{Matthew~B Dwyer}.} \bibinfo{year}{2021}\natexlab{}.
\newblock \showarticletitle{Distribution Models for Falsification and
  Verification of DNNs}. In \bibinfo{booktitle}{\emph{IEEE/ACM Int. Conf. on
  Automated Software Engineering (ASE'21)}}.
\newblock


\bibitem[\protect\citeauthoryear{Wang, Webb, and Rainforth}{Wang
  et~al\mbox{.}}{2021c}]%
        {wang2021statistically}
\bibfield{author}{\bibinfo{person}{Benjie Wang}, \bibinfo{person}{Stefan Webb},
  {and} \bibinfo{person}{Tom Rainforth}.} \bibinfo{year}{2021}\natexlab{c}.
\newblock \showarticletitle{Statistically robust neural network
  classification}. In \bibinfo{booktitle}{\emph{Uncertainty in Artificial
  Intelligence}}. PMLR, \bibinfo{pages}{1735--1745}.
\newblock


\bibitem[\protect\citeauthoryear{Wang, Chen, Sun, Ma, Wang, Sun, and
  Cheng}{Wang et~al\mbox{.}}{2021a}]%
        {wang2021robot}
\bibfield{author}{\bibinfo{person}{Jingyi Wang}, \bibinfo{person}{Jialuo Chen},
  \bibinfo{person}{Youcheng Sun}, \bibinfo{person}{Xingjun Ma},
  \bibinfo{person}{Dongxia Wang}, \bibinfo{person}{Jun Sun}, {and}
  \bibinfo{person}{Peng Cheng}.} \bibinfo{year}{2021}\natexlab{a}.
\newblock \showarticletitle{Robot: robustness-oriented testing for deep
  learning systems}. In \bibinfo{booktitle}{\emph{IEEE/ACM 43rd International
  Conference on Software Engineering}}. \bibinfo{pages}{300--311}.
\newblock


\bibitem[\protect\citeauthoryear{Wang, Chen, Sun, Ma, Wang, Sun, and
  Cheng}{Wang et~al\mbox{.}}{2021b}]%
        {wang_robot_2021}
\bibfield{author}{\bibinfo{person}{Jingyi Wang}, \bibinfo{person}{Jialuo Chen},
  \bibinfo{person}{Youcheng Sun}, \bibinfo{person}{Xingjun Ma},
  \bibinfo{person}{Dongxia Wang}, \bibinfo{person}{Jun Sun}, {and}
  \bibinfo{person}{Peng Cheng}.} \bibinfo{year}{2021}\natexlab{b}.
\newblock \showarticletitle{{RobOT}: {Robustness}-{Oriented} {Testing} for
  {Deep} {Learning} {Systems}}. In \bibinfo{booktitle}{\emph{{IEEE}/{ACM} 43rd
  {Int.} {Conf.} on {Softw.} {Engineering}}}. \bibinfo{pages}{300--311}.
\newblock


\bibitem[\protect\citeauthoryear{Wang, Ma, Bailey, Yi, Zhou, and Gu}{Wang
  et~al\mbox{.}}{2019}]%
        {DBLP:conf/icml/WangM0YZG19}
\bibfield{author}{\bibinfo{person}{Yisen Wang}, \bibinfo{person}{Xingjun Ma},
  \bibinfo{person}{James Bailey}, \bibinfo{person}{Jinfeng Yi},
  \bibinfo{person}{Bowen Zhou}, {and} \bibinfo{person}{Quanquan Gu}.}
  \bibinfo{year}{2019}\natexlab{}.
\newblock \showarticletitle{On the Convergence and Robustness of Adversarial
  Training}. In \bibinfo{booktitle}{\emph{Proc. of the 36th Int. Conf.on
  Machine Learning}}, Vol.~\bibinfo{volume}{97}. \bibinfo{publisher}{{PMLR}},
  \bibinfo{pages}{6586--6595}.
\newblock


\bibitem[\protect\citeauthoryear{Wang, Bovik, Sheikh, and Simoncelli}{Wang
  et~al\mbox{.}}{2004}]%
        {DBLP:journals/tip/WangBSS04}
\bibfield{author}{\bibinfo{person}{Zhou Wang}, \bibinfo{person}{Alan~C. Bovik},
  \bibinfo{person}{Hamid~R. Sheikh}, {and} \bibinfo{person}{Eero~P.
  Simoncelli}.} \bibinfo{year}{2004}\natexlab{}.
\newblock \showarticletitle{Image quality assessment: from error visibility to
  structural similarity}.
\newblock \bibinfo{journal}{\emph{{IEEE} Trans. Image Process.}}
  \bibinfo{volume}{13}, \bibinfo{number}{4} (\bibinfo{year}{2004}),
  \bibinfo{pages}{600--612}.
\newblock


\bibitem[\protect\citeauthoryear{Webb, Rainforth, Teh, and Kumar}{Webb
  et~al\mbox{.}}{2019}]%
        {webb_statistical_2019}
\bibfield{author}{\bibinfo{person}{Stefan Webb}, \bibinfo{person}{Tom
  Rainforth}, \bibinfo{person}{Yee~Whye Teh}, {and} \bibinfo{person}{M.~Pawan
  Kumar}.} \bibinfo{year}{2019}\natexlab{}.
\newblock \showarticletitle{A statistical approach to assessing neural network
  robustness}. In \bibinfo{booktitle}{\emph{ICLR'19}}. \bibinfo{address}{New
  Orleans, LA, USA}.
\newblock


\bibitem[\protect\citeauthoryear{Weiss and Tonella}{Weiss and Tonella}{2022}]%
        {weiss2022simple}
\bibfield{author}{\bibinfo{person}{Michael Weiss} {and} \bibinfo{person}{Paolo
  Tonella}.} \bibinfo{year}{2022}\natexlab{}.
\newblock \showarticletitle{Simple techniques work surprisingly well for neural
  network test prioritization and active learning (replicability study)}. In
  \bibinfo{booktitle}{\emph{Proc. of the 31st ACM SIGSOFT Int. Symp. on
  Software Testing and Analysis}}. \bibinfo{pages}{139--150}.
\newblock


\bibitem[\protect\citeauthoryear{Weng, Chen, Nguyen, Squillante, Boopathy,
  Oseledets, and Daniel}{Weng et~al\mbox{.}}{2019}]%
        {weng_proven_2019}
\bibfield{author}{\bibinfo{person}{Lily Weng}, \bibinfo{person}{Pin-Yu Chen},
  \bibinfo{person}{Lam Nguyen}, \bibinfo{person}{Mark Squillante},
  \bibinfo{person}{Akhilan Boopathy}, \bibinfo{person}{Ivan Oseledets}, {and}
  \bibinfo{person}{Luca Daniel}.} \bibinfo{year}{2019}\natexlab{}.
\newblock \showarticletitle{{PROVEN}: Verifying Robustness of Neural Networks
  with a Probabilistic Approach}. In \bibinfo{booktitle}{\emph{International
  Conference on Machine Learning (ICML'19)}}, Vol.~\bibinfo{volume}{97}.
  \bibinfo{publisher}{PMLR}, \bibinfo{pages}{6727--6736}.
\newblock


\bibitem[\protect\citeauthoryear{Weng, Zhang, Chen, Yi, Su, Gao, Hsieh, and
  Daniel}{Weng et~al\mbox{.}}{2018}]%
        {DBLP:conf/iclr/WengZCYSGHD18}
\bibfield{author}{\bibinfo{person}{Tsui{-}Wei Weng}, \bibinfo{person}{Huan
  Zhang}, \bibinfo{person}{Pin{-}Yu Chen}, \bibinfo{person}{Jinfeng Yi},
  \bibinfo{person}{Dong Su}, \bibinfo{person}{Yupeng Gao},
  \bibinfo{person}{Cho{-}Jui Hsieh}, {and} \bibinfo{person}{Luca Daniel}.}
  \bibinfo{year}{2018}\natexlab{}.
\newblock \showarticletitle{Evaluating the Robustness of Neural Networks: An
  Extreme Value Theory Approach}. In \bibinfo{booktitle}{\emph{6th Int.
  Conference on Learning Representations, {ICLR} 2018, Vancouver, BC, Canada,
  April 30 - May 3, 2018, Conference Track Proceedings}}.
  \bibinfo{publisher}{OpenReview.net}.
\newblock


\bibitem[\protect\citeauthoryear{Wu, Luo, Zhou, Xu, and Zhu}{Wu
  et~al\mbox{.}}{2021}]%
        {DBLP:conf/cec/WuLZXZ21}
\bibfield{author}{\bibinfo{person}{Chenwang Wu}, \bibinfo{person}{Wenjian Luo},
  \bibinfo{person}{Nan Zhou}, \bibinfo{person}{Peilan Xu}, {and}
  \bibinfo{person}{Tao Zhu}.} \bibinfo{year}{2021}\natexlab{}.
\newblock \showarticletitle{Genetic Algorithm with Multiple Fitness Functions
  for Generating Adversarial Examples}. In \bibinfo{booktitle}{\emph{{IEEE}
  Congress on Evolutionary Computation, {CEC} 2021, Krak{\'{o}}w, Poland, June
  28 - July 1, 2021}}. \bibinfo{publisher}{{IEEE}},
  \bibinfo{pages}{1792--1799}.
\newblock


\bibitem[\protect\citeauthoryear{Xie, Li, Wang, Ma, Guo, Juefei-Xu, and
  Liu}{Xie et~al\mbox{.}}{2022}]%
        {xie_npc_2022}
\bibfield{author}{\bibinfo{person}{Xiaofei Xie}, \bibinfo{person}{Tianlin Li},
  \bibinfo{person}{Jian Wang}, \bibinfo{person}{Lei Ma}, \bibinfo{person}{Qing
  Guo}, \bibinfo{person}{Felix Juefei-Xu}, {and} \bibinfo{person}{Yang Liu}.}
  \bibinfo{year}{2022}\natexlab{}.
\newblock \showarticletitle{{NPC}: {Neuron} {Path} {Coverage} via
  {Characterizing} {Decision} {Logic} of {Deep} {Neural} {Networks}}.
\newblock \bibinfo{journal}{\emph{ACM Trans. Softw. Eng. Methodol.}}
  \bibinfo{volume}{31}, \bibinfo{number}{3} (\bibinfo{year}{2022}).
\newblock
\showISSN{1049-331X}


\bibitem[\protect\citeauthoryear{Yan, Tao, Liu, Zhai, Ma, Xu, and Zhang}{Yan
  et~al\mbox{.}}{2020}]%
        {yan_correlations_2020}
\bibfield{author}{\bibinfo{person}{Shenao Yan}, \bibinfo{person}{Guanhong Tao},
  \bibinfo{person}{Xuwei Liu}, \bibinfo{person}{Juan Zhai},
  \bibinfo{person}{Shiqing Ma}, \bibinfo{person}{Lei Xu}, {and}
  \bibinfo{person}{Xiangyu Zhang}.} \bibinfo{year}{2020}\natexlab{}.
\newblock \showarticletitle{Correlations between {Deep} {Neural} {Network}
  {Model} {Coverage} {Criteria} and {Model} {Quality}}. In
  \bibinfo{booktitle}{\emph{The 28th {ACM} {Joint} {Meeting} on {European}
  {Software} {Engineering} {Conference} and {Symposium} on the {Foundations} of
  {Software} {Engineering}}} \emph{(\bibinfo{series}{{ESEC}/{FSE} 2020})}.
  \bibinfo{publisher}{ACM}, \bibinfo{pages}{775--787}.
\newblock


\bibitem[\protect\citeauthoryear{Yang, Rashtchian, Zhang, Salakhutdinov, and
  Chaudhuri}{Yang et~al\mbox{.}}{2020}]%
        {yang_closer_2020}
\bibfield{author}{\bibinfo{person}{Yao-Yuan Yang}, \bibinfo{person}{Cyrus
  Rashtchian}, \bibinfo{person}{Hongyang Zhang}, \bibinfo{person}{Russ~R
  Salakhutdinov}, {and} \bibinfo{person}{Kamalika Chaudhuri}.}
  \bibinfo{year}{2020}\natexlab{}.
\newblock \showarticletitle{A {Closer} {Look} at {Accuracy} vs. {Robustness}}.
  In \bibinfo{booktitle}{\emph{Advances in {Neural} {Information} {Processing}
  {Systems}}} \emph{(\bibinfo{series}{{NeurIPS}'20},
  Vol.~\bibinfo{volume}{33})},
  \bibfield{editor}{\bibinfo{person}{H.~Larochelle},
  \bibinfo{person}{M.~Ranzato}, \bibinfo{person}{R.~Hadsell},
  \bibinfo{person}{M.~F. Balcan}, {and} \bibinfo{person}{H.~Lin}} (Eds.).
  \bibinfo{publisher}{Curran Associates, Inc.}, \bibinfo{pages}{8588--8601}.
\newblock


\bibitem[\protect\citeauthoryear{Yu, Qi, Guo, Juefei-Xu, Xie, Ma, and Zhao}{Yu
  et~al\mbox{.}}{2022}]%
        {9508369}
\bibfield{author}{\bibinfo{person}{Bing Yu}, \bibinfo{person}{Hua Qi},
  \bibinfo{person}{Qing Guo}, \bibinfo{person}{Felix Juefei-Xu},
  \bibinfo{person}{Xiaofei Xie}, \bibinfo{person}{Lei Ma}, {and}
  \bibinfo{person}{Jianjun Zhao}.} \bibinfo{year}{2022}\natexlab{}.
\newblock \showarticletitle{DeepRepair: Style-Guided Repairing for Deep Neural
  Networks in the Real-World Operational Environment}.
\newblock \bibinfo{journal}{\emph{IEEE Transactions on Reliability}}
  \bibinfo{volume}{71}, \bibinfo{number}{4} (\bibinfo{year}{2022}),
  \bibinfo{pages}{1401--1416}.
\newblock


\bibitem[\protect\citeauthoryear{Zhang, Yu, Jiao, Xing, Ghaoui, and
  Jordan}{Zhang et~al\mbox{.}}{2019}]%
        {zhang2019theoretically}
\bibfield{author}{\bibinfo{person}{Hongyang Zhang}, \bibinfo{person}{Yaodong
  Yu}, \bibinfo{person}{Jiantao Jiao}, \bibinfo{person}{Eric Xing},
  \bibinfo{person}{Laurent~El Ghaoui}, {and} \bibinfo{person}{Michael Jordan}.}
  \bibinfo{year}{2019}\natexlab{}.
\newblock \showarticletitle{Theoretically Principled Trade-off between
  Robustness and Accuracy}. In \bibinfo{booktitle}{\emph{Proceedings of the
  36th International Conference on Machine Learning}},
  Vol.~\bibinfo{volume}{97}. \bibinfo{publisher}{PMLR},
  \bibinfo{pages}{7472--7482}.
\newblock


\bibitem[\protect\citeauthoryear{Zhang, Harman, Ma, and Liu}{Zhang
  et~al\mbox{.}}{2022}]%
        {zhang_machine_2020}
\bibfield{author}{\bibinfo{person}{Jie~M. Zhang}, \bibinfo{person}{Mark
  Harman}, \bibinfo{person}{Lei Ma}, {and} \bibinfo{person}{Yang Liu}.}
  \bibinfo{year}{2022}\natexlab{}.
\newblock \showarticletitle{Machine Learning Testing: Survey, Landscapes and
  Horizons}.
\newblock \bibinfo{journal}{\emph{IEEE Transactions on Software Engineering}}
  \bibinfo{volume}{48}, \bibinfo{number}{1} (\bibinfo{year}{2022}),
  \bibinfo{pages}{1--36}.
\newblock


\bibitem[\protect\citeauthoryear{Zhao, Banks, Sharp, Robu, Flynn, Fisher, and
  Huang}{Zhao et~al\mbox{.}}{2020}]%
        {zhao_safety_2020}
\bibfield{author}{\bibinfo{person}{Xingyu Zhao}, \bibinfo{person}{Alec Banks},
  \bibinfo{person}{James Sharp}, \bibinfo{person}{Valentin Robu},
  \bibinfo{person}{David Flynn}, \bibinfo{person}{Michael Fisher}, {and}
  \bibinfo{person}{Xiaowei Huang}.} \bibinfo{year}{2020}\natexlab{}.
\newblock \showarticletitle{A {Safety} {Framework} for {Critical} {Systems}
  {Utilising} {Deep} {Neural} {Networks}}. In
  \bibinfo{booktitle}{\emph{Computer {Safety}, {Reliability}, and {Security}}}
  \emph{(\bibinfo{series}{{LNCS}}, Vol.~\bibinfo{volume}{12234})}.
  \bibinfo{publisher}{Springer Int. Publishing}, \bibinfo{address}{Cham},
  \bibinfo{pages}{244--259}.
\newblock


\bibitem[\protect\citeauthoryear{Zhao, Huang, Banks, Cox, Flynn, Schewe, and
  Huang}{Zhao et~al\mbox{.}}{2021a}]%
        {zhao_assessing_2021}
\bibfield{author}{\bibinfo{person}{Xingyu Zhao}, \bibinfo{person}{Wei Huang},
  \bibinfo{person}{Alec Banks}, \bibinfo{person}{Victoria Cox},
  \bibinfo{person}{David Flynn}, \bibinfo{person}{Sven Schewe}, {and}
  \bibinfo{person}{Xiaowei Huang}.} \bibinfo{year}{2021}\natexlab{a}.
\newblock \showarticletitle{Assessing the {Reliability} of {Deep} {Learning}
  {Classifiers} {Through} {Robustness} {Evaluation} and {Operational}
  {Profiles}}. In \bibinfo{booktitle}{\emph{{AISafety}'21 {Workshop} at
  {IJCAI}'21}}, Vol.~\bibinfo{volume}{2916}.
\newblock


\bibitem[\protect\citeauthoryear{Zhao, Huang, Schewe, Dong, and Huang}{Zhao
  et~al\mbox{.}}{2021b}]%
        {zhao_detecting_2021}
\bibfield{author}{\bibinfo{person}{Xingyu Zhao}, \bibinfo{person}{Wei Huang},
  \bibinfo{person}{Sven Schewe}, \bibinfo{person}{Yi Dong}, {and}
  \bibinfo{person}{Xiaowei Huang}.} \bibinfo{year}{2021}\natexlab{b}.
\newblock \showarticletitle{Detecting Operational Adversarial Examples for
  Reliable Deep Learning}. In \bibinfo{booktitle}{\emph{51th Annual IEEE-IFIP
  Int. Conf. on Dependable Systems and Networks (DSN'21)}},
  Vol.~\bibinfo{volume}{Fast Abstract}.
\newblock


\bibitem[\protect\citeauthoryear{Zhao, Dua, and Singh}{Zhao
  et~al\mbox{.}}{2018}]%
        {DBLP:conf/iclr/ZhaoDS18}
\bibfield{author}{\bibinfo{person}{Zhengli Zhao}, \bibinfo{person}{Dheeru Dua},
  {and} \bibinfo{person}{Sameer Singh}.} \bibinfo{year}{2018}\natexlab{}.
\newblock \showarticletitle{Generating Natural Adversarial Examples}. In
  \bibinfo{booktitle}{\emph{6th Int. Conference on Learning Representations,
  {ICLR} 2018, Conference Track Proceedings}}.
  \bibinfo{publisher}{OpenReview.net}.
\newblock


\bibitem[\protect\citeauthoryear{Zhong, Liu, Zhao, and Li}{Zhong
  et~al\mbox{.}}{2020}]%
        {zhong2020generative}
\bibfield{author}{\bibinfo{person}{Yue Zhong}, \bibinfo{person}{Lizhuang Liu},
  \bibinfo{person}{Dan Zhao}, {and} \bibinfo{person}{Hongyang Li}.}
  \bibinfo{year}{2020}\natexlab{}.
\newblock \showarticletitle{A generative adversarial network for image
  denoising}.
\newblock \bibinfo{journal}{\emph{Multimedia Tools and Applications}}
  \bibinfo{volume}{79}, \bibinfo{number}{23} (\bibinfo{year}{2020}),
  \bibinfo{pages}{16517--16529}.
\newblock


\bibitem[\protect\citeauthoryear{Zhu, Bi, Liu, Ma, Li, and Wu}{Zhu
  et~al\mbox{.}}{2020}]%
        {zhu-etal-2020-batch}
\bibfield{author}{\bibinfo{person}{Qile Zhu}, \bibinfo{person}{Wei Bi},
  \bibinfo{person}{Xiaojiang Liu}, \bibinfo{person}{Xiyao Ma},
  \bibinfo{person}{Xiaolin Li}, {and} \bibinfo{person}{Dapeng Wu}.}
  \bibinfo{year}{2020}\natexlab{}.
\newblock \showarticletitle{A Batch Normalized Inference Network Keeps the {KL}
  Vanishing Away}. In \bibinfo{booktitle}{\emph{Proceedings of the 58th Annual
  Meeting of the Association for Computational Linguistics}}.
  \bibinfo{publisher}{Association for Computational Linguistics},
  \bibinfo{address}{Online}, \bibinfo{pages}{2636--2649}.
\newblock


\end{thebibliography}

\appendix

\end{document}